\documentclass[aps,prb,twocolumn,floatfix,showpacs,citeautoscript,superscriptaddress]{revtex4}

\usepackage{amsmath}
\usepackage{amssymb}
\usepackage{amsfonts}
\usepackage{graphicx,graphics}
\usepackage{bm}

\usepackage[normalem]{ulem}

\newcommand{\ket}[1]{\lvert #1\rangle}
\newcommand{\bra}[1]{\langle#1 \rvert}
\newcommand{\abs}[1]{\lvert #1 \rvert}
\newcommand{\expect}[1]{\langle #1\rangle}

\newcommand{\br}{\mathbf{r}}
\newcommand{\bk}{\mathbf{k}}
\newcommand{\bq}{\mathbf{q}}

\begin{document}

\title{Acoustic phonon limited mobility in two-dimensional semiconductors:
  Deformation potential and piezoelectric scattering in monolayer MoS$_2$ from
  first principles}
\author{Kristen Kaasbjerg}
\email{cosby@fys.ku.dk}
\affiliation{School of Chemistry, The Sackler Faculty of Exact Sciences, Tel Aviv
  University, Tel Aviv 69978, Israel}
\author{Kristian S. Thygesen}
\affiliation{Center for Nanostructured Graphene (CNG), Department of Micro- and Nanotechnology, DTU Nanotech, \\
  Technical University of Denmark, DK-2800 Kongens Lyngby, Denmark}
\affiliation{Center for Atomic-scale Materials Design (CAMD), Department of Physics, \\
  Technical University of Denmark, DK-2800 Kongens Lyngby, Denmark}
\author{Antti-Pekka Jauho}
\affiliation{Center for Nanostructured Graphene (CNG), Department of Micro- and Nanotechnology, DTU Nanotech, \\
  Technical University of Denmark, DK-2800 Kongens Lyngby, Denmark}
\date{\today}

\begin{abstract}
  We theoretically study the acoustic phonon limited mobility in $n$-doped
  two-dimensional MoS$_2$ for temperatures $T<100$~K and high carrier densities
  using the Boltzmann equation and first-principles calculations of the acoustic
  electron-phonon (el-ph) interaction. In combination with a continuum elastic
  model, analytic expressions and the coupling strengths for the deformation
  potential and piezoelectric interactions are established. We furthermore show
  that the deformation potential interaction has contributions from both normal
  and umklapp processes and that the latter contribution is only weakly affected
  by carrier screening. Consequently, the calculated mobilities show a
  transition from a high-temperature $\mu \sim T^{-1}$ behavior to a stronger
  $\mu \sim T^{-4}$ behavior in the low-temperature Bloch-Gr{\"u}neisen regime
  characteristic of unscreened deformation potential scattering. Intrinsic
  mobilities in excess of $10^5$~cm$^2$~V$^{-1}$~s$^{-1}$ are predicted at
  $T<10$~K and high carrier densities ($n \gtrsim 10^{11}$~cm$^{-2}$). At
  $100$~K, the mobility does not exceed $\sim 7\times
  10^3$~cm$^2$~V$^{-1}$~s$^{-1}$. Our findings provide new and important
  understanding of the acoustic el-ph interaction and its screening by free
  carriers, and is of high relevance for the understanding of acoustic
  phonon limited mobilities in general.
\end{abstract}

\maketitle

\section{Introduction}

Two-dimensional (2D) atomic crystals~\cite{Geim:2D} such as
graphene~\cite{Geim:Graphene,RMP:Graphene,Sarma:RMP} are promising candidates
for future electronic applications. Monolayers of semiconducting transition
metal dichalcogenides (MX$_2$) constitute a new family of 2D
materials~\cite{Strano:NNanoReview, Zhang:NChemReview} which have interesting
electronic and optical
properties~\cite{Bucher:High,Heinz:ThinMoS2,Wang:Emerging,Kis:MoS2Transistor,Schuller:Photocarrier,Iwasa:Ambipolar,Yao:SpinValley,Feng:ValleySelective,Heinz:Control,Cui:ValleyPolarization}.
In conjunction with the excellent gate control inherent to atomically thin
materials their finite gap makes them desirable materials for various electronic
applications. However, in spite of the recent progress in sample fabrication and
transport measurements on gated single to few-layer
samples~\cite{Geim:2D,Fuhrer:Ultrathin,Pati:Analogues,Kis:MoS2Transistor,Iwasa:Ambipolar,Li:LargeArea,Peide:Dual,Kim:HighMobility,Tutuc:FieldEffect},
little is so far known about the intrinsic carrier properties such as factors
limiting the achievable mobilities.

Experimentally, monolayer MoS$_2$ has been demonstrated to be a direct-gap
semiconductor with a band gap of $\sim$1.8~eV~\cite{Heinz:ThinMoS2} and a
room-temperature mobility in $n$-type samples ranging from $\sim$1 to
$\sim$200~cm$^2$~V$^{-1}$~s$^{-1}$ depending on the device
structure~\cite{Geim:2D,Fuhrer:Ultrathin,Pati:Analogues,Kis:MoS2Transistor,Hone:Measurement,Kis:Reply}.
The highest values have been obtained in top-gated samples with a high-$\kappa$
gate dielectric~\cite{Kis:MoS2Transistor,Kis:Reply}, indicating that impurity
scattering can be strongly suppressed by dielectric
engineering~\cite{Konar:Engineering}, and mobilities close to our theoretically
predicted intrinsic phonon limited mobility of
$\sim410$~cm$^2$~V$^{-1}$~s$^{-1}$ can be achieved~\cite{Kaasbjerg:MoS2}. Other
theoretical studies have addressed different issues related to the performance
of monolayer MoS$_2$ transistors~\cite{Salahuddin:HowGood,Tomanek:Designing}.

At low temperatures where optical phonon scattering is suppressed, scattering by
acoustic phonons can be expected to become an important limiting factor for the
mobility of the two-dimensional electron gas (2DEG) confined to the atomic layer
of an extrinsic 2D semiconductor. This is the case in conventional
heterostructure-based 2DEGs where impurity and acoustic phonon scattering are
dominating scattering mechanisms at low temperatures~\cite{Sarma:100}. In
contrast to impurity scattering which can be suppressed by e.g. dielectric
engineering, scattering by acoustic phonons is intrinsic to the semiconductor
and cannot be eliminated. The intrinsic mobility determined by acoustic phonon
scattering alone therefore provides an important upper limit for the achievable
mobilities.

\begin{figure}[!b]
  \centering
  \begin{minipage}{1.0\linewidth}
    \hspace{.4cm}
    \includegraphics[width=0.9\linewidth]{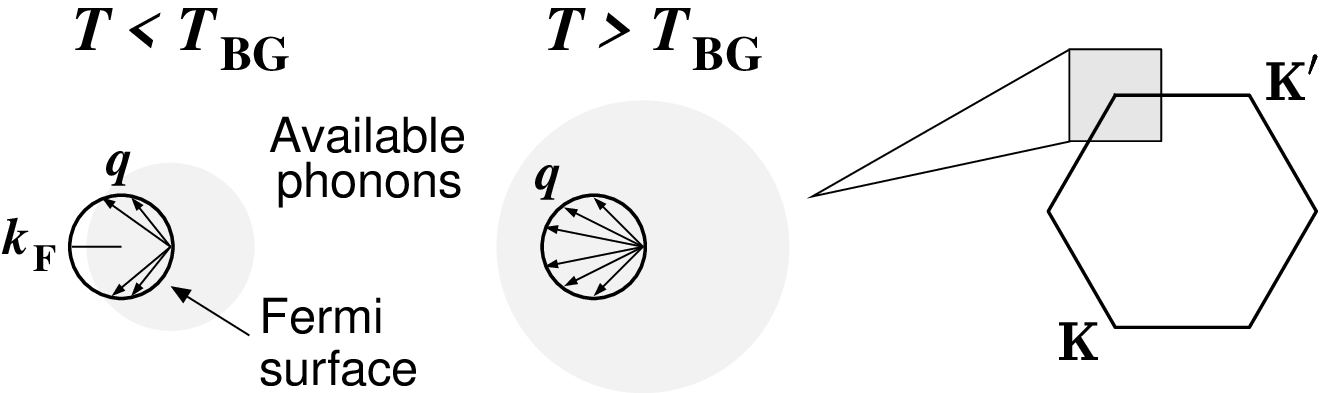}
    \vspace{0.6cm}
  \end{minipage}
  \includegraphics[width=0.9\linewidth]{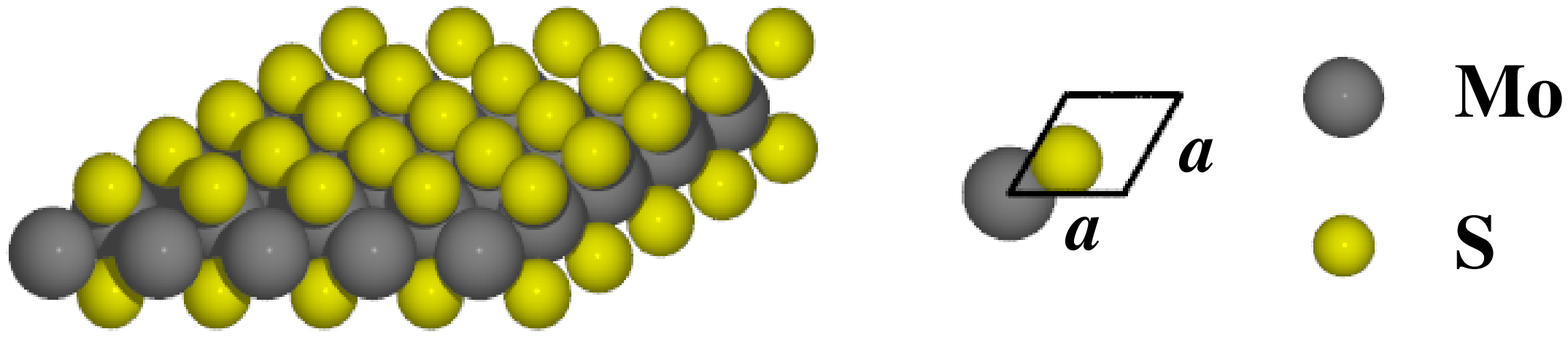}
  \caption{(Color online) Top: Illustration of acoustic phonon scattering in
    the $K,K'$ valleys of the Brillouin zone showing the phase space available
    for scattering below and above the Bloch-Gr{\"u}neisen temperature
    $T_\text{BG}$. The part of the hexagonal Brillouin zone marked by the
    gray-shaded box indicates the plotting range for the contour plots in
    Fig.~\ref{fig:M}. Bottom: Lattice and primitive unit cell of 2D hexagonal
    MoS$_2$.}
  \label{fig:mos2}
\end{figure}

In the low-temperature regime, acoustic phonon-dominated transport manifests
itself in a strong change in the temperature dependence of the carrier mobility
once the temperature is lowered below the Bloch-Gr{\"u}neisen (BG) temperature
$T_\text{BG}$. It is given by $k_\text{B}T_\text{BG} = 2\hbar c_\lambda k_F$,
where $k_F$ is the Fermi wave vector, $c_\lambda$ is the acoustic sound
velocity, and $k_\text{B}$ is the Boltzmann constant, and marks the temperature
below which full backscattering at the Fermi surface by acoustic phonons is
frozen out (see Fig.~\ref{fig:mos2}). For heterostructure-based 2DEGs the BG
regime is well established~\cite{Price:BG,West:ObservationOfBG}, and recently,
transport in the BG regime has been studied in graphene both
experimentally~\cite{Kim:Controlling} and
theoretically~\cite{Sarma:Acoustic,Guinea:Flexural,Kaasbjerg:Unraveling}.

In a 2DEG the Fermi wave vector $k_F$ scales with the carrier density as
$\sqrt{n}$ and the Bloch-Gr{\"u}neisen temperature acquires a similar density
dependence $k_\text{B}T_\text{BG} = 2\hbar c_\lambda \sqrt{4\pi n/g_sg_v}$ with
$g_s$ and $g_v$ denoting the spin and valley degeneracy,
respectively~\cite{footnote1}. For monolayer MoS$_2$ ($g_v=2$) this results in
BG temperatures
\begin{equation}
  \label{eq:T_BG}
  T_\text{BG}^\text{TA} \approx 11 \sqrt{\tilde{n}} \; \mathrm{K}
  \quad \text{and} \quad
  T_\text{BG}^\text{LA} \approx 18 \sqrt{\tilde{n}} \; \mathrm{K} ,
\end{equation}
for the transverse (TA) and longitudinal (LA) acoustic phonon, respectively,
with the carrier density $\tilde{n}=n/10^{12}$~cm$^{-2}$ measured in units of
$10^{12}$~cm$^{-2}$. These numbers are on the same order of magnitude as those
for graphene~\cite{Guinea:Flexural}, and transport in the high-mobility BG
regime should be achievable in monolayer MoS$_2$ (and other 2D transition metal
dichalcogenides). However, the above considerations also emphasize the
importance of high extrinsic carrier densities $n \gtrsim 10^{12}$~cm$^{-2}$ in
order for the BG transition to occur at sufficiently high temperatures where
acoustic phonon scattering is significant~\cite{Sarma:100}. Such large carrier
densities can be achieved with e.g. advanced electrolytic gating where densities
on the order of $n\sim 10^{14}$~cm$^{-2}$ have been reached in 2D samples of
graphene and
MoS$_2$~\cite{Sood:Monitoring,Kim:Controlling,Iwasa:Ambipolar,Sood:Symmetry}.

In the present work, we study the acoustic phonon limited mobility of $n$-type
2D MoS$_2$ at low temperatures ($T<100$~K) taking into account both deformation
potential (DP) and piezoelectric (PE) scattering. The flexural phonon couples
weakly to charge carriers and is here neglected. In our previous work
considering scattering of both acoustic and optical
phonons~\cite{Kaasbjerg:MoS2}, only the deformation potential interaction was
taken into account in the coupling to the acoustic phonons. There we found that
the mobility at higher temperatures ($T>100$~K) was dominated by optical phonon
scattering. With piezoelectric interaction included this is still the case,
however, with a slightly lower room-temperature mobility of
320~cm$^2$~V$^{-1}$~s$^{-1}$ at $n=10^{11}$~cm$^{-2}$. Otherwise the conclusions
of Ref.~\onlinecite{Kaasbjerg:MoS2} remain unaffected. For the temperatures
considered in this work, scattering by intervalley acoustic phonons and optical
phonons is strongly suppressed and can be neglected~\cite{Kaasbjerg:MoS2}.

Using a first-principles approach, we calculate the deformation potential and
piezoelectric interactions in 2D MoS$_2$. Supported by continuum model
calculations of the acoustic el-ph interaction in 2D hexagonal lattices, this
allows us to establish analytic expressions and the individual coupling
strengths for the two scattering mechanisms. The calculated intrinsic
low-temperature mobility provides a platform for comparison with future
measurements of the carrier mobility in monolayer MoS$_2$ which can (i) lead to
an experimental verification of the theoretical deformation potentials and
piezoelectric constant reported here~\cite{Sarma:Hetero1}, (ii) reveal to what
extent the mobility is affected by extrinsic surface acoustic/optical
phonons~\cite{Knabchen:SurfaceAcousticII,Ando:ElphHetero} which have turned out
to be important in substrate-supported graphene
samples~\cite{Fuhrer:GrapheneSiO2,Guinea:SubstrateLimited,Jena:HighKappa,Peeters:PiezoelectricSurface},
and (iii) address the importance of the interplay between scattering of acoustic
phonons and impurities which results in a complex temperature and density
dependence of the mobility~\cite{Sarma:Interplay}. In this context previous
studies have emphasized the importance of including both the TA and LA phonon in
order to obtain good agreement with experiment~\cite{West:ObservationOfBG}. With
the present work we uncover new important aspects of the acoustic el-ph
interaction and how it is affected by carrier screening. These are issues of
high relevance for the understanding of acoustic phonon limited 2DEG mobilities
in semiconductors and, in particular, monolayers of transition metal
dichalcogenides.

The paper is organized as follows. Section~\ref{sec:II} briefly summarizes the
Boltzmann transport theory for acoustic phonon scattering. In Sec.~\ref{sec:III}
the first-principles results for the acoustic electron-phonon (el-ph)
interaction are presented and the deformation potential and piezoelectric
interactions are discussed in closer detail along with a microscopic description
of carrier screening. Finally, the results for temperature and density
dependence of the acoustic phonon limited mobility are presented in
Sec.~\ref{sec:IV}.

\section{Boltzmann Theory}
\label{sec:II}

Two-dimensional MoS$_2$ has a hexagonal lattice structure like graphene with the
bottom of the conduction band residing in the $K,K'$ points at the corners of
the Brillouin zone~\cite{Eriksson:2D,Lambrecht:Quasiparticle}. The two
$K,K'$ valleys are perfectly isotropic with an effective electron mass of $m^* =
0.48\; m_e$~\cite{Kaasbjerg:MoS2}. The satellite valleys located at the
$\Gamma$-$K$ path inside the Brillouin zone are well separated in energy from the
$K,K'$-valleys and therefore not important for the low-field transport
properties~\cite{Lambrecht:Quasiparticle,Kaasbjerg:MoS2}. The conduction band
spin splitting of a few meV due to the intrinsic spin-orbit interaction in 2D
transition metal dichalcogenides~\cite{Schwing:GiantSO,Lambrecht:Quasiparticle}
can be safely neglected here. At the same time we note that a Rashba-type
spin-orbit interaction can affect the phonon limited 2DEG
mobility~\cite{Zhang:2DSemiconductorSOC,Ghosh:SO2DEG}.

In Boltzmann theory, the drift mobility $\mu_{xx} = \sigma_{xx} / ne$, where
$\sigma_{xx}$ is the conductivity, is in the presence of (quasi) elastic
scattering given by the Drude-like expression~\cite{Sarma:Hetero2}
\begin{equation}
  \label{eq:mobility}
  \mu_{xx} = \frac{e \expect{\tau_k}}{m^*} ,
\end{equation}
where $\tau_k$ is the energy-dependent relaxation time and the energy-weighted
average $\expect{\cdot}$ is defined by
\begin{equation} 
  \label{eq:tau_averaged}
  \expect{A} = \frac{1}{n}
  \int \! d\varepsilon_\bk \; \rho(\varepsilon_\bk) \varepsilon_\bk
  A(\varepsilon_\bk)
 \left(- \frac{\partial f}{\partial \varepsilon_\bk} \right).
\end{equation}
Here, $n$ is the two-dimensional carrier density, $\rho = g_s g_v m^* /
2\pi\hbar^2$ is the density of states in 2D, $g_s=2$ and $g_v = 2$ are the spin
and valley degeneracy, respectively, $\varepsilon_\bk = \hbar^2 k^2 / 2 m^*$ is
the carrier energy, $f(\varepsilon_\bk) = \left\{1 + \exp{[ (\varepsilon_\bk -
    \mu)/ k_\text{B} T ]} \right\}^{-1}$ is the equilibrium Fermi-Dirac
distribution function, and $\mu$ is the chemical potential. For a degenerate
electron gas, only scattering within a shell of width $k_\text{B} T$ around the
Fermi level is relevant and $\mu_{xx} \approx e \tau_{k_F} / m^*$ applies.

In valley-degenerate semiconductors scattering in inequivalent valleys is not
necessarily identical. In such cases the Boltzmann equation must be solved
explicitly in all inequivalent valleys. In the absence of intervalley scattering
this amounts to replacing the relaxation time in~\eqref{eq:mobility} with a
valley-averaged relaxation time: $\tau = \sum_v \tau_v / N_v$, where $v$ denotes
the valley index, $N_v$ is the number of inequivalent valleys, and $\tau_v$
denotes the individual valley relaxation times.

For acoustic phonon scattering, which to a good approximation can be treated as
a quasielastic scattering process, the relaxation time for the individual
acoustic phonons is given by~\cite{Sarma:Hetero2,Kaasbjerg:MoS2}
\begin{equation}
  \label{eq:tau_acoustic}
  \frac{1}{\tau_{\lambda}(\varepsilon_\bk)} = \sum_{\bk'} 
      \left( 1 - \cos{\theta_{\bk\bk'}} \right) P_{\bk\bk'}^\lambda 
       \frac{1 - f_{\bk'}}{1 - f_\bk}   ,
\end{equation}
where $\lambda$ denotes the branch index ($\lambda=$TA, LA), $\theta_{\bk\bk'}$
is the scattering angle, and $f_\bk=f(\varepsilon_\bk)$ is understood. The
transition matrix element is given by
\begin{align}
  \label{eq:P_acoustic}
  P_{\bk\bk'}^\lambda & =  \frac{2\pi}{\hbar} 
     \left\vert \frac{ g_{\bk\bq}^\lambda} {\epsilon(q, T)} \right \vert^2
    \bigg[  
    N_{\bq\lambda} 
    \delta(\varepsilon_{\bk'} - \varepsilon_\bk - \hbar\omega_{\bq\lambda}) \bigg. \nonumber \\
    & \quad + \bigg. \left(1 + N_{\bq\lambda} \right) 
     \delta(\varepsilon_{\bk'} - \varepsilon_{\bk} + \hbar\omega_{\bq\lambda})
      \bigg]  ,
\end{align}
where $\bq=\bk'-\bk$, $g_{\bk\bq}^\lambda$ is the el-ph coupling,
$\epsilon(q,T)$ is the wave vector and temperature dependent static dielectric
function of the 2DEG, and $\hbar\omega_{\bq\lambda}=\hbar c_\lambda q$ is the
acoustic phonon energy. The phonons are assumed to be in equilibrium and
populated according to the Bose-Einstein distribution function
$N_{\bq\lambda}=N(\hbar\omega_{\bq\lambda})$.

Screening of the el-ph interaction by the carriers themselves is accounted for
by the dielectric function $\epsilon(q, T)$. As we here show, the presence of
both normal and umklapp processes in the acoustic deformation potential
interaction requires a microscopic description of carrier screening. The
consequence of this is a central result of this work, and will be discussed in
further detail in Sec.~\ref{sec:screening}.

In the present work the expression for the relaxation time in
Eq.~\eqref{eq:tau_acoustic} is evaluated numerically assuming quasielastic
scattering; i.e., the phonon energies are omitted in the $\delta$ functions of
Eq.~\eqref{eq:P_acoustic} (implying $q=2k\sin{\theta_{\bk\bk'}/2}$) but included
in the Fermi function $f_{\bk'}=f(\varepsilon_\bk \pm \hbar\omega_{\bq\lambda})$
of Eq.~\eqref{eq:tau_acoustic}. This is particularly important in the BG regime
where the phonon energy becomes comparable to the thermal smearing at the Fermi
level; i.e., $\hbar\omega_{\bq\lambda} \sim k_\text{B} T$.

\section{Interaction with acoustic phonons}
\label{sec:III} 

In the following, we present first-principles calculations of the el-ph
interaction in 2D MoS$_2$ obtained with the density-functional based method
outlined in Ref.~\onlinecite{Kaasbjerg:MoS2} and implemented in the GPAW
electronic structure package~\cite{GPAW,GPAW1,GPAW2}. As a complement to our
first-principles calculations, we calculate in App.~\ref{app:continuum} the
acoustic el-ph interaction in 2D materials using an elastic continuum model.

\subsection{First-principles calculations}

The interaction with the acoustic phonons can be written in the general form
\begin{equation}
  \label{eq:g}
  g_{\bk\bq}^\lambda = \sqrt{\frac{\hbar}{2A\rho\omega_{\bq\lambda}}}
                       M_{\bk\bq}^\lambda ,
\end{equation}
where $A$ is the area of the sample, $\rho$ is the mass density,
$M_{\bk\bq}^\lambda = \bra{\bk+\bq} \delta V_{\bq\lambda} \ket{\bk}$ is the
matrix element between the Bloch states with wave vectors $\bk$ and $\bk + \bq$,
and $\delta V_{\bq\lambda}$ is the change in the crystal potential due to a
phonon with wave vector $\bq$ and branch index $\lambda$. The couplings in the
$K,K'$ valleys are related through time-reversal symmetry as
$\abs{M_{\bq\lambda}^K} = \abs{M_{-\bq\lambda}^{K'}}$. As the hexagonal lattice
of two-dimensional MoS$_2$ lacks a center of symmetry, charge carriers in
monolayer MoS$_2$ interact with acoustic phonons through both the deformation
potential and the piezoelectric interaction. The coupling matrix element
therefore has contributions from both coupling mechanisms, i.e.
\begin{equation}
  \label{eq:M}
  M_{\bq\lambda} = M_{\bq\lambda}^\text{DP} + M_{\bq\lambda}^\text{PE}  .
\end{equation}
The two coupling mechanisms are often assumed to be out of phase, i.e. one is
real and the other imaginary~\cite{Mahan} (see also
App.~\ref{app:continuum}). This implies that piezoelectric and deformation
potential interactions do not interfere in lowest-order perturbation theory,
i.e. $\abs{M_{\bq\lambda}}^2 = \abs{M_{\bq\lambda}^\text{DP} +
  M_{\bq\lambda}^\text{PE}}^2 = \abs{M_{\bq\lambda}^\text{DP}}^2 +
\abs{M_{\bq\lambda}^\text{PE}}^2$, and can therefore be treated as separate
scattering mechanisms.
\begin{figure}[!t]
  \vspace{0.2cm}
  \begin{minipage}{1.0\linewidth}
    \begin{centering}
      \textsf{\textbf{Deformation potential interaction}}
    \end{centering}
    \vspace{0.2cm}
  \end{minipage}
  \includegraphics[width=0.49\linewidth]{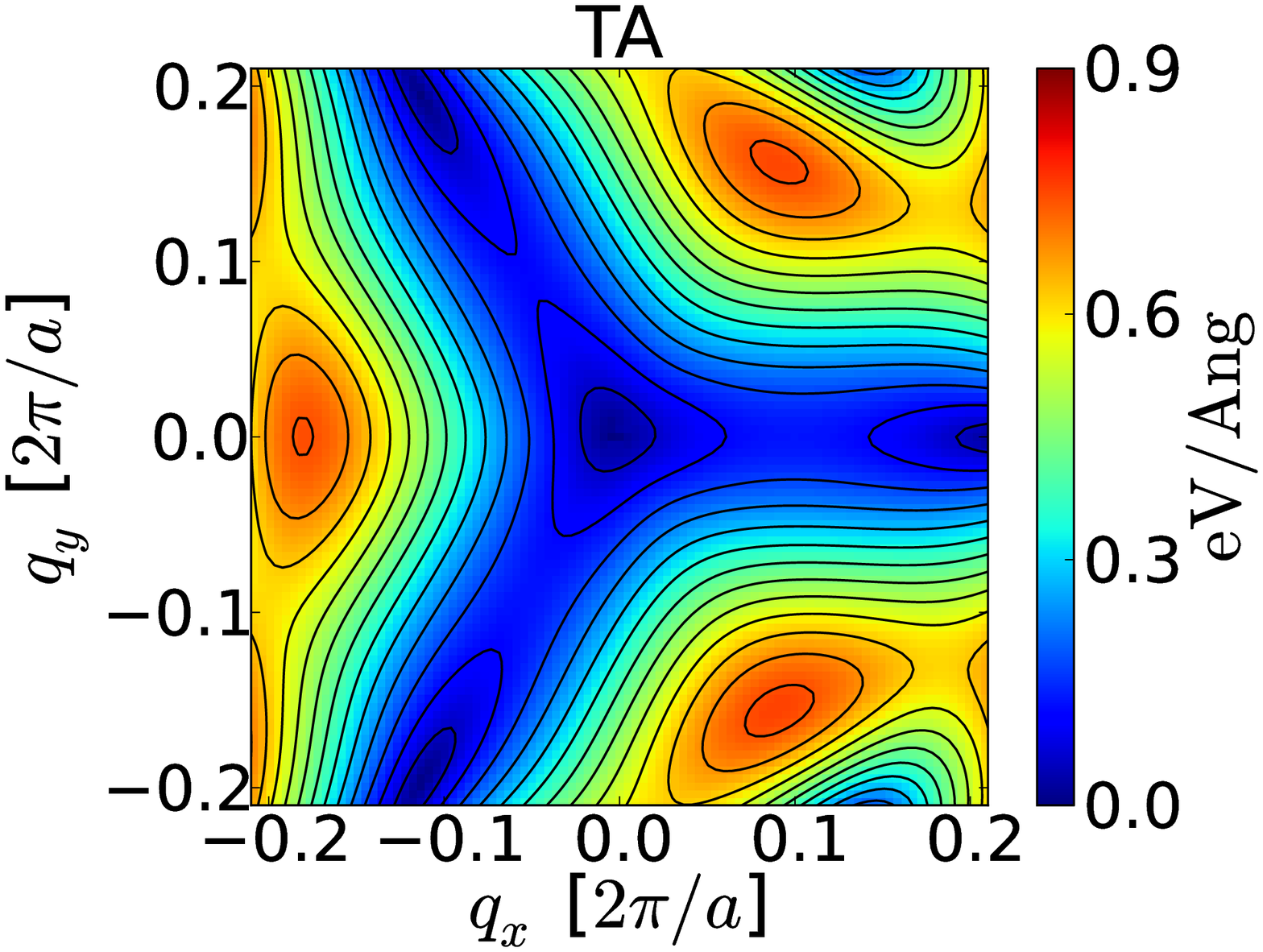}
  \includegraphics[width=0.49\linewidth]{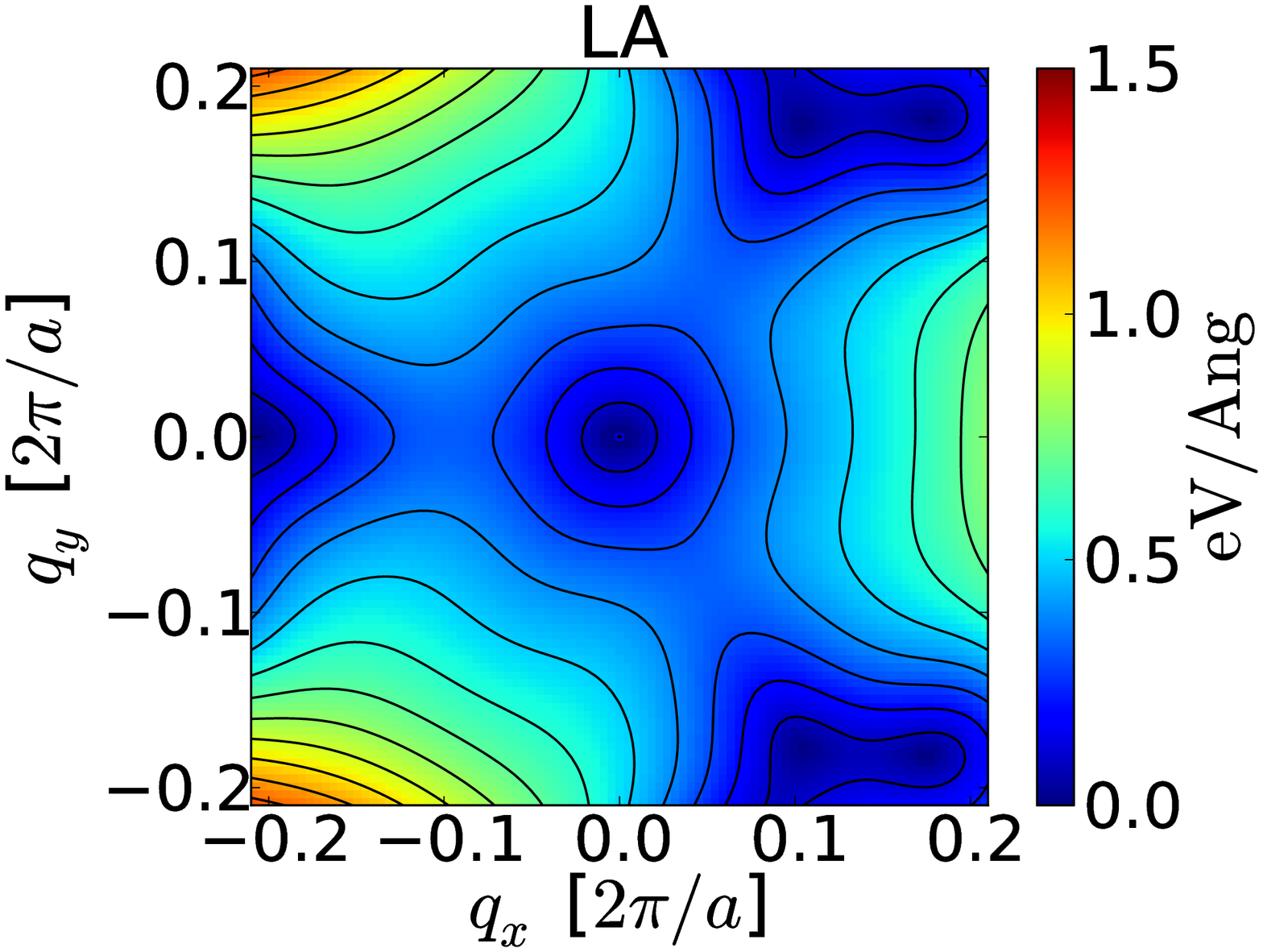}
  \begin{minipage}{1.0\linewidth}
  \vspace{0.4cm}
    \begin{centering}
      \textsf{\textbf{Piezoelectric interaction}}
    \end{centering}
    \vspace{0.2cm}
  \end{minipage}
  \includegraphics[width=0.49\linewidth]{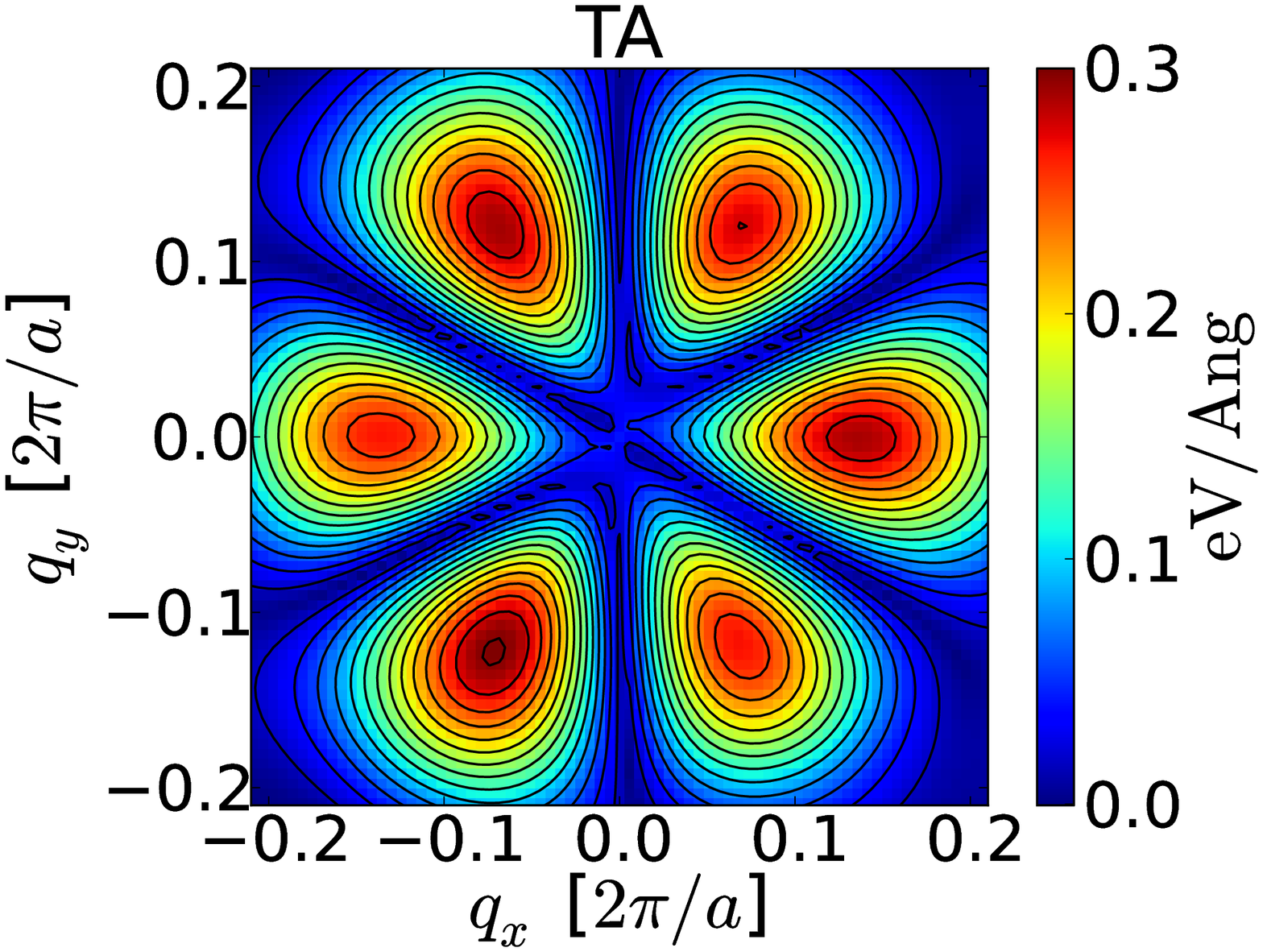}
  \includegraphics[width=0.49\linewidth]{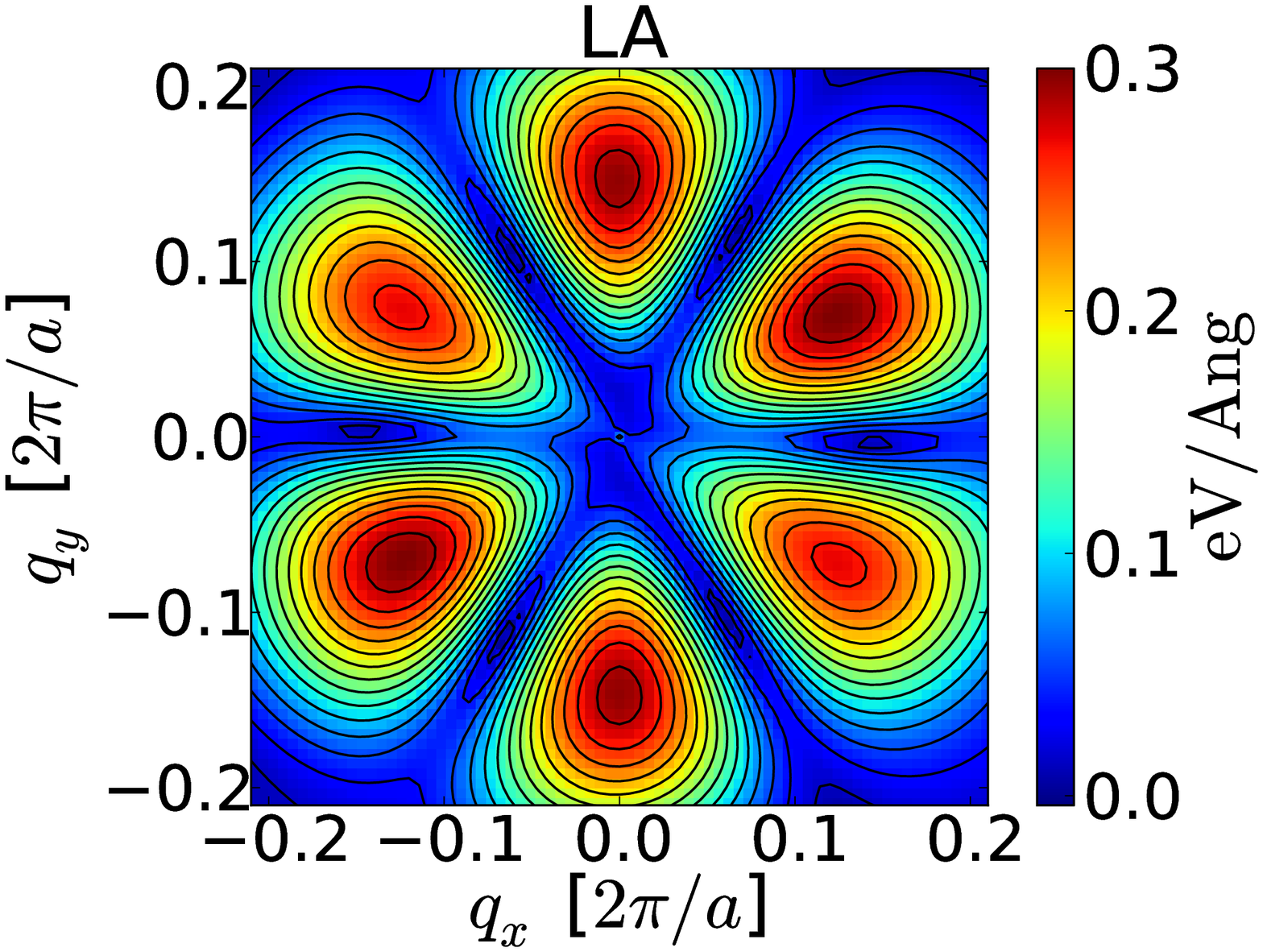}
  \begin{minipage}{1.0\linewidth}
  \vspace{0.4cm}
    \begin{centering}
      \textsf{\textbf{Relative phase $\abs{\phi}$}}
    \end{centering}
    \vspace{0.2cm}
  \end{minipage}
  \includegraphics[width=0.49\linewidth]{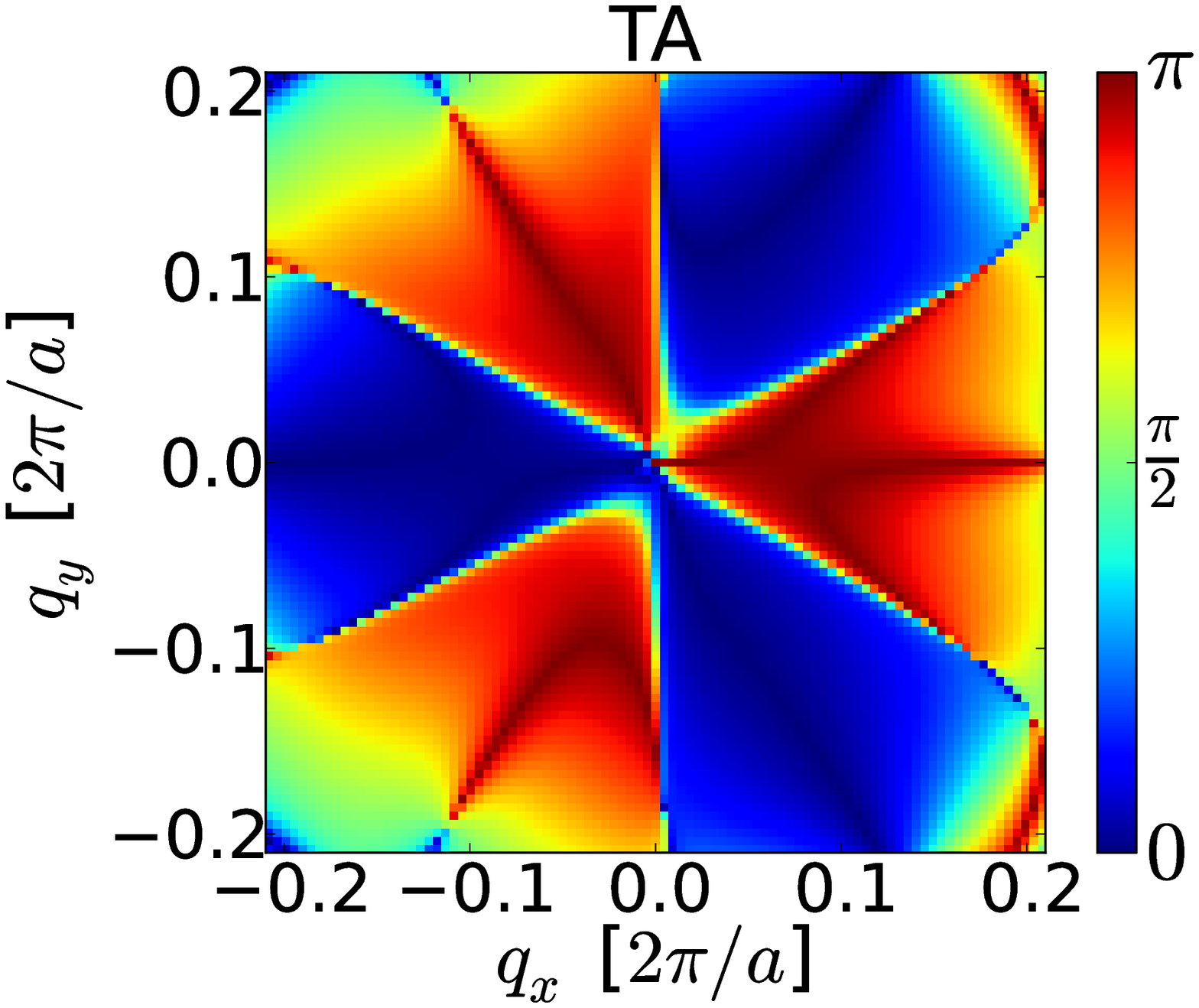}
  \includegraphics[width=0.49\linewidth]{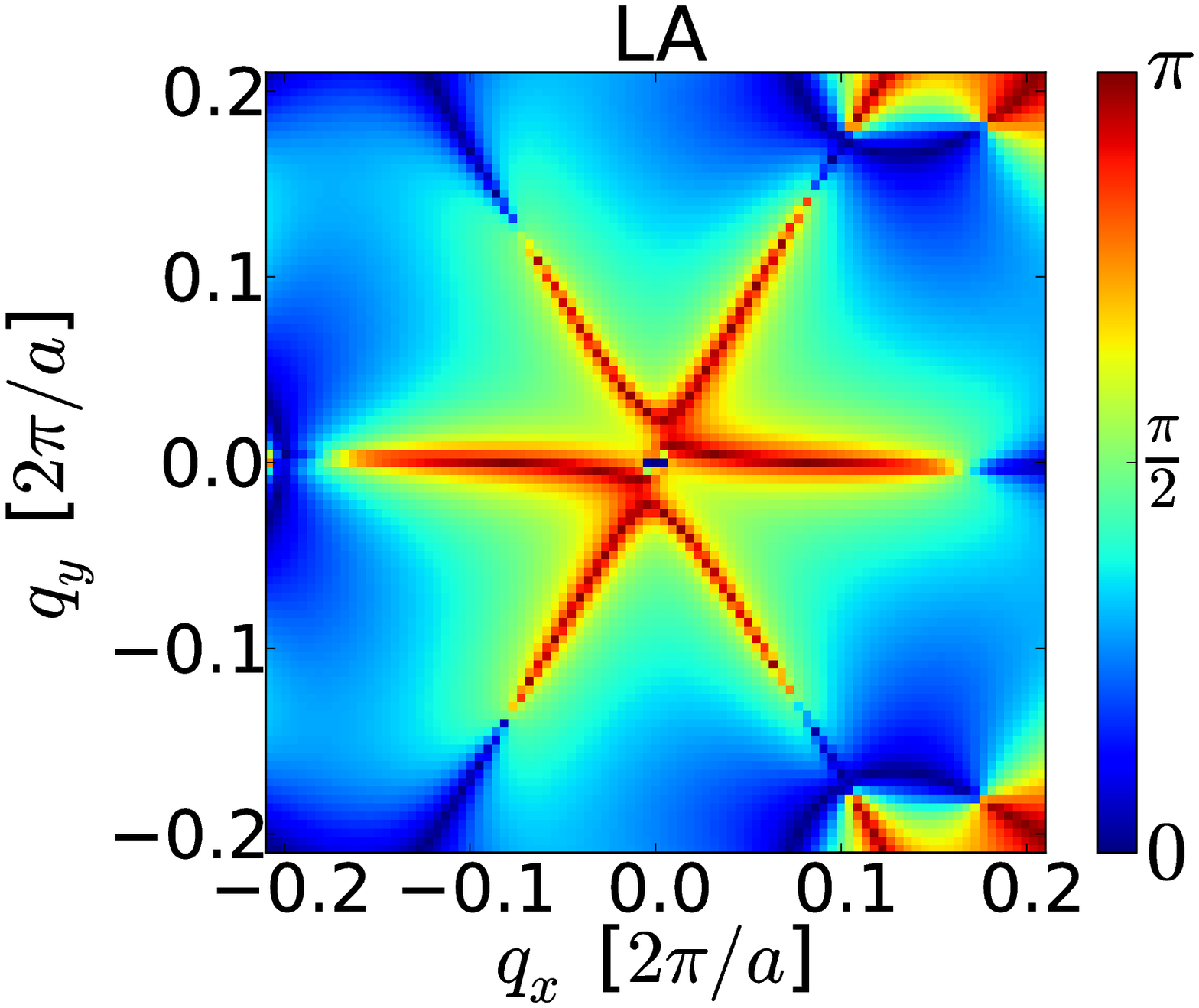}
  \caption{(Color online) Calculated deformation potential and piezoelectric
    interactions in the $K$-valley of the conduction band in monolayer
    MoS$_2$. The contour plots show the absolute value of the calculated
    coupling matrix elements $M_{\bq\lambda}^\text{DP/PE}$ at $\bk=\mathbf{K}$
    for the TA (left) and LA (right) phonons as a function of the
    two-dimensional phonon wave vector $\bq$. The absolute value of the relative
    phase $\phi$ between the two interactions (see Eq.~\eqref{eq:interference})
    is shown in the bottom plots.}
  \label{fig:M}
\end{figure}

In Fig.~\ref{fig:M} we show our first-principles results for the deformation
potential and piezoelectric interactions with the TA and LA phonons in 2D
MoS$_2$ for $\bk = \mathbf{K}$~\cite{footnote2}. The two coupling mechanisms
have been obtained from the total coupling matrix element in Eq.~\eqref{eq:M}
using the real-space partitioning scheme outlined in
App.~\ref{app:partition}. The scheme is based on the observation that the
deformation potential is short range while the piezoelectric interaction is
long range, and can therefore be separated in real space. While the deformation
potential couplings have the three fold rotational symmetry of the conduction
band in the vicinity of the $K,K'$ points, the six-fold rotational symmetry of
the piezoelectric couplings stems from the hexagonal crystal lattice.

The interference between the deformation potential and piezoelectric interaction
can be inferred from the relative phase $\phi$ between their complex-valued
matrix elements,
\begin{equation}
  \label{eq:interference}
  \phi=\text{Im}[\ln{(M_{\bq\lambda}^\text{PE} / M_{\bq\lambda}^\text{DP})}] .
\end{equation}
The absolute value of the relative phase is shown in the bottom row of
Fig.~\ref{fig:M}. At long wavelengths, we find that the above-mentioned
out-of-phase property ($\phi=\pi/2$) holds for the LA phonon only. For the TA
phonon the two coupling mechanisms interfere ($\phi=0, \pi$) and must hence be
considered together. Deviations from this behavior occur at short wavelengths
where the strictly transverse and longitudinal character of the TA and LA
phonons vanishes. However, for long-wavelength acoustic phonon scattering, the
interaction with the TA and LA phonons is given by fully interfering and
noninterfering couplings, respectively.

\subsubsection{Normal and umklapp contributions}
\label{sec:normalumklapp}

In order to gain further understanding of the deformation potential interaction,
we here quantify the contributions from normal and umklapp processes. Formally,
the two are associated with terms in the Fourier expansion of the short-range
phonon-induced change in the crystal potential,
\begin{equation}
  \label{eq:deltaV_Fourier}
  \delta V_{\bq\lambda}(\br) = \sum_{\mathbf{G}} e^{i (\bq + \mathbf{G})\cdot \br}
                        \delta V_{\bq + \mathbf{G}}^\lambda ,
\end{equation}
with $\mathbf{G}=\mathbf{0}$ and $\mathbf{G}\neq \mathbf{0}$ for normal and
umklapp processes, respectively ($\mathbf{G}$ is a reciprocal lattice vector,
see also App.~\ref{app:normalumklapp}). Since the coupling to the TA phonon
vanishes when umklapp processes are neglected altogether~\cite{Madelung}, they
are essential for a correct description of the acoustic el-ph interaction.
\begin{figure}[!b]
  \includegraphics[width=0.49\linewidth]{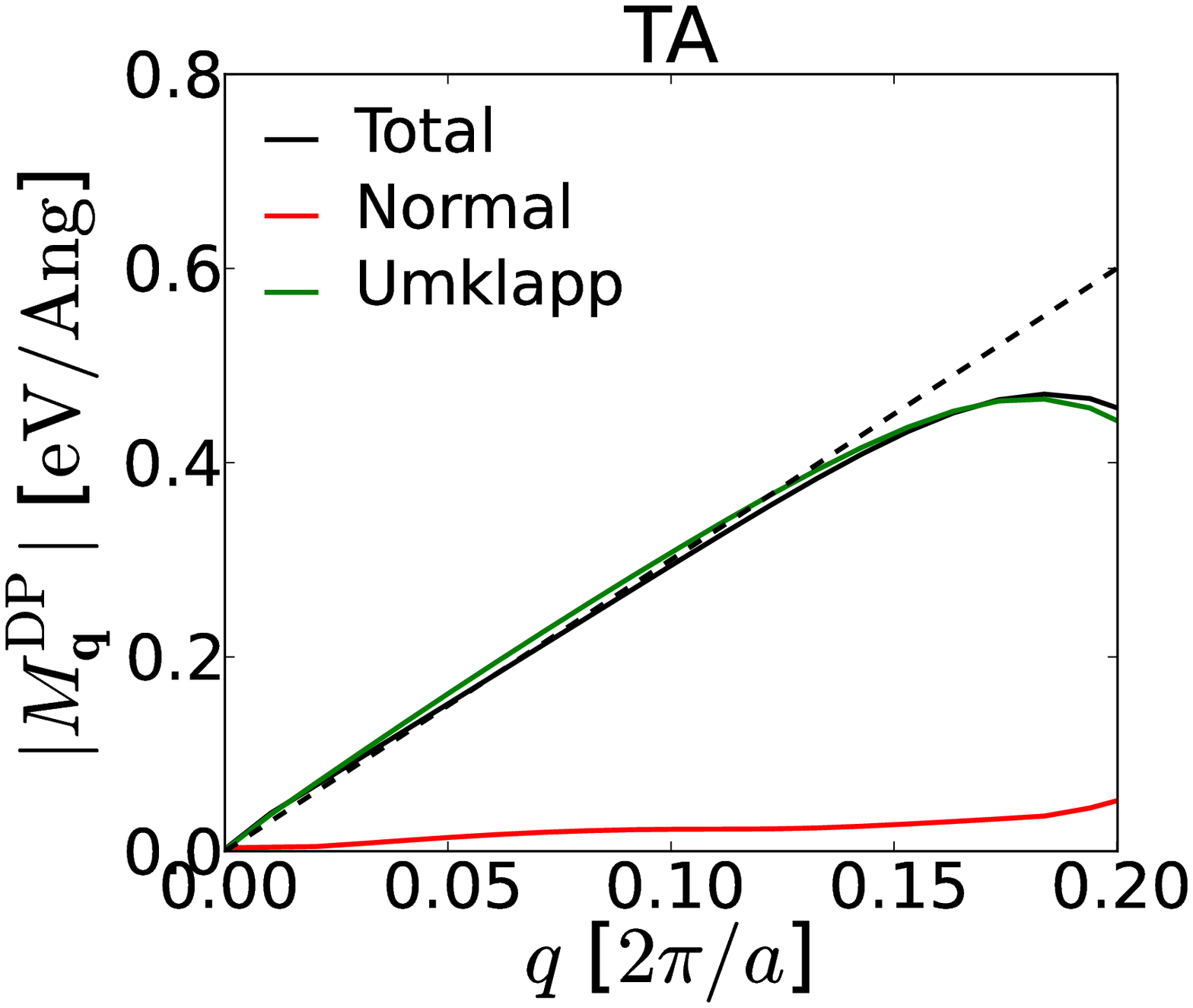}
  \includegraphics[width=0.49\linewidth]{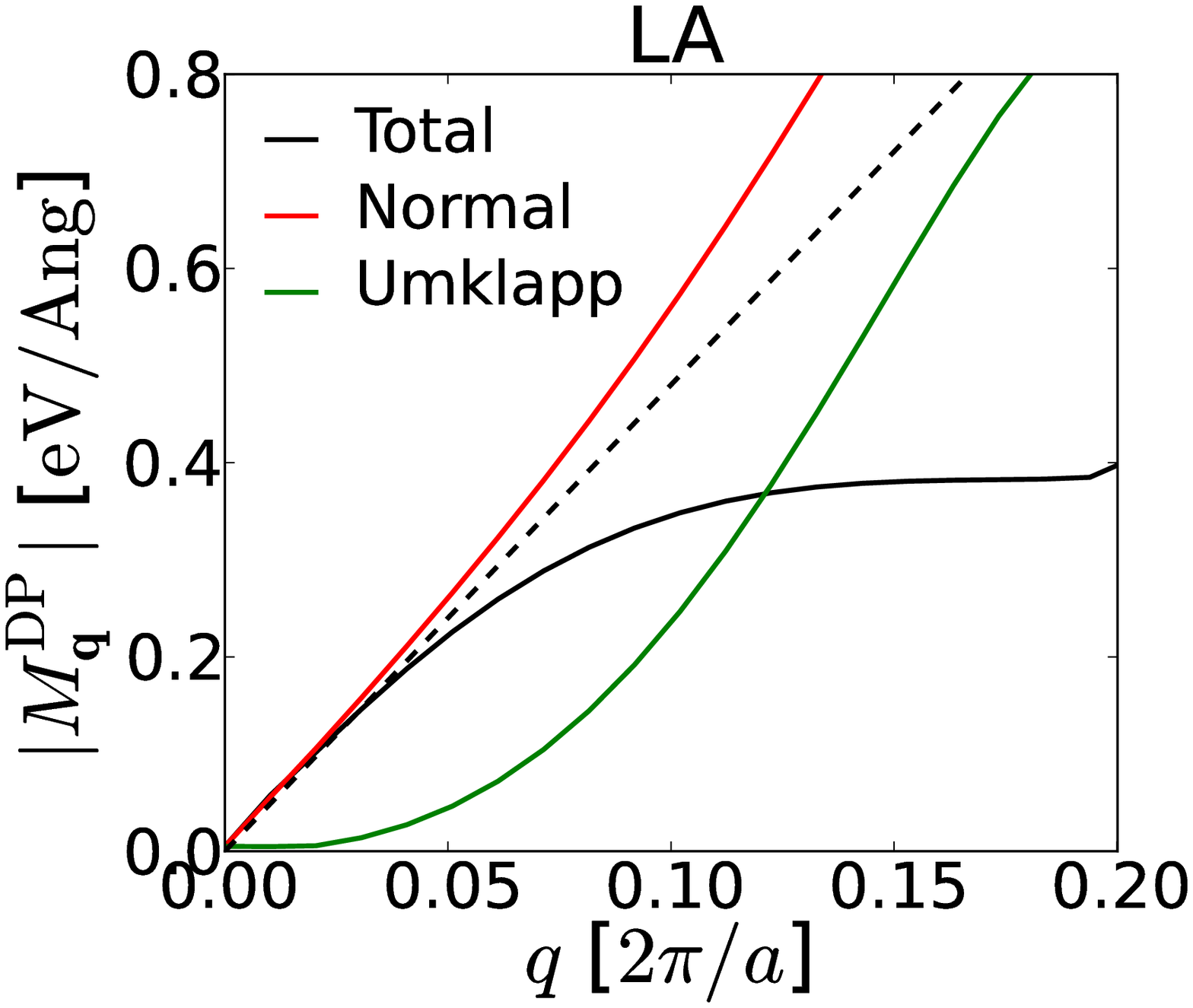}
  \caption{(Color online) Normal and umklapp contributions to the deformation
    potential interaction for the TA (left) and LA (right) phonon. The plots
    show the angular average of the deformation potential interactions in
    Fig.~\ref{fig:M} and their contributions from normal and umklapp processes
    given by Eq.~\eqref{eq:M_normalumklapp} (full lines). The dashed lines show
    the analytic deformation potential interaction in Eq.~\eqref{eq:M_DP}.}
  \label{fig:M_df_normalvsumklapp}
\end{figure}

In Fig.~\ref{fig:M_df_normalvsumklapp} we show the normal and umklapp
contributions to the acoustic deformation potential interactions obtained with
the Fourier filtering method outlined in App.~\ref{app:normalumklapp}. The plots
show the absolute value of the $\bq$-direction average of the matrix element
\begin{equation}
  \label{eq:M_normalumklapp}
  M_{\bq\lambda}^{\text{DP},X} = 
       \bra{\bk+\bq} \delta V_{\bq\lambda}^X \ket{\bk}, \quad X=N, U, 
\end{equation}
where $\delta V_{\bq\lambda}^X$ denotes the phonon-induced potential with normal
($N$) and umklapp ($U$) processes included, respectively. Due to the
complex-valued matrix elements, the absolute values of the normal and umklapp
contributions do not add up to the total matrix element. In accordance with the
statement below Eq.~\eqref{eq:deltaV_Fourier}, we find that in the
long-wavelength limit the deformation potential interactions for the TA and LA
phonons are completely dominated by umklapp and normal processes,
respectively. At shorter wavelengths both processes contribute.

The separation of the deformation potential interaction into contributions from
normal and umklapp processes is not only of technical character. As we show
below in Sec.~\ref{sec:screening}, it has important consequences for the
screening of the deformation potential interaction.

\subsection{Analytic expressions for the acoustic el-ph interaction}

In the following, the analytic expressions for the deformation potential and
piezoelectric couplings obtained in App.~\ref{app:continuum} are introduced and
the coupling strengths are determined from the first-principles el-ph
couplings.

\subsubsection{Deformation potential interaction}
\label{sec:DP}

The deformation potential originates from the local changes of the crystal
potential caused by the atomic displacements due to an acoustic phonon. The
determination of the interaction strength thus requires a microscopic
calculation, such as the first-principles approach used in this work.

The interaction with acoustic phonons via the deformation potential interaction
is most commonly assumed to be isotropic and linear in the phonon wave vector,
i.e.
\begin{equation}
  \label{eq:M_DP}
  \abs{M_{\bq\lambda}^{\text{DP}}} = \Xi_\lambda q ,
\end{equation}
where $\Xi_\lambda$ is the acoustic deformation potential~\cite{Mahan}. Since
the \emph{true} deformation potential couplings in Fig.~\ref{fig:M} are
anisotropic and show a more complex $q$-dependence at shorter wavelengths, the
deformation potential of Eq.~\eqref{eq:M_DP} must be regarded as an
\emph{effective} coupling parameter. In valley-degenerate semiconductors where
the couplings in the different valleys are related through time-reversal
symmetry, the \emph{effective} deformation potential furthermore accounts for
the variation in the angular dependence of the coupling between inequivalent
valleys. We have here recalculated the acoustic deformation potentials from
Ref.~\onlinecite{Kaasbjerg:MoS2} in order to avoid undesired contributions from
the piezoelectric interaction that potentially were included there. The new
deformation potentials are given in Tab.~\ref{tab:parameters} and the couplings
are shown in Fig.~\ref{fig:M_df_normalvsumklapp} (dashed lines) together with
the angular average of the first-principles couplings.

For deformation potential scattering above the BG temperature where the
equipartition approximation $N_\bq \sim k_\text{B}T / \hbar\omega_\bq \gg 1$
applies and with screening neglected, the relaxation time in
Eq.~\eqref{eq:tau_acoustic} becomes independent of the carrier energy and is
given by~\cite{Kaasbjerg:MoS2}
\begin{equation}
  \label{eq:tau_DP_highT}
  \frac{1}{\tau_\lambda (\varepsilon_\bk)}
  = \frac{m^*\Xi_\lambda^2k_\text{B}T}{\hbar^3\rho c_\lambda^2} .
\end{equation}
This results in a $\mu \sim T^{-1}$ temperature dependence of the 2DEG mobility
characteristic of acoustic deformation potential scattering in the
high-temperature regime.

\subsubsection{Piezoelectric interaction}
\label{sec:PE}

Piezoelectric coupling to acoustic phonons occurs in crystals lacking an
inversion center and originates from the macroscopic polarization that
accompanies an applied strain $\epsilon_{ij}$. The strength of the interaction
is given by the piezoelectric tensor $e_{ij}$ here given in Voigt notation.
\begin{table}[!b]
  \caption{Material parameters for monolayer MoS$_2$ used in the present
    work. Apart from the acoustic deformation potentials and the piezoelectric
    constant all parameters have been adopted from 
    Ref.~\onlinecite{Kaasbjerg:MoS2}.}
\begin{ruledtabular}
\begin{tabular}{lcc}
Parameter &  Symbol  & Value  \\ 
\hline                                     
Lattice constant               &   $a$                     &   3.14 \AA            \\
Ion mass density               &   $\rho$                  &   $3.1\times 10^{-7}$ g/cm$^2$ \\
Effective electron mass        &   $m^*$                   &   0.48 $m_e$          \\
Transverse sound velocity      &   $c_\text{TA}$           &   $4.2 \times 10^3$ m/s  \\
Longitudinal sound velocity    &   $c_\text{LA}$           &   $6.7 \times 10^3$ m/s  \\
Acoustic deformation potentials&                           &                  \\
TA                             &   $\Xi_\text{TA}$         &   $1.5$ eV       \\
LA                             &   $\Xi_\text{LA}$         &   $2.4$ eV       \\
Piezoelectric constant         &   $e_{11}$                &   $3.0 \times 10^{-11}$~C/m   \\
Effective layer thickness      &   $\sigma$                &   5.41 \AA            \\
\end{tabular}
\end{ruledtabular}
\label{tab:parameters}
\end{table}

In App.~\ref{app:piezo} we obtain the piezoelectric interaction in a 2D
hexagonal lattice using continuum theory. We find that the piezoelectric
interaction is given by
\begin{equation}
  \label{eq:M_PE}
   \abs{M_{\bq\lambda}^{\text{PE}}} = \frac{e_{11} e}{\epsilon_0}
   q \times \text{erfc}(q \sigma / 2) \abs{A_\lambda(\hat{\bq})} ,
\end{equation}
where $e_{11}$ (units of C/m) is the only independent component of the
piezoelectric tensor of the 2D hexagonal lattice~\cite{Mele:NTPiezo},
$\epsilon_0$ is the vacuum permeability, erfc is the complementary error
function, $\sigma$ is an effective width of the electronic wave functions, and
$A_\lambda(\hat{\bq})$ is an anisotropy factor that accounts for the angular
dependence of the piezoelectric interaction. It is given by
\begin{equation}
  \label{eq:A_theta}
  A_\text{TA}(\hat{\bq}) = - \sin{3 \theta_\bq} \quad \text{and} \quad
  A_\text{LA}(\hat{\bq}) = \cos{3 \theta_\bq}
\end{equation}
for the TA and LA phonon, respectively, and results in a highly
anisotropic piezoelectric interaction.

The first-principles results for the piezoelectric interaction in
Fig.~\ref{fig:M} are in overall good agreement with the analytic expression in
Eq.~\eqref{eq:M_PE} (see also Figs.~\ref{fig:M_analytic}
and~\ref{fig:M_fit})~\cite{footnote3}. From a fit to the first-principles
results, the piezoelectric constant of 2D MoS$_2$ is estimated to be $e_{11}
\sim 3.0 \times 10^{-11}$~C/m ($\sim$0.01~$e$/bohr). This is an order of
magnitude smaller than a recently reported value ($e_{11} = 3.06 \times
10^{-10}$) obtained with a Berry's phase
approach~\cite{Reed:Piezoelectricity}. We are at present, however, not able to
clarify the origin of this disagreement.

The $q$-dependence of the 2D piezoelectric interaction in Eq.~\eqref{eq:M_PE} is
qualitatively different from the one in 3D bulk systems where
$M_{\bq\lambda}^{\text{PE}}\sim\text{constant}$~\cite{Mahan}. In the
long-wavelength limit where $\text{erfc}(q \sigma / 2) \sim 1 - q\sigma /
\sqrt{2\pi}$, the 2D piezoelectric interaction acquires a linear $q$-dependence
$M_{\bq\lambda}^{\text{PE}} \sim q$. Hence, the deformation potential and
piezoelectric interaction in a 2D lattice behave qualitatively the same in the
long-wavelength limit. Assuming that the linear long-wavelength behavior holds,
the high-temperature relaxation time for piezoelectric scattering is given by
Eq.~\eqref{eq:tau_DP_highT} with the replacement
\begin{equation}
  \label{eq:tau_PE_highT}
  \Xi_\lambda^2 \rightarrow \frac{1}{2} 
      \left( \frac{e_{11} e}{\epsilon_0} \right)^2,
\end{equation}
where the factor $1/2 = \expect{A_\lambda^2}$ stems from the angular mean of the
piezoelectric interaction~\cite{footnote4}. This corresponds to an effective
isotropic piezoelectric coupling with $A_\lambda(\hat{\bq}) = 1/\sqrt{2}$ in
Eq.~\eqref{eq:M_PE}. The relative strength of the deformation potential and
piezoelectric interactions is thus governed by the ratio $(e_{11} e /
\epsilon_0) / \Xi_\lambda$ of the prefactors in Eqs.~\eqref{eq:M_DP}
and~\eqref{eq:M_PE}. Since this is of the order of unity with the parameters for
2D MoS$_2$ listed in Table~\ref{tab:parameters}, both coupling mechanisms must
be taken into account.

\subsection{Screening of the acoustic el-ph interaction}
\label{sec:screening}

The first-principles el-ph interactions presented above have been obtained for
the neutral material and therefore do not take into account screening by the
2DEG in extrinsic 2D MoS$_2$. In the following we apply a microscopic theory for
carrier screening and show that the normal and umklapp contributions to the
deformation potential interaction are screened differently.

Formally, the screened el-ph interaction can be obtained by replacing the
phonon-induced potential $\delta V_{\bq\lambda}$ in the matrix element
$M_{\bk\bq}^\lambda$ of Eq.~\eqref{eq:g} with its screened counterpart
\begin{equation}
  \delta V^\text{scr}_{\bq\lambda} (\br) = 
  \int \! d\br' \,  \epsilon^{-1}(\br,\br') \delta V_{\bq\lambda} (\br') ,
\end{equation}
where $\epsilon^{-1}$ is the (static) microscopic dielectric function of the
2DEG. Inserting~\eqref{eq:deltaV_Fourier}, this can be recast in Fourier space
in terms of the $\bq$-dependent dielectric matrix
$\epsilon_{\mathbf{G}\mathbf{G}'}^{-1}$~\cite{Louie:Dielectric} as
\begin{align}
  \label{eq:deltaV_screen}
  \delta V_{\bq\lambda}^\text{scr}(\br) & = \sum_{\mathbf{G}} e^{i (\bq + \mathbf{G})\cdot \br}
        \sum_{\mathbf{G}'} \epsilon_{\mathbf{G}\mathbf{G}'}^{-1} (\bq) 
        \delta V_{\bq + \mathbf{G}'}^\lambda \nonumber \\
    & \simeq \sum_{\mathbf{G}} e^{i (\bq + \mathbf{G})\cdot \br}
         \epsilon_{\mathbf{G}\mathbf{G}}^{-1} (\bq)
         \delta V_{\bq + \mathbf{G}}^\lambda   ,
\end{align}
where $\mathbf{G}$ and $\mathbf{G}'$ are reciprocal lattice vectors and the
second equality holds in the diagonal approximation
$\epsilon_{\mathbf{G}\mathbf{G}'}^{-1} = \delta_{\mathbf{G}\mathbf{G}'}
\epsilon_{\mathbf{G}\mathbf{G}}^{-1}$. To a good approximation, the screened
phonon-induced potential thus follows by dividing the individual Fourier
components in Eq.~\eqref{eq:deltaV_Fourier} by the components
$\epsilon_{\mathbf{G}\mathbf{G}}$ of the diagonal dielectric matrix. The latter
is related to the 2DEG polarizability $\chi^0_{\mathbf{G}\mathbf{G}}(\bq)$
through the expression~\cite{Louie:Dielectric}
\begin{equation}
  \label{eq:epsilon_G}
  \epsilon_{\mathbf{G}\mathbf{G}}(\bq) = 
     1 - \frac{e^2}{2\epsilon_0 \abs{\bq + \mathbf{G}}} 
     \chi^0_{\mathbf{G}\mathbf{G}}(\bq) ,
\end{equation}
which is similar to the standard long-wavelength expression for the dielectric
function in Eq.~\eqref{eq:epsilon_rpa}, however, with the important difference
that the denominator in the second term of Eq.~\eqref{eq:epsilon_G} contains a
factor $\abs{\bq + \mathbf{G}}$ instead of a factor $q$. For intravalley
scattering where $q \ll \abs{\mathbf{G}}$, this implies that
$\epsilon_{\mathbf{G}\mathbf{G}}(\bq)$ behaves differently at long
($\mathbf{G}=\mathbf{0}$) and short ($\mathbf{G}\neq\mathbf{0}$) wavelengths;
while it diverges as $1/q$ in the $q\rightarrow 0$ limit in the former case, it
approaches a finite value in the latter. As an immediate consequence, normal and
umklapp components of the el-ph interaction are renormalized differently with a
significantly stronger screening of the former.

While a correct description of the screened el-ph interaction can only be
obtained from Eq.~\eqref{eq:deltaV_screen}, the calculation of the microscopic
dielectric function from first principles is, however, beyond the scope of the
present study. Instead, we adopt the following \emph{ad hoc} approach to carrier
screening. 

\subsubsection{Effective screening scheme}

In order to account for the qualitative difference between the screening of
normal and umklapp processes, the dielectric function of the 2DEG in
Eq.~\eqref{eq:epsilon_G} is approximated as follows.

For the long-wavelength component ($\mathbf{G}=\mathbf{0}$) of the dielectric
function, well-established approximations exist in the
literature~\cite{Stern:2D}. We here apply the finite-temperature RPA theory due
to Maldague~\cite{Maldague:ManyBody},
\begin{equation}
  \label{eq:epsilon_rpa}
  \epsilon(q, T, \mu) = 1 - \frac{e^2}{2\epsilon_0q} \chi^0(q,T, \mu) ,
\end{equation}
where $\mu$ is the chemical potential and the static polarizability at finite
temperatures is obtained as
\begin{equation}
  \label{eq:maldague}
  \chi^0(q, T,\mu) = \int_0^\infty \! d\mu' \, 
      \frac{\chi^0(q, 0, \mu')}{4k_\text{B}T \cosh^2{\frac{\mu -
            \mu'}{2k_\text{B}T}}}  .
\end{equation}
Here, $\chi^0(q, 0, \mu)$ is the zero-temperature RPA polarizability given by
the density of states $\chi^0(q, 0, \mu)=-\rho$ for $q<2k_F$ (see also
App.~\ref{app:screening})~\cite{Stern:2D}. The integral in
Eq.~\eqref{eq:maldague} is evaluated numerically using the approach of
Ref.~\onlinecite{Flensberg:Plasmon}. We have here neglected the form factor in
the polarizability arising from the finite thickness of the electronic Bloch
functions. First-principles calculations of the RPA dielectric function (see
e.g. Refs.~\onlinecite{Louie:Dielectric,Jun:Response}) could be helpful in
clarifying to which extent this leads to an overestimation of the screening
strength.

For the short-wavelength part ($\mathbf{G}\neq\mathbf{0}$) of the dielectric
function, we introduce an \emph{effective} dielectric constant
$\epsilon_\text{eff}$ which acts as a simple scaling parameter for the
$\mathbf{G}\neq\mathbf{0}$ components of the potential in
Eq.~\eqref{eq:deltaV_screen}. From the relations $\abs{\mathbf{G}} \gtrsim
q_\text{TF} \gg e^2/2\epsilon_0 \abs{\chi^0_{\mathbf{G}\mathbf{G}}}$, where
$q_\text{TF}$ is the Thomas-Fermi screening wave vector (see
App.~\ref{app:screening}) and the latter inequality follows from the expression
for the microscopic polarizability~\cite{Louie:Dielectric}, we observe that the
short-wavelength screening efficiency is relatively weak;
i.e. $\epsilon_{\mathbf{G}\mathbf{G}}\sim 1$. We can hence to a good
approximation set $\epsilon_\text{eff}=1$ thus leaving the umklapp contribution
to the deformation potential interaction unscreened.

\subsubsection{Total screened el-ph couplings}
\label{sec:M_screened}

With our findings above, the screened matrix element for the acoustic el-ph
interaction can be written as
\begin{equation}
  \label{eq:M_screen}
  M_{\bq\lambda}^\text{scr}(n, T)  = 
       \frac{ M_{\bq\lambda}^{\text{DP},N}}{\epsilon (q,T,\mu)}
     + \frac{M_{\bq\lambda}^{\text{DP},U}}{\epsilon_\text{eff}} 
     + \frac{ M_{\bq\lambda}^{\text{PE}}}{\epsilon (q,T,\mu)}  .
\end{equation}
Here, the normal contribution to the deformation potential interaction and the
long-range piezoelectric interaction are screened by the long-wavelength
dielectric function in Eq.~\eqref{eq:epsilon_rpa}, while the umklapp
contribution to the deformation potential interaction is screened by
$\epsilon_\text{eff}$. It should be noted that this differs from conventional
descriptions of the screened acoustic el-ph interaction where the deformation
potential interaction is either screened with a long-wavelength dielectric
function or left unscreened (see
e.g. Refs.~\onlinecite{Gatos:Heterostructures,Sarma:Hetero2}).

As the deformation potential interactions for the TA and LA phonons are largely
dominated by umklapp and normal processes, they can to a good approximation be
screened by $\epsilon_\text{eff}$ and $\epsilon$ in Eq.~\eqref{eq:epsilon_rpa},
respectively, thus leaving the deformation potential interaction for the TA
phonon unscreened. Taking into account the interference between the deformation
potential and piezoelectric interaction, we can hence approximate the coupling
matrix elements for the TA and LA phonon as
\begin{equation}
  \label{eq:M_TA_screen}
  M_{\bq \text{TA}}^{\text{scr}}(n, T) \approx 
        \frac{ M_{\bq \text{TA}}^\text{DP}}{\epsilon_\text{eff}}
      + \frac{ M_{\bq \text{TA}}^\text{PE}}{\epsilon (q,T,\mu)}  
\end{equation}
and 
\begin{equation}
  \label{eq:M_LA_screen}
  M_{\bq \text{LA}}^{\text{scr}}(n, T) \approx i
      \frac{M_{\bq \text{LA}}^\text{DP}}{\epsilon (q,T,\mu)}
      + \frac{ M_{\bq \text{LA}}^\text{PE}}{\epsilon (q,T,\mu)}  ,
\end{equation}
respectively, where $M_{\bq\lambda}^\text{DP/PE}$ are given by
Eqs.~\eqref{eq:M_DP} and~\eqref{eq:M_PE}.

\subsubsection{Efficiency of long-wavelength screening}

In the following we provide a qualitative estimate of the efficiency of
long-wavelength carrier screening given by the dielectric function in
Eq.~\eqref{eq:epsilon_rpa}. The screening strength is in this case governed by
the dimensionless parameter
\begin{equation}
  \label{eq:q_s}
  q_s(T) = \frac{q_\text{TF}(T)}{{\tilde{q}}(T)} ,
\end{equation}
where $q_\text{TF}(T)$ is the finite-temperature Thomas-Fermi wave vector and
$\tilde{q}$ denotes a typical scattering wave vector. In the case of acoustic
phonon scattering $\tilde{q}$ is given by
\begin{equation}   
  \label{eq:q_tilde}
  \tilde{q} =
  \bigg\{           
    \begin{array}{l}
      \mathrm{min}(k_F,q_\text{th}),           \quad T \lesssim T_F \\
      \mathrm{min}(k_\text{avg}, q_\text{th}), \quad T \gtrsim  T_F  ,
    \end{array}
\end{equation}
in the degenerate ($T \lesssim T_F$) and nondegenerate ($T \gtrsim T_F$) regime,
respectively, and where $T_F$ is the Fermi temperature. Here $q_\text{th} =
k_\text{B} T / 2 \hbar c_\text{ph}$ is a typical scattering wave vector in the
BG regime where the accessible phase space is restricted by the availability of
thermally excited phonons. Above the BG temperature where scattering on the full
Fermi surface is possible, $k_F$ becomes a typical scattering wave vector. In
the case of a nondegenerate 2DEG where the average carrier energy is
$\expect{\varepsilon_\bk} = k_\text{B}T$, $k_\text{avg}=\sqrt{2 m^* k_\text{B}T
  / \hbar^2}$ is a typical wave vector.
\begin{figure}[!t]
  \includegraphics[width=0.6\linewidth]{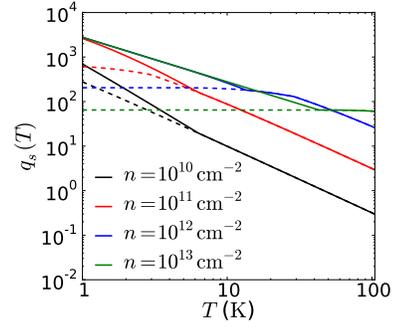}
  \caption{(Color online) Temperature and density dependence of the
    dimensionless screening parameter $q_s(T)$ in Eq.~\eqref{eq:q_s} for
    acoustic phonon (full lines) and charged impurity (dashed lines) scattering
    in 2D MoS$_2$.}
\label{fig:q_s}
\end{figure}

In the low-temperature limit, the screening parameter for acoustic phonon
scattering is given by
\begin{equation}
  \label{eq:q0_ac_s}
  q_s(T \rightarrow 0) \approx \frac{q_\text{TF}}{q_\text{th}} =
  \frac{g_s g_v e^2 m^* c_\text{ph}}{2\pi\epsilon_0 \hbar k_\text{B}T } ,
\end{equation}
which is independent of the carrier density. The $T^{-1}$ divergence of the
low-temperature screening parameter implies that scattering of acoustic phonons
via normal process deformation potential and piezoelectric interaction is
strongly suppressed for a degenerate 2DEG in the BG regime.

It is interesting to compare this with the screening parameter for charged
impurity scattering. In this case, scattering on the full Fermi surface is
possible at low temperatures. Hence, $\tilde{q} = k_{F/\text{avg}}$ in the
degenerate/nondegenerate regime and the low-temperature limit of the screening
parameter becomes
\begin{equation}
  \label{eq:q0_s}
  q_s(T\rightarrow 0) \approx \frac{q_\text{TF}}{k_F} =
  \frac{(g_s g_v)^{3/2}e^2 m^*}{4\pi\epsilon_0 \hbar^2 \sqrt{4\pi n}} ,
\end{equation}
which is independent of temperature and decreases with the carrier density. This
results in a less efficient screening of impurity scattering compared to
acoustic phonon scattering at high densities and low temperatures.

The full temperature dependence of the screening parameter for scattering of
acoustic phonons (full lines) and impurities (dashed lines) is shown in
Fig.~\ref{fig:q_s} for different carrier densities and $c_\text{ph} = 5\times
10^{3}$~m/s representative of the acoustic sound velocities in 2D MoS$_2$. In
the low temperature regime, the limits in Eqs.~\eqref{eq:q0_ac_s}
and~\eqref{eq:q0_s} are approached. Due to the density and temperature
dependence of the Debye-H{\"u}ckel wave vector, $q_\text{D}\propto
n/k_\text{B}T$, the screening strength becomes independent of the scattering
mechanism and increases (decreases) with the carrier density (temperature) in
the nondegenerate high-temperature regime.

The overall large values of the screening parameter ($q_s(T) \gg 1$) in
Fig.~\ref{fig:q_s} follow from a large effective mass and valley
degeneracy. Carrier screening in monolayer MoS$_2$ and other 2D transition metal
dichalcogenides is therefore inherently strong and the screening strength
exceeds that of e.g. Si and GaAs based 2DEGs~\cite{Sarma:RMP}. For scattering of
acoustic phonons this has the important consequence that normal process
deformation potential and piezoelectric interaction is strongly reduced already
at relative low carrier densities $n\gtrsim 10^{11}$~cm$^{-2}$.

\section{Results}
\label{sec:IV}

In the following we use the Boltzmann equation approach outlined in
Section~\ref{sec:II} to study the temperature and density dependence of the
acoustic phonon limited mobility in 2D MoS$_2$ for temperatures $T< 100$~K and
high carrier densities $10^{10}$ to $3\times 10^{13}$~cm$^{-2}$. The mobility
limited by scattering of acoustic phonons follows a generic temperature
dependence $\mu \sim T^{-\gamma}$ where the exponent $\gamma$ depends on
temperature, carrier density, and the dominating scattering mechanism. The same
holds for the resistivity $\rho = (ne \mu)^{-1}$ with a change in the sign of
the exponent. 

\begin{figure}[!b]
  \includegraphics[width=0.49\linewidth]{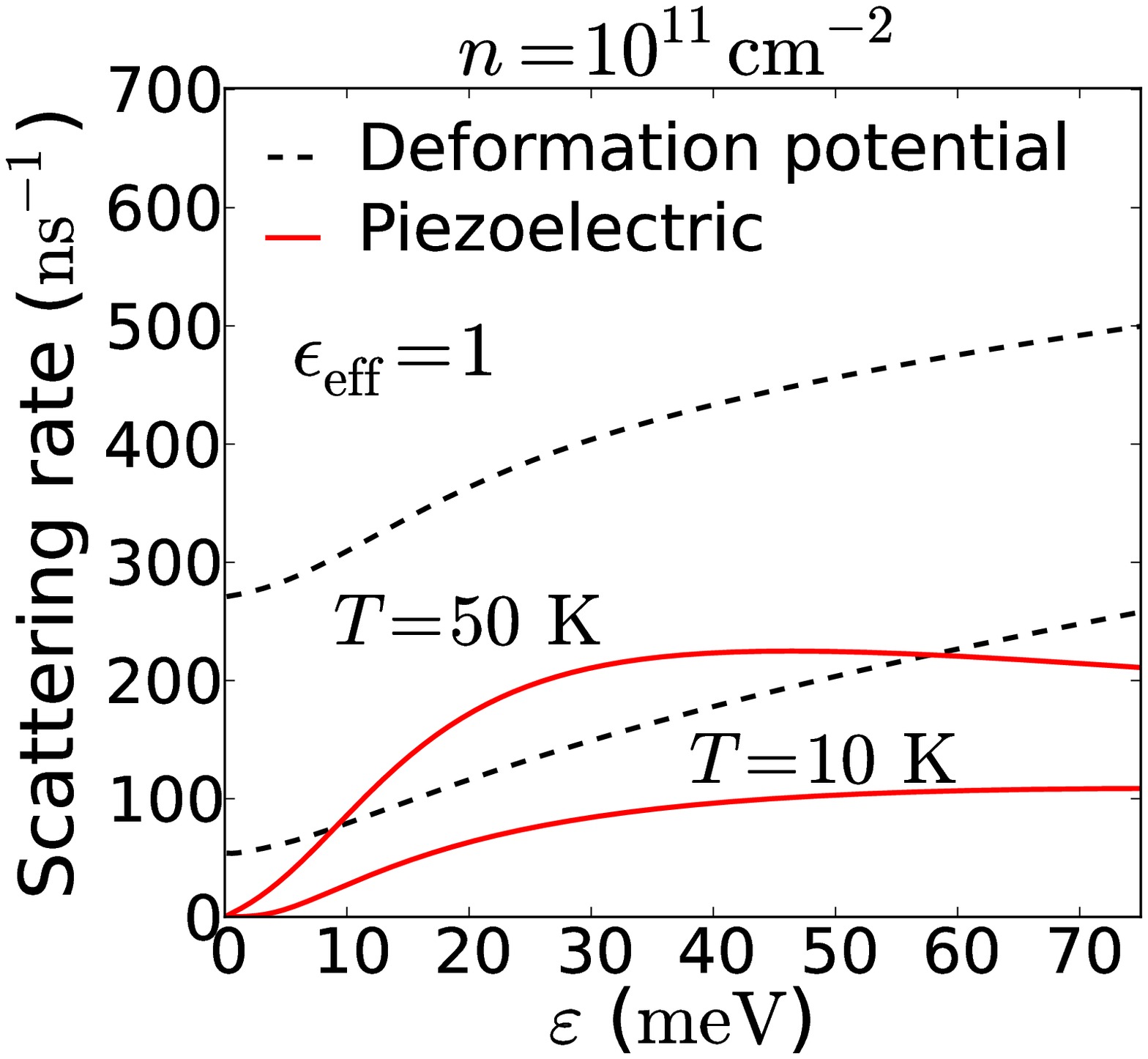}
  \includegraphics[width=0.49\linewidth]{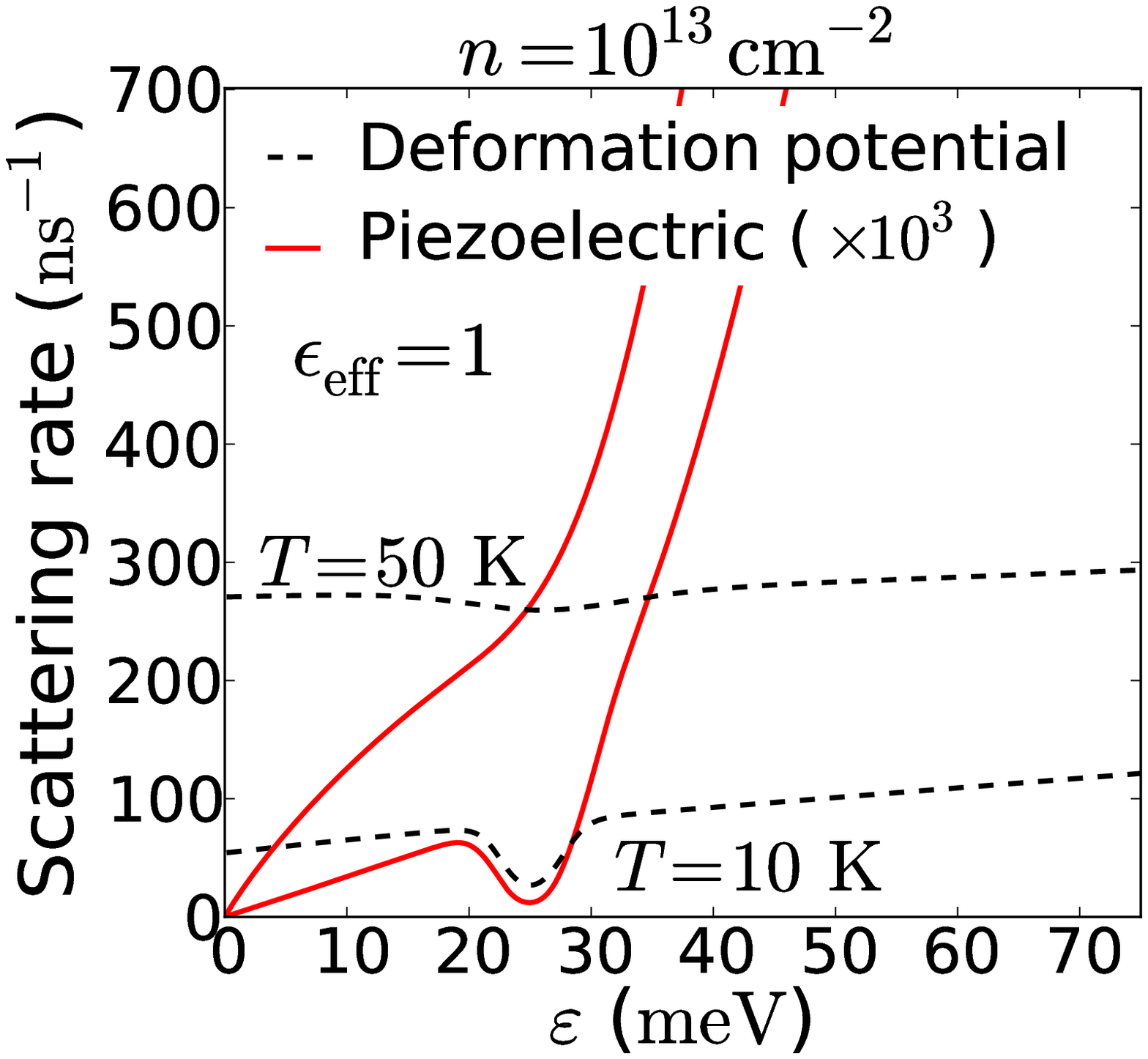}
  \caption{(Color online) Scattering rate for deformation potential (dashed
    lines) and piezoelectric (full lines) interaction in 2D MoS$_2$ at
    temperatures $T=10$~K and $T=50$~K and carrier densities
    $n=10^{11}$~cm$^{-2}$ (left) and $n=10^{13}$~cm$^{-2}$ (right). This
    corresponds to Fermi energies of $E_F \approx 0.25$~meV and $E_F \approx
    25$~meV, respectively. In the right plot, the BG temperature is $T_\text{BG}
    \approx 36$ $(57)$~K for the TA (LA) phonon. At $T=10$~K, the dip in the
    scattering rate that appears at the Fermi energy is a clear fingerprint of
    Bloch-Gr{\"u}neisen physics.}
\label{fig:tau_vs_e}
\end{figure}

In order to establish the relative strength of deformation potential and
piezoelectric scattering in 2D MoS$_2$, we start by considering the scattering
rate given by the expression for the inverse relaxation time in
Eq.~\eqref{eq:tau_acoustic} with the replacement 
$(1 - \cos{\theta_{\bk\bk'}})\rightarrow 1$. Figure~\ref{fig:tau_vs_e} shows
energy dependence of the individual scattering rates due to deformation
potential (dashed lines) and piezoelectric (full lines) scattering for carrier
densities $n=10^{11}$~cm$^{-2}$ (left) and $n=10^{13}$~cm$^{-2}$ (right) and
temperatures $T=10$~K and $T=50$~K. With the Fermi temperature given by $T_F
\approx 29\,\tilde{n}$~K ($E_F \approx 2.5\,\tilde{n}$~meV), the two plots
correspond to a nondegenerate and degenerate carrier distribution,
respectively. The BG temperatures for the TA and LA phonons are in the two
plots: (left) $<10$~K, and (right) $\sim$36~K and $\sim$57~K, respectively.

In the nondegenerate regime shown in the left plot of Fig.~\ref{fig:tau_vs_e},
carrier screening is weak implying that deformation potential and piezoelectric
scattering are of the same order of magnitude. At low energies, however,
deformation potential scattering of the LA phonon and piezoelectric scattering
are strongly screened and unscreened deformation potential scattering of the TA
phonon dominates the scattering rate. The saturation of the piezoelectric
scattering rate at high energies is a consequence of the nonmonotonic
$q$-dependence of the matrix element in Eq.~\eqref{eq:M_PE} (see also
Fig.~\ref{fig:M_fit}).

In the degenerate high-density regime shown in the right plot of
Fig.~\ref{fig:tau_vs_e}, carrier screening is so strong that the piezoelectric
scattering rate is diminished by almost three orders of magnitude relative to
the low-density scattering rate (note the scaling factor in the legend of the
right plot in Fig.~\ref{fig:tau_vs_e}). In this regime, unscreened deformation
potential scattering of the TA phonon therefore completely dominates. The dip in
the scattering rate that develops at the Fermi level with decreasing temperature
is a signature of transport in the BG regime. In this temperature regime the
freezing out of short-wavelength acoustic phonons and the sharpening of the
Fermi surface strongly limit the phase space available for acoustic phonon
scattering resulting in a strong suppression of scattering at the Fermi level.

Next, we consider the temperature and density dependence of the mobility. Due to
the strong anisotropy of the piezoelectric interaction, the mobility in 2D
MoS$_2$ is slightly anisotropic. Along the different high-symmetry directions of
the hexagonal lattice we find that the variation in the mobility is less than
$\sim$10\%. In the following we shall focus on a single direction of the applied
electric field~\cite{footnote5}.

\begin{figure}[!t]
  \includegraphics[width=0.58\linewidth]{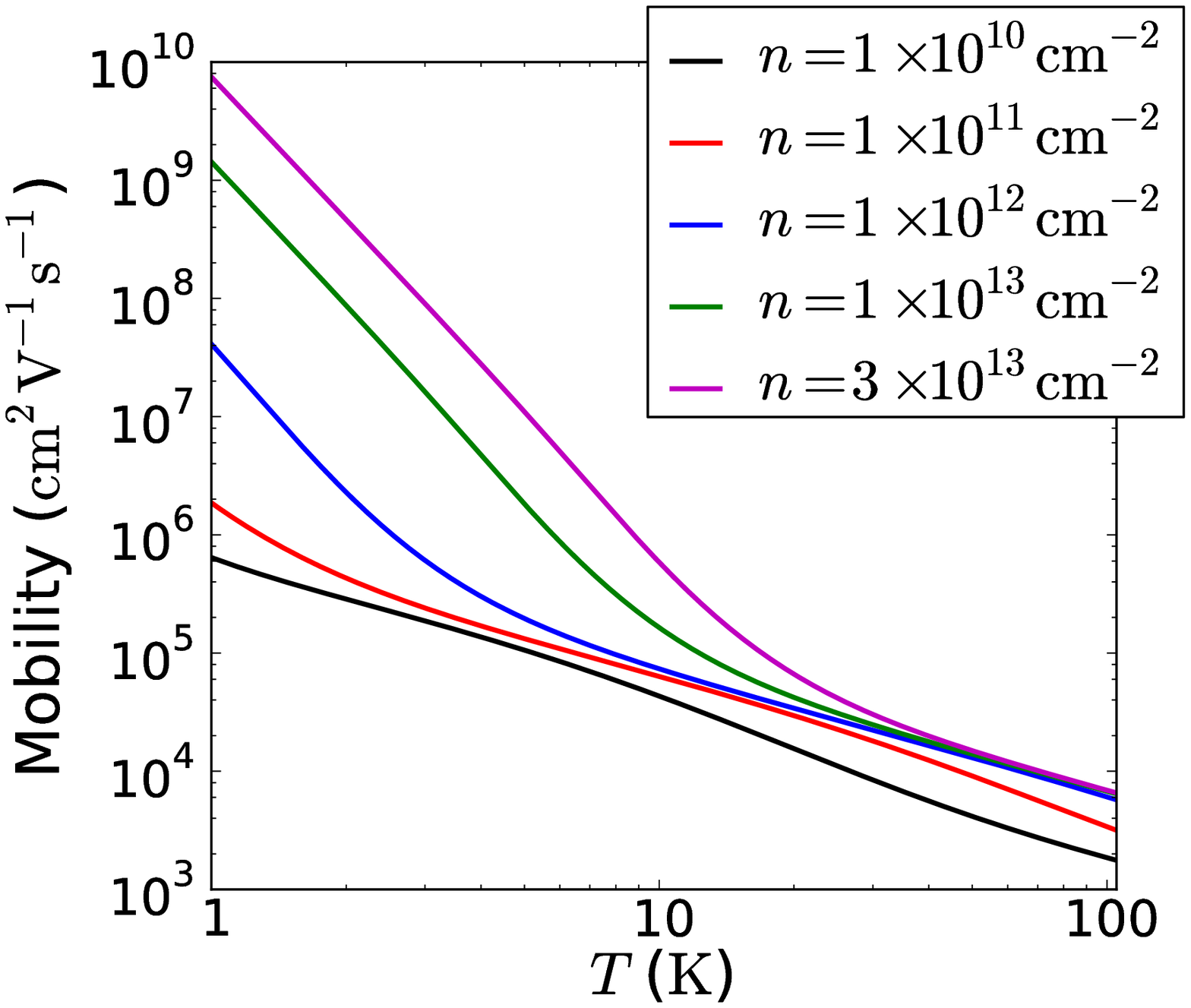}
  \hfill
  \includegraphics[width=0.58\linewidth]{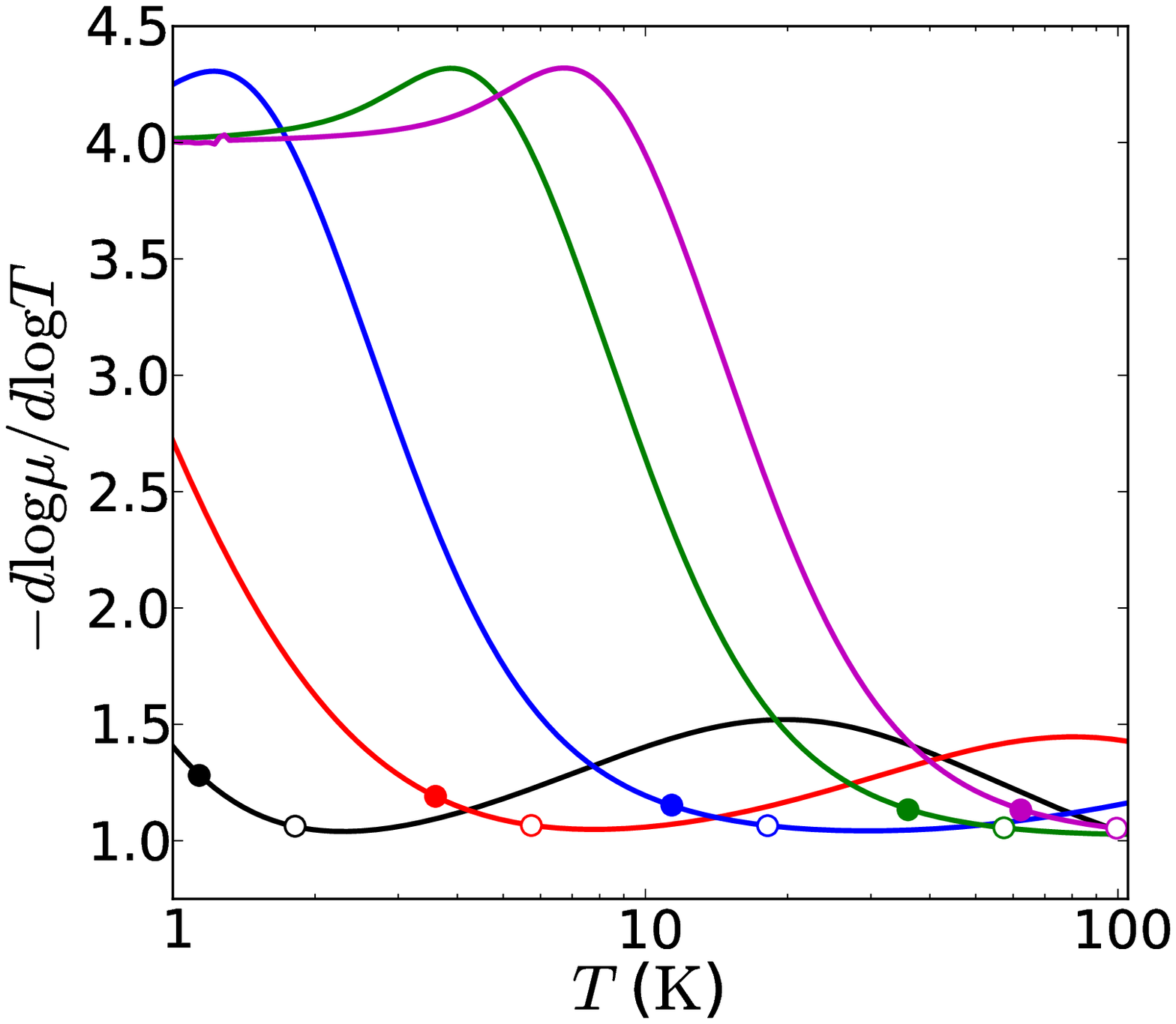}
  \caption{(Color online) Acoustic phonon limited mobility vs
    temperature. Top: Mobility vs temperature for different carrier
    densities. Bottom: Temperature dependence of the exponent $\gamma$ in
    $\mu\sim T^{-\gamma}$ for the same set of carrier densities. The dots mark
    the BG temperatures of the TA ($\bullet$) and LA ($\circ$) phonons,
    respectively.}
\label{fig:mobility_vs_T}
\end{figure}

The temperature dependence of the mobility is shown in
Fig.~\ref{fig:mobility_vs_T} for carrier densities $10^{10}$ to $3\times
10^{13}$~cm$^{-2}$ corresponding to BG temperatures up to $\sim$62~K
($\sim$99~K) for the TA (LA) phonon. Both the mobility (upper) and the exponent
$\gamma = -d\log{\mu}/d\log{T}$ (lower) of its power-law dependence $\mu \sim
T^{-\gamma}$ are shown. At the lowest densities the characteristic temperatures
$T_\text{BG}$ and $T_F$ are comparable while $T_F > T_\text{BG}$ for $n \gtrsim
10^{12}$~cm$^{-2}$. As a consequence, the crossover to the high-mobility BG
regime at $T \sim T_\text{BG}$, marked by the dots in the lower plot, appears
clearly for all carrier densities.

In the high-temperature regime $T \gtrsim T_\text{BG}$, the mobility shows an
approximate linear temperature dependence with $1 \lesssim \gamma \lesssim 1.5$.
This is in good agreement with the individual high-temperature limits for
unscreened deformation potential and piezoelectric scattering which we find to
be $\gamma=1$ and $\gamma \lesssim 1$, respectively. At the lowest carrier
densities, the larger value of $\gamma \sim 1.5$ appearing at $T>T_\text{BG}$
originates from the temperature dependence of the dielectric function for a
nondegenerate carrier distribution.

In the low-temperature BG regime $T<T_\text{BG}$, a stronger temperature
dependence with $1 < \gamma \lesssim 4$ appears and the mobility approaches a
$\mu\sim T^{-4}$ limiting behavior at $T \ll T_\text{BG}$. Numerically, we find
that the $\gamma = 4$ limiting behavior is characteristic of unscreened
deformation potential scattering. Screened deformation potential and
piezoelectric scattering share the same $\gamma = 6$ limit due the identical
long-wavelength limits of their respective couplings in Eqs.~\eqref{eq:M_DP}
and~\eqref{eq:M_PE}. The mobility at $T \ll T_\text{BG}$ is thus completely
dominated by unscreened deformation potential scattering of the TA phonon. Our
findings for the low-temperature limits of the mobility due to scattering of 2D
phonons differ from the usual low-temperature limits of the 2DEG mobility with
scattering of bulk 3D phonons in heterostructures. In this case, the limits are
given by $\gamma=5$ for unscreened deformation potential scattering and
$\gamma=7$ and $\gamma=5$ for screened deformation potential and piezoelectric
scattering, respectively~\cite{Price:BG,West:ObservationOfBG}. As the
low-temperature limits of the mobility are only realized deep inside the BG
regime $T \ll T_\text{BG}$, they may, however, be difficult to observe
experimentally.

\begin{figure}[!b]
  \includegraphics[width=0.59\linewidth]{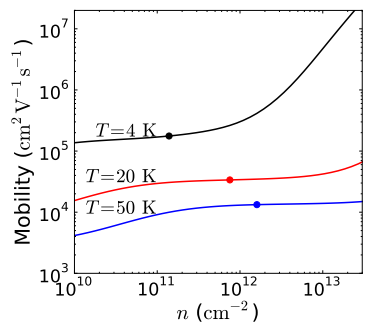}
  \caption{(Color online) Acoustic phonon limited mobility vs carrier density
    for temperatures $T=4$, 20, 50~K. The dots mark the quantum-classical
    crossover from a nondegenerate to a degenerate carrier distribution at
    $T=T_F$.}
\label{fig:mobility_vs_n}
\end{figure}
In Fig.~\ref{fig:mobility_vs_n} we show the calculated density dependence of the
mobility for temperatures $T=4$, 20, 50~K. In the nondegenerate low-density
regime, the density dependence of Debye-H{\"u}ckel screening results in a
mobility that increases with the density. For the lowest temperature where
carrier screening becomes strong, unscreened deformation potential scattering of
the TA phonon dominates the mobility resulting in a weaker density
dependence. The same holds for densities in the vicinity of the
quantum-classical crossover at $T \sim T_F$ (marked by dots in
Fig.~\ref{fig:mobility_vs_n}) where the mobility shows almost no density
dependence. In the degenerate high-density regime, the mobility approaches a
$\mu\sim n^{1.5}$ behavior at low temperatures. The strong density dependence of
the mobility in this regime can be ascribed to two factors: (i) a Fermi velocity
that increases with the carrier density as $v_F\sim\sqrt{n}$, and (ii) transport
in the Bloch-Gr{\"u}neisen regime where the scattering rate for fixed
temperature $T$ decreases with the density. The latter is a consequence of a
reduction in the fraction of the Fermi surface that is probed by acoustic phonon
scattering when the density (Fermi wave vector) increases.

We end by briefly commenting on the applicability of the expression
$1/\mu=1/\mu_0 + \alpha T$ often used to fit experimental
mobilities~\cite{West:ObservationOfBG,Sarma:Hetero1,Sarma:Hetero2} in the linear
high-temperature regime. Here, $\mu_0$ is the residual mobility due to
e.g. impurity scattering and $\alpha$ is the density-dependent slope of the
linear temperature dependence. In Fig.~\ref{fig:mobility_vs_T} it is seen to
apply at $T > T_\text{BG}$ except for the lowest densities where the inverse
mobility becomes slightly nonlinear. From the density dependence of the mobility
in Fig.~\ref{fig:mobility_vs_n} we conclude that the temperature coefficient
$\alpha$ is a monotonically decreasing function of the carrier density.

\section{Conclusions}

In this work we have combined analytic and first-principles calculations of the
el-ph interaction with an semianalytic solution of the Boltzmann equation to
study the temperature and density dependence of the acoustic phonon limited
mobility in 2D $n$-type MoS$_2$. The acoustic deformation potentials for the TA
and LA phonons and the piezoelectric constant in 2D MoS$_2$ were extracted from
the first-principles el-ph interaction and from a microscopic description of
carrier screening it was shown that the umklapp contribution to the deformation
potential interaction is not affected by screening.

Due to strong screening of deformation potential scattering of the LA phonon and
piezoelectric scattering of both the TA and LA phonon, the mobility was found to
be dominated by unscreened deformation potential scattering of the TA phonon at
high densities. At low carrier densities $10^{10}$--$10^{11}$~cm$^{-2}$
deformation potential and piezoelectric scattering were found to be
comparable. For $T<10$~K and moderate to high carrier densities $n\gtrsim
10^{11}$~cm$^{-2}$, intrinsic mobilities in excess of
$~10^5$~cm$^2$~V$^{-1}$~s$^{-1}$ were predicted. At temperatures $T\sim 100$~K,
the acoustic phonon limited mobility does not exceed $\sim7\times
10^3$~cm$^2$~V$^{-1}$~s$^{-1}$. In the low-temperature BG regime $T \ll
T_\text{BG}$, the mobility acquires a $\mu\sim T^{-4}$ dependence characteristic
of unscreened deformation potential scattering of 2D phonons. The mobility was
furthermore found to increase monotonically with the carrier density. Similar
conclusions can be expected to hold for monolayers of other transition metal
dichalcogenides which have similar atomic and electronic structures.

Apart from our findings for the mobility, we here list a few other key results
of our work: (i) In the long-wavelength limit the acoustic deformation potential
interactions for the TA and LA phonons were found to be completely dominated by
umklapp and normal processes, respectively; (ii) from a microscopic treatment of
carrier screening we showed that normal and umklapp processes in the el-ph
interaction are screened differently, and that the deformation
potential interaction with the TA phonon to a good approximation can be left
unscreened; (iii) our further developments for first-principles calculations of
the el-ph interaction included in App.~\ref{app:firstprinciples}.

As a final remark, we note that our conclusion regarding the screening of the
acoustic el-ph interaction can be verified experimentally. For a purely
long-wavelength treatment of carrier screening, we find that the theoretically
predicted mobility is substantially higher and shows a much richer temperature
dependence with higher values of $\gamma$ and no linear temperature dependence
(see also Ref.~\onlinecite{Sarma:Interplay}). Experimental low-temperature
mobility data on high-mobility monolayer MoS$_2$ samples matching the
predictions of this work will hence provide strong support for our findings. As
our findings for the acoustic el-ph interaction must be expected to be relevant
in other semiconductors, we believe that the present study is of high importance
for an improved understanding of the acoustic el-ph interaction and
phonon limited mobilities in semiconductor based 2DEGs.

\begin{acknowledgments}
  We thank O.~Hod and A.~Konar for fruitful discussions. The Center for
  Nanostructured Graphene (CNG) is sponsored by the Danish National Research
  Foundation, Project No. DNRF58. This work was supported by the Villum Kann
  Rasmussen Foundation.
\end{acknowledgments}

\appendix

\section{2D Thomas-Fermi screening}
\label{app:screening}

In the Thomas-Fermi (TF) approach to screening, the finite-temperature
dielectric function of a 2DEG is in the long-wavelength limit given
by~\cite{Stern:2D}
\begin{equation}
  \epsilon(q, T) = 1 + \frac{q_\text{TF}(T)}{q} \\
\end{equation}
where 
\begin{equation}
  q_\text{TF}(T) = q_\text{TF}
  \left[ 
    1 - \exp{\left( \frac{-E_F}{k_\text{B}T} \right) }
  \right]
\end{equation}
is the temperature-dependent screening wave vector, $q_\text{TF} =
e^2/(2\epsilon_0)\rho$ the zero-temperature TF wave vector, $\rho = g_s g_v
m^*/2\pi\hbar^2$ the constant density of states in 2D, and $E_F=n/\rho$ the
Fermi level. This reproduces the RPA and Debye-H{\"u}ckel results
\begin{align}
  q_\text{TF}(T\rightarrow 0) & = q_\text{TF} , \\
  q_\text{TF}(T\rightarrow \infty) & = q_\text{D} = \frac{ne^2}{2\epsilon_0 k_\text{B}T} 
\end{align}
for the screening wave vector in the low and high-temperature limits,
respectively.

In a degenerate 2DEG, Thomas-Fermi screening overestimates the screening
strength at $q > 2 k_F$ where 2DEG screening becomes less efficient. RPA
corrections to the dielectric function are required to cure this
problem~\cite{Stern:2D}. However, for quasi-elastic scattering with $q \lesssim
2 k_F$, TF theory provides a good approximation to the dielectric function.

\section{Continuum theory for the acoustic el-ph interaction in 2D materials}
\label{app:continuum}

In this appendix, we calculate the acoustic el-ph interaction in 2D materials
using continuum theory. For this purpose, the electronic states are described by
plane-wave solutions $\psi_\bk(\br) = 1/\sqrt{A} \chi_\bk(z)
e^{i\bk\cdot\br_\parallel}$ where $A$ is the area of the sample, $\bk$ is the
two-dimensional electronic wave vector, $\br=(\br_\parallel, z)$, and
$\chi_\bk(z)$ is the normalized envelope of the electronic wave functions
accounting for its confinement in the direction perpendicular to the material
layer.

The Hamiltonian for the el-ph interaction takes the usual form
\begin{equation}
  \label{eq:H_elph}
  H_\text{el-ph}  = \sum_{\bk\bq\lambda} g_{\bk\bq}^\lambda
     c_{\bk+\bq}^\dagger c_{\bk}^{\phantom\dagger} (a_{\bq\lambda}^\dagger +
     a_{-\bq\lambda}) ,
\end{equation}
where $\bq=(q_x,q_y)$ is the two-dimensional phonon wave vector, $\lambda$ is
the acoustic branch index and
$g_{\bk\bq}^\lambda=\sqrt{\hbar/2A\rho\omega_{\bq\lambda}} M_{\bk\bq}^\lambda$
is the coupling constant. In the following, the matrix element
$M_{\bk\bq}^\lambda$ is obtained by applying an elastic continuum model for the
acoustic phonons.

In a lattice without an inversion center, the acoustic el-ph is a sum of
deformation potential (DP) and piezoelectric (PE) interactions. The unscreened
el-ph interaction which couples to the carrier density is in real space given
by~\cite{Madelung,Mahan}
\begin{align}
  \label{eq:H_elph_realspace}
  H_\text{el-ph}(\br) & = H_\text{DP}(\br) + H_\text{PE}(\br) \nonumber \\
                      & = \Xi \nabla\cdot \mathbf{u}(\br) - e \phi(\br)  ,
\end{align}
where $\mathbf{u}$ is the displacement field due to the acoustic phonons, $\Xi$
is the deformation potential, and $\phi$ is the electrostatic potential from the
piezoelectric polarization of the lattice.

Considering an isolated 2D material sheet, the in-plane acoustic phonons can be
described by the quantized two-dimensional displacement field
\begin{equation}
  \label{eq:displacement}
  \mathbf{u}(\br) = \sum_{\bq\lambda} \mathbf{u}_{\bq\lambda}(z) e^{i \bq\cdot\br_\parallel}
  , \quad \mathbf{u}_{\bq\lambda}(z) = \hat{\mathbf{e}}_\lambda f_\bq(z) Q_{\bq\lambda}
\end{equation}
where $\bq=(q_x,q_y)$ is the two-dimensional phonon wave vector,
$\hat{\mathbf{e}}_\lambda$ is a unit vector describing the polarization of the
acoustic branch $\lambda$, $f_\bq$ is the $z$-profile of the displacement field
in the direction perpendicular to the material sheet, and $Q_{\bq\lambda} =
\sqrt{\hbar/2A \rho \omega_{\bq\lambda}} (a_{\bq\lambda}^\dagger +
a_{-\bq\lambda})$ is the vibrational normal coordinate. In the long-wavelength
limit the polarization vectors for the TA and LA phonons are perpendicular
($\hat{\mathbf{e}}_\text{TA} \perp \bq$) and parallel
($\hat{\mathbf{e}}_\text{TA} \parallel \bq$) to the phonon wave vector $\bq$,
respectively, and can to a good approximation be assumed independent on
$q$~\cite{Mahan}.

With the displacement field written in the form in Eq.~\eqref{eq:displacement},
the Hamiltonian~\eqref{eq:H_elph_realspace} can be recast as a sum over terms
from the individual phonons,
\begin{align}
  \label{eq:H_z}
  H_\text{el-ph}(\br) & = \sum_{\bq\lambda} H_\text{el-ph}^{\bq\lambda}(\br) 
  \nonumber \\
  & = \sum_{\bq\lambda} \left[ 
    H_{\bq\lambda}^\text{DP}(z) + H_{\bq\lambda}^\text{PE}(z)
  \right] e^{i \bq\cdot \br_\parallel} ,
\end{align}
where $H_{\bq\lambda}^\text{DP/PE}$ are to be determined below. The el-ph
coupling constant is given by the matrix element
\begin{align}
  \label{eq:g_continuum}
  g_{\bk\bq}^\lambda & = \int \! d\br \; \psi_{\bk+\bq}^*(\br) 
      H_\text{el-ph}^{\bq\lambda}(\br) \psi_\bk (\br)  \nonumber \\
      & = \int \! dz \; \chi_{\bk+\bq}^*(z) 
      H_\text{el-ph}^{\bq\lambda}(z) \chi_\bk (z)  .
\end{align}
In the following, the envelope function $\chi_\bk$ is assumed independent of the
electronic wave vector $\bk$.

\subsection{Deformation potential interaction}
\label{app:deformation}

Taking the divergence of the displacement field in Eq.~\eqref{eq:displacement},
the deformation potential interaction is found to be
\begin{equation}
  H_{\bq\lambda}^\text{DP}(z) = i \Xi 
      \bq\cdot\hat{\mathbf{e}}_\lambda  f_\bq(z) Q_{\bq\lambda} .
\end{equation}
Because of the dot product between the phonon wave vector and the polarization
vector, the interaction with the TA phonon vanishes. The coupling matrix element
for the LA phonon is given by
\begin{equation}
  \label{eq:M_DP_continuum}
  M_{\bq\lambda}^\text{DP} = i \Xi q  , 
\end{equation}
where the result of the $z$-integral in Eq.~\eqref{eq:g_continuum} has been
absorbed in the deformation potential constant. The fact the TA phonon does not
couple, illustrates the limitation of the often assumed form for deformation
potential interaction in Eq.~\eqref{eq:H_elph_realspace} and underlines the
importance of more involved descriptions~\cite{Dery:SymmetryBased}.

\subsection{Piezoelectric interaction in 2D hexagonal lattices}
\label{app:piezo}

Piezoelectric interaction with acoustic phonons appears in lattices which lack a
center of symmetry. In this case, the displacement field $\mathbf{u}(\br)$
associated with the acoustic phonons leads to a polarization $\mathbf{P}$ of the
lattice given by~\cite{Mahan}
\begin{equation}
  \label{eq:P_piezo}
  P_i = \sum_{jk} e_{i,jk} \epsilon_{jk},
  \quad i,j,k = x,y
\end{equation}
where 
\begin{equation}
  \label{eq:strain}
  \epsilon_{ij} = \frac{1}{2} 
           \left( 
             \frac{\partial u_i}{\partial x_j}  + \frac{\partial u_j}{\partial x_i}
           \right) 
\end{equation}
is the strain tensor and $e$ is the tensor of piezoelectric moduli having
symmetry $e_{i,jk} = e_{i,kj}$. In 2D materials the piezoelectric coupling has
units of C/m (C/m$^2$ in 3D)---the displacement field can be thought to have a
normalized $z$-profile with units of m$^{-1}$. For a 2D hexagonal lattice with a
basis there is only one independent piezoelectric component $e_{11}$ (Voigt
notation) which is related to the other nonzero components
as~\cite{Mele:NTPiezo}
\begin{equation}
  \label{eq:piezo_hexagonal}
  e_{11} = - e_{12} = - e_{26} .
\end{equation}
Here, the primitive lattice vectors of the hexagonal lattice have been chosen as
$\mathbf{a}_{1,2} = a (\sqrt{3}/2, \pm 1/2)$ where $a$ is the lattice constant.

Expanding the strain tensor as in Eq.~\eqref{eq:displacement}, its
($\bq$,$\lambda$)-components follow directly from Eq.~\eqref{eq:strain} as
\begin{align}
  \epsilon_{xx}^{\bq\lambda}(z) & = i q_x \hat{e}_{\lambda,x} f_\bq(z) Q_{\bq\lambda}\\
  \epsilon_{xy}^{\bq\lambda}(z) & = \epsilon_{yx} = \frac{i}{2} \left( 
    q_y \hat{e}_{\lambda,x} + q_x \hat{e}_{\lambda,y} 
  \right) f_\bq(z) Q_{\bq\lambda},
\end{align}
and the associated polarization in Eq.~\eqref{eq:P_piezo} is given by
\begin{align}
  \label{eq:P_hexagonal}
  P_{\bq,x}^\lambda(z ) & = e_{11} 
  \left( 
    \epsilon_{xx} - \epsilon_{yy} 
  \right)   \nonumber \\ 
  & = i e_{11} \left( 
     q_x \hat{e}_{\lambda,x} - q_y \hat{e}_{\lambda,y} 
  \right) f_\bq(z) Q_{\bq\lambda} \\ 
  P_{\bq,y}^\lambda (z) & = - 2 e_{11} \epsilon_{xy} \nonumber \\ 
  & = - i e_{11} 
    \left( q_y \hat{e}_{\lambda,x} + q_x \hat{e}_{\lambda,y}
  \right) f_\bq(z) Q_{\bq\lambda}.
\end{align}
The potential $\phi$ resulting from the piezoelectric polarization field is
given by Poisson's equation $-\epsilon_0 \nabla^2 \phi(\br) = \rho$ where $\rho
= - \nabla \cdot \mathbf{P}$ is the polarization charge. Since we are
considering an isolated material sheet, the only boundary condition that applies
is $\phi\rightarrow 0$ for $z \rightarrow\pm\infty$. Fourier transforming in all
three directions, we find
\begin{equation}
  \label{eq:poisson_ft}
  \epsilon_0 (q^2 + k^2) \phi_\bq (k) = \rho_\bq f_\bq(k) ,
\end{equation}
where $k$ is the Fourier variable in the direction perpendicular to the plane of
the layer. The Fourier components of the branch-resolved piezoelectric
polarization charge are given by 
\begin{align}
  \rho_{\bq\lambda} & = -i \bq \cdot \mathbf{P}_{\bq\lambda} \nonumber \\ 
  & = e_{11} 
  \left( 
    q_x^2 \hat{e}_{\lambda,x} - q_y^2 \hat{e}_{\lambda,x} 
    - 2 q_x q_y \hat{e}_{\lambda,y}
  \right) Q_{\bq\lambda} \nonumber \\
  & \equiv e_{11}  q^2 A_\lambda(\hat{\bq}) Q_{\bq\lambda}
\end{align}
where the angular dependencies have been collected in the anisotropy factor
$A_\lambda$. It is given by
\begin{equation}
  \label{eq:anisotropy}
  A_\text{TA}(\hat{\bq}) = - \sin{3 \theta_\bq} \quad \text{and} \quad
  A_\text{LA}(\hat{\bq}) = \cos{3 \theta_\bq}
\end{equation}
for the TA and LA phonon, respectively.

The $z$-dependence of the piezoelectric potential is given by the inverse Fourier
transform of~\eqref{eq:poisson_ft} with respect to $k$ which yields
\begin{align}
  \label{eq:phi_PE}
  \phi_{\bq\lambda}(z) & = \frac{e_{11}}{\epsilon_0} A_\lambda(\hat{\bq})Q_{\bq\lambda}
                \int \! dk \; e^{ikz} f_\bq(k)
                \frac{q^2}{q^2 + k^2} \nonumber \\
              & = \frac{e_{11}}{\epsilon_0}  q e^{-q \abs{z}}
                  A_\lambda(\hat{\bq}) Q_{\bq\lambda} ,
\end{align}
where in the last equality, a $\delta$-function $z$-profile with $f_\bq(k) = 1$
has been assumed. For atomically thin materials, this should be a reasonable
approximation. 
\begin{figure}[!t]
  \includegraphics[width=0.49\linewidth]{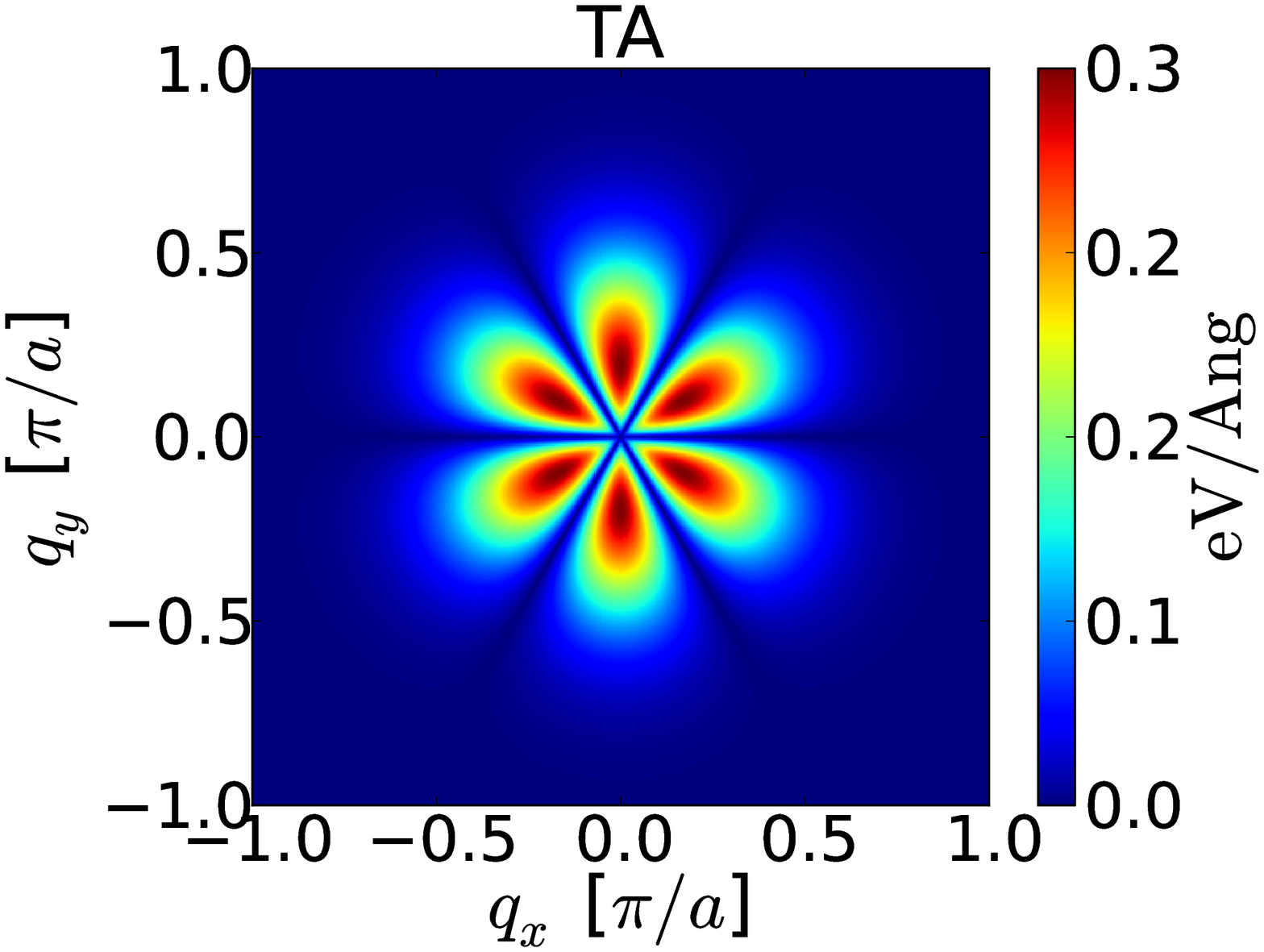}
  \includegraphics[width=0.49\linewidth]{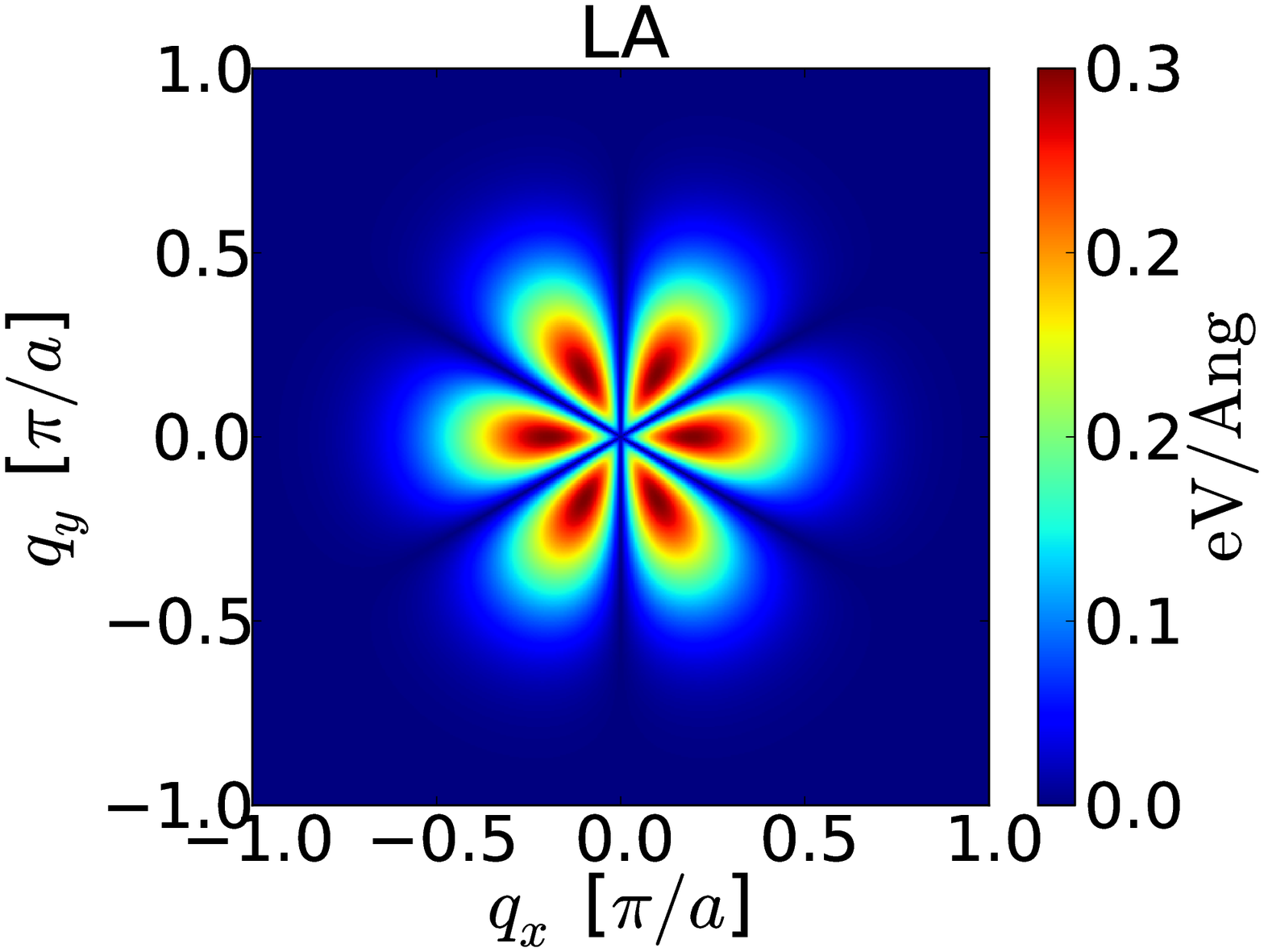}
  \caption{(Color online) Piezoelectric interaction in a 2D hexagonal
    lattice. The plots show the absolute value of the coupling matrix element
    $M_{\bq\lambda}^\text{PE}$ in Eq.~\ref{eq:M_PE_continuum} for the TA (left)
    and LA (right) phonon as a function of the phonon wave vector $\bq$. The
    parameters for MoS$_2$ in Tab.~\ref{tab:parameters} have been used.}
  \label{fig:M_analytic}
\end{figure}

The piezoelectric el-ph interaction in Eq.~\eqref{eq:H_z} is now given by
$H_{\bq\lambda}^\text{PE}(z) = -e \phi_{\bq\lambda}(z)$, and assuming, for
simplicity, a Gaussian envelope in Eq.~\eqref{eq:g_continuum}, the
long-wavelength limit of the piezoelectric coupling matrix element
becomes~\cite{Kaasbjerg:MoS2}
\begin{equation}
  \label{eq:M_PE_continuum}
   M_{\bq\lambda}^\text{PE} = \frac{e_{11} e}{\epsilon_0}
                    q \times \text{erfc}(q \sigma / 2) A_\lambda(\hat{\bq}) ,
\end{equation}
where $\sigma$ is the effective width of the electronic envelope function. The
piezoelectric interaction has the general property that
$M_{-\bq\lambda}^\text{PE} = -M_{\bq\lambda}^\text{PE}$~\cite{Mahan}. The
absolute value of the piezoelectric interaction~\eqref{eq:M_PE_continuum} is
shown in Fig.~\ref{fig:M_analytic}. Here, the six fold rotational symmetry stems
from the hexagonal crystal lattice and is accounted for by the anisotropy
factors in Eq.~\eqref{eq:anisotropy}.

Contrary to the 3D bulk case where the piezoelectric interaction is independent
of $q$~\cite{Mahan}, the 2D interaction in Eq.~\eqref{eq:M_PE_continuum} depends
on the magnitude of the phonon wave vector and acquires a linear $q$-dependence
for $q\rightarrow 0$. The latter is a consequence of the 2D crystal lattice
which does not support a piezoelectric potential from the acoustic phonons in
the long-wavelength limit (see Eq.~\eqref{eq:phi_PE}).

\section{First-principles calculation of the el-ph interaction}
\label{app:firstprinciples}

In this appendix, we present the first-principles method applied in the
calculation of the el-ph interaction. In order to support the new developments
included in this work, we start by briefly highlighting the most important
aspects of the main method presented in full detail in
Ref.~\onlinecite{Kaasbjerg:MoS2}.

The coupling matrix element $M_{\bk\bq}^\lambda$ in the el-ph interaction in
Eq.~\eqref{eq:g} of the main text involves the change in the crystal potential
$\delta V_{\bq\lambda}$ due to a phonon with wave vector $\bq$ and branch index
$\lambda$. Under the assumption that the atomic displacements are small, the
phonon-induced change in the potential can be constructed as a sum over
individual atomic gradients,
\begin{equation}
  \label{eq:deltaV}
  \delta V_{\bq\lambda}(\br) = \frac{1}{N} \sum_{\alpha l} 
      e^{i \bq\cdot \mathbf{R}_l} 
      \hat{\mathbf{e}}_{\bq\lambda}^\alpha \cdot
      \nabla_{\alpha l} V(\br)  .
\end{equation}
Here, $\alpha$ is an atomic index in the primitive unit cell, $\mathbf{R}_l$ is
the lattice vector of unit cell $l$ (relative to the reference unit cell at
$\mathbf{R}_0$), $\hat{\mathbf{e}}_{\bq\lambda}$ is the mass-scaled phonon
polarization vector, $\nabla_{\alpha l}$ denotes the gradient with respect to
displacements of atom $(\alpha, l)$ in the $x,y,z$ directions, $V$ is the
crystal potential, and $N$ is the number of unit cells in the lattice.

\begin{figure}[!t]
  \includegraphics[width=0.6\linewidth]{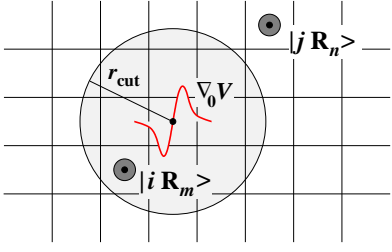}
  \hfill
  \caption{(Color online) Schematic illustration of the LCAO supercell matrix
    involved in the calculation of the el-ph coupling in
    Eq.~\eqref{eq:M_first}. The square lattice indicates the unit cells of the
    crystal lattice. The real-space cutoff $r_\text{cut}$ measured from the
    position of the atomic site where the gradient of the potential is taken is
    used to separate out the short- and long-range part of the el-ph
    interaction. Matrix elements involving LCAO orbitals located beyond the
    cutoff as the one shown are defined as long range (see also
    Eq.~\eqref{eq:separation}).}
\label{fig:supercell_matrix}
\end{figure}

The matrix element of the phonon-induced potential between the Bloch states with
wave vectors $\bk$ and $\bk+\bq$ are evaluated by expanding the Bloch function
in an LCAO basis. The resulting expression for the matrix element follows
by exploiting the periodicity of the crystal lattice and takes the
form~\cite{Kaasbjerg:MoS2} 
\begin{align}
  \label{eq:M_first}
  M_{\bk\bq}^\lambda & = \bra{\mathbf{k+q}} \delta V_{\bq\lambda}(\br) \ket{\bk}
  \nonumber \\
  & = \frac{1}{N} \sum_{ij} c_i^{*} c_j  \sum_{mn}
      e^{i \bk \cdot (\mathbf{R}_n - \mathbf{R}_m) -
         i \bq \cdot \mathbf{R}_m}
       \nonumber \\
  & \quad \times  \bra{i \mathbf{R}_m} \hat{\mathbf{e}}_{\bq\lambda} \cdot
      \nabla_0 V(\br) \ket{j \mathbf{R}_n}  ,
\end{align}
where $i=(\alpha, \mu)$ is a composite atomic ($\alpha$) and orbital ($\mu$)
index, $\ket{i\mathbf{R}_m}$ denotes the atomic orbital $\mu$ on atom $\alpha$
in the primitive unit cell $m$, $c_i$ are the LCAO expansion coefficients, and
$\nabla_0 V$ is the gradient of the crystal potential with respect to atomic
displacements in the reference unit cell. The quantity in the last line of
Eq.~\eqref{eq:M_first} is the LCAO supercell matrix of the potential
gradient. The real-space structure of its matrix elements is illustrated
schematically in Fig.~\ref{fig:supercell_matrix}.

\subsection{Real-space separation of the short and long-range part of the
  electron-phonon interaction}
\label{app:partition}

In first-principles calculations of the el-ph interaction both short range
(deformation potential) and long range (piezoelectric and Fr{\"o}hlich
interaction) are included in the coupling in Eq.~\eqref{eq:g}. However, due to
their different origin it may sometimes be desirable to consider them
separately. In the following we outline a real-space partitioning scheme to
separate the short-range and long-range contributions to the el-ph interaction.

The central quantity of the partitioning scheme is the LCAO supercell matrix
illustrated schematically in Fig.~\ref{fig:supercell_matrix}. The partitioning
scheme consists in splitting up the summations in Eq.~\eqref{eq:M_first} into
two contributions; (i) a short-range part which neglects all matrix elements
involving LCAO orbitals beyond a chosen real-space cutoff $r_\text{cut}$, and
(ii) a long-range part which includes the remaining matrix elements; i.e.
\begin{widetext}
\begin{equation}
  \label{eq:separation}
  \bra{i \mathbf{R}_m} \nabla_{\alpha 0} V(\br) \ket{j \mathbf{R}_n}
  \rightarrow 
  \left\{
    \begin{array}{l}
      \text{short-range,}\;  \abs{\mathbf{R}_{im} - \mathbf{R}_{k0}} < r_\text{cut} 
      \; \text{and} \; \abs{\mathbf{R}_{jn} - \mathbf{R}_{k0}} < r_\text{cut} \\
      \quad \\
      \text{long-range,} \; 
      \abs{\mathbf{R}_{im/jn} - \mathbf{R}_{k0}} > r_\text{cut} 
    \end{array}
  \right.
\end{equation}
\end{widetext}
where $\mathbf{R}_{mi}=\mathbf{R}_m + \mathbf{R}_\alpha$ denotes the atomic
positions of the LCAO orbitals and the potential gradients.
 
In general, the real-space cutoff $r_\text{cut}$ must be chosen small enough
that the short-range part does not include contributions from truly long-range
effects in the relevant range of phonon wave vectors; i.e. $\pi / r_\text{cut} >
q_\text{max}$ where $q_\text{max}$ is the maximum phonon wave vector of
interest. At the same time, the cutoff cannot be chosen too small that
short-range effects are cut off. As these guidelines do not provide a unique way
to choose the real-space cutoff, it should be verified in an actual calculation
that the results do not change significantly for different values of the cutoff.
\begin{figure}[!t]
  \begin{minipage}{1.0\linewidth}
    \begin{centering}
      \textsf{\textbf{Deformation potential interaction}}
    \end{centering}
    \vspace{0.2cm}
  \end{minipage}
  \includegraphics[width=0.49\linewidth]{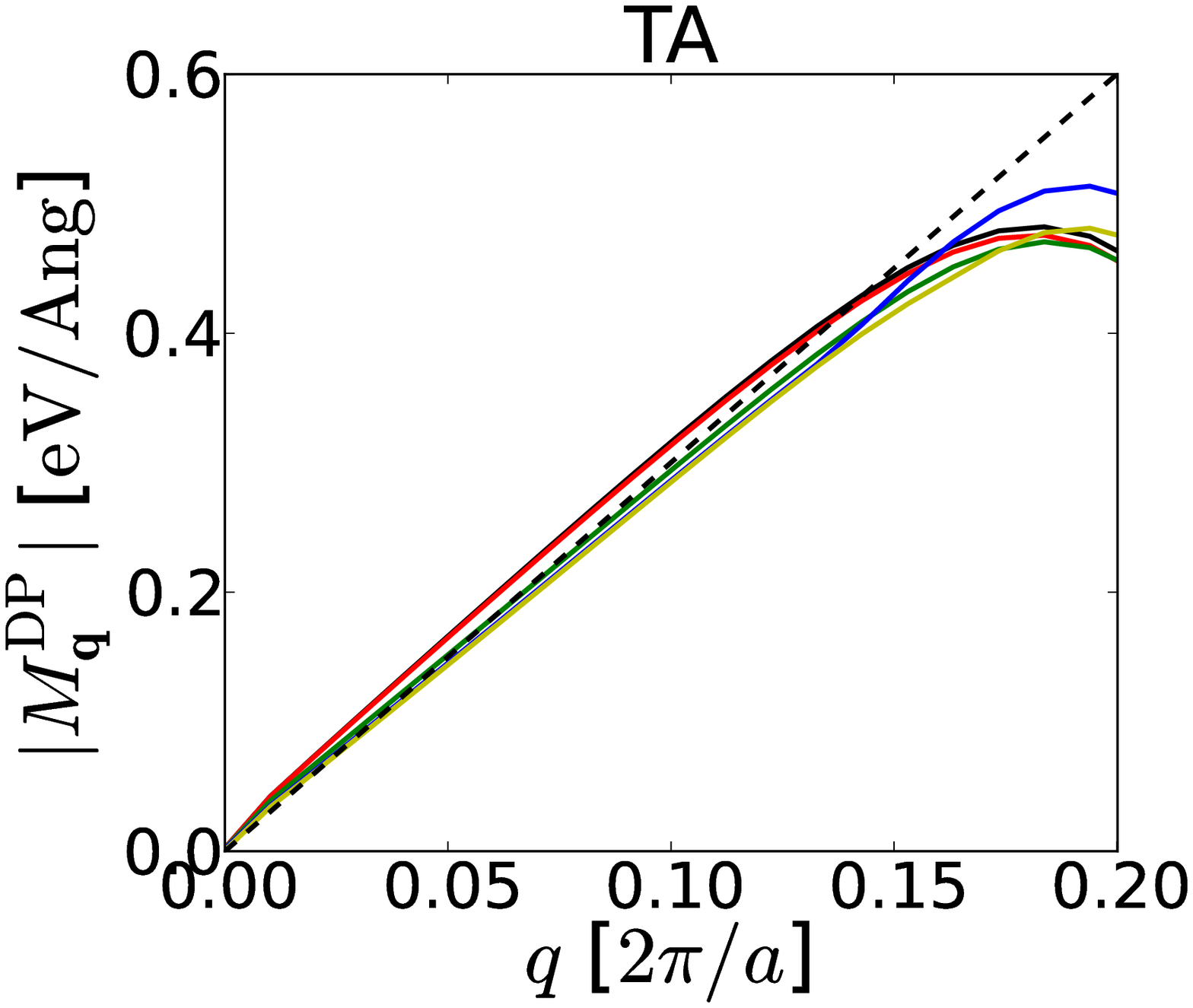}
  \includegraphics[width=0.49\linewidth]{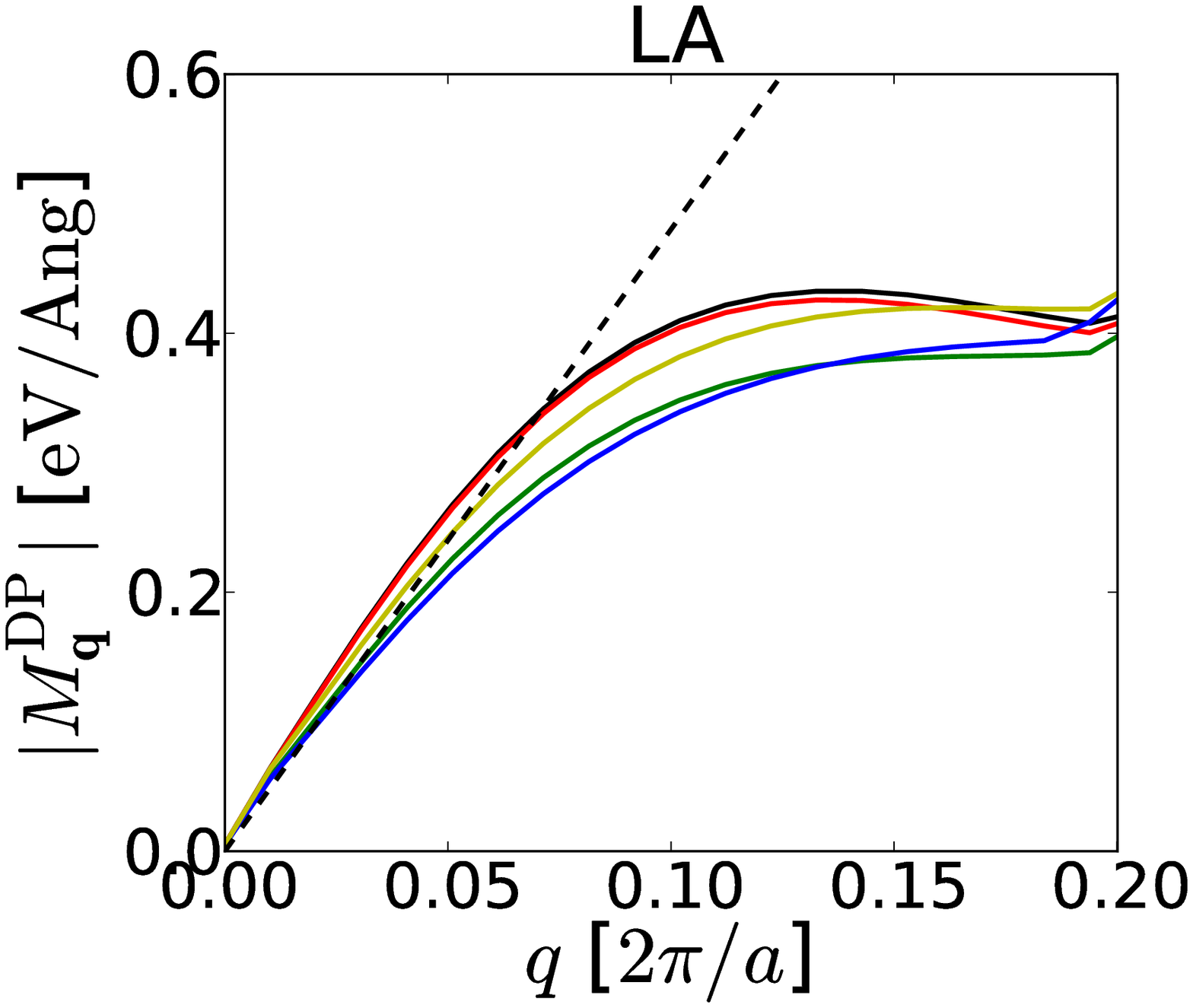}
  \begin{minipage}{1.0\linewidth}
  \vspace{0.4cm}
    \begin{centering}
      \textsf{\textbf{Piezoelectric interaction}}
    \end{centering}
    \vspace{0.2cm}
  \end{minipage}
  \includegraphics[width=0.49\linewidth]{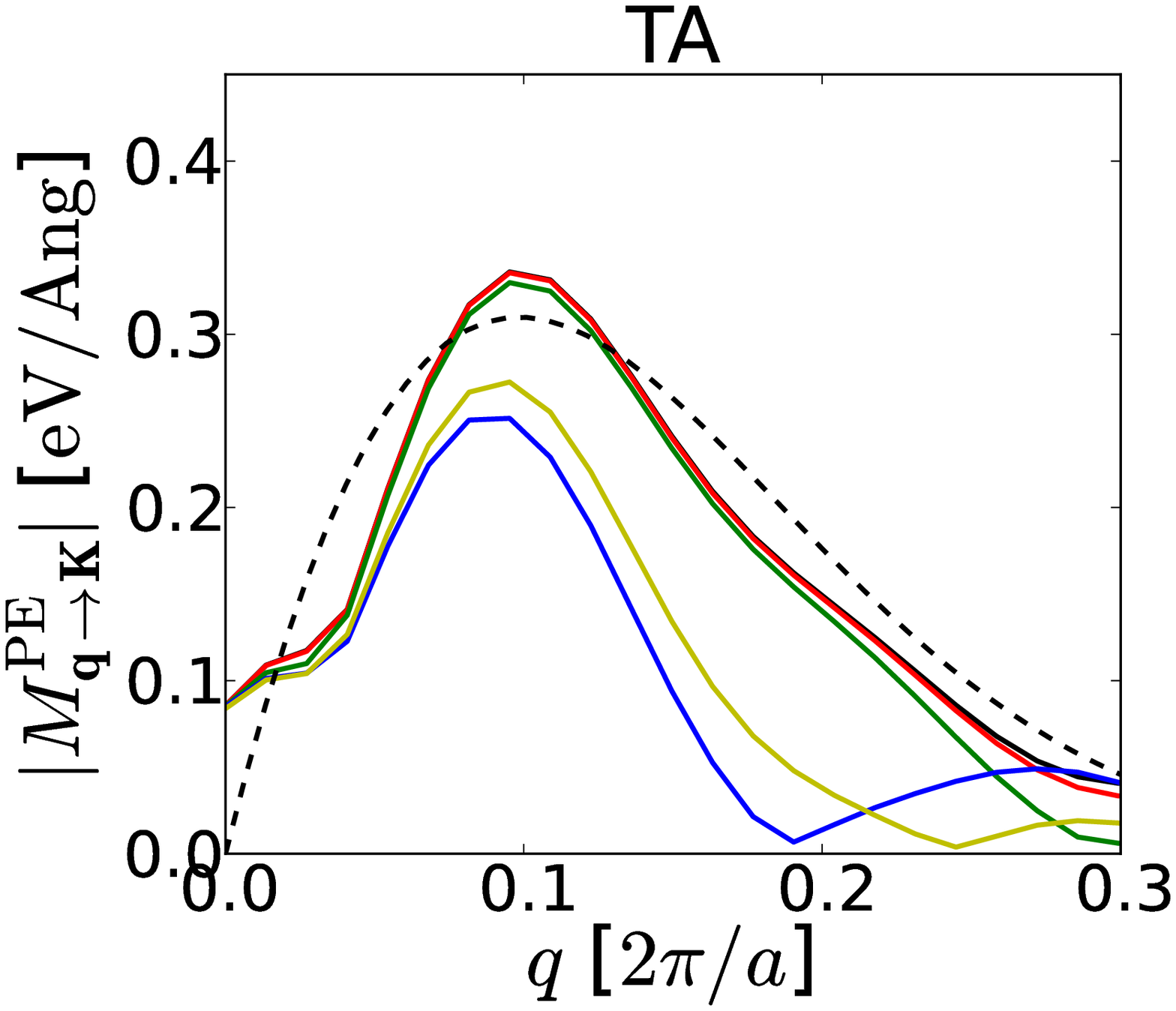}
  \includegraphics[width=0.49\linewidth]{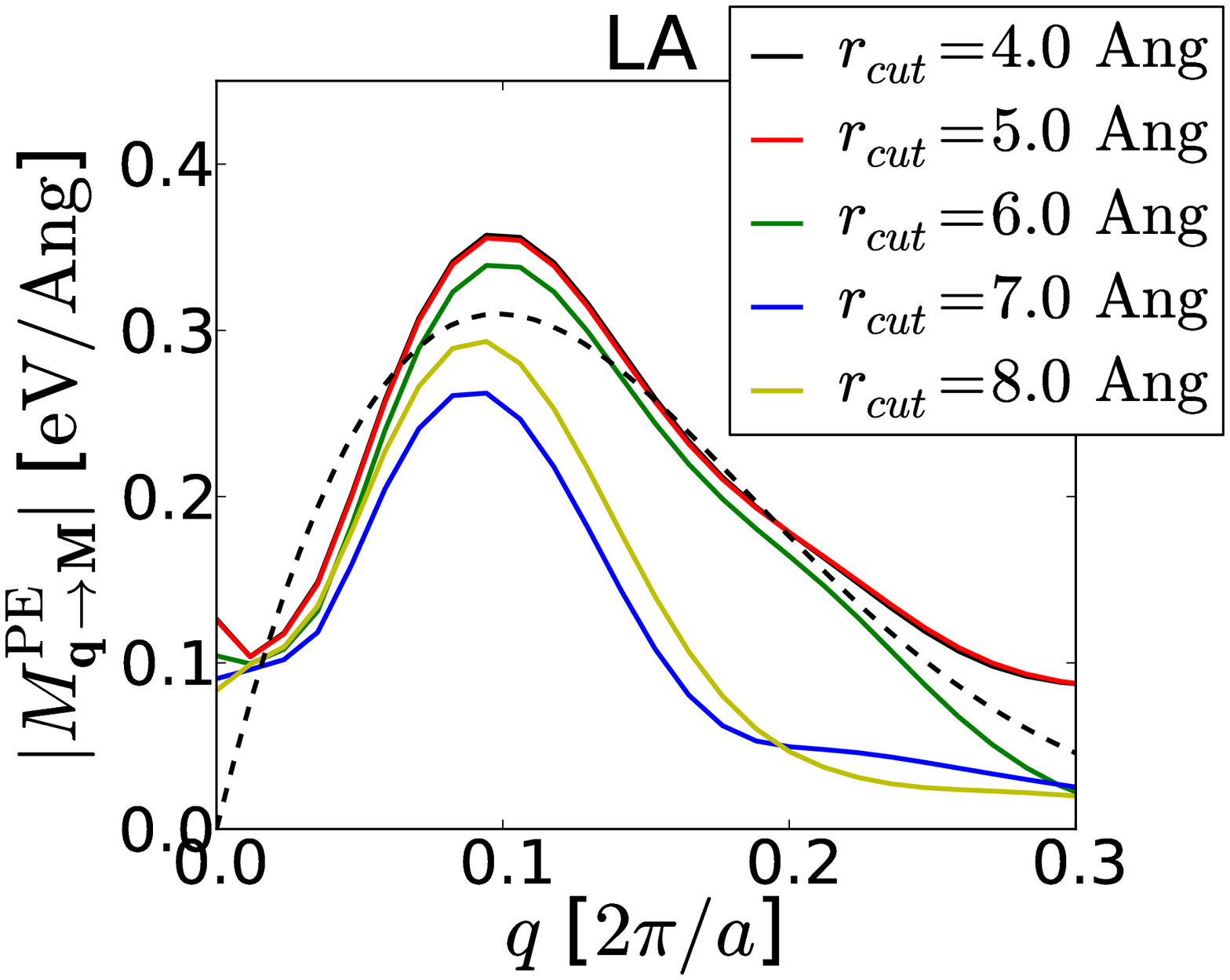}
  \caption{(Color online) Deformation potential (upper row) and piezoelectric
    (lower row) interactions in monolayer MoS$_2$ obtained with the real-space
    partitioning scheme for different values of the real-space cutoff
    $r_\text{cut}$. The plots show the absolute value of the matrix elements
    $M_{\bq\lambda}^\text{DP/PE}$ averaged over the high-symmetry directions of
    the hexagonal lattice. The analytic couplings in Eqs.~\eqref{eq:M_DP}
    and~\eqref{eq:M_PE} are shown with dashed lines for the MoS$_2$ parameters
    listed in Tab.~\ref{tab:parameters}.}
  \label{fig:M_fit}
\end{figure}

In Fig.~\ref{fig:M_fit} we show the deformation potential (top row) and
piezoelectric (bottom row) interactions for the TA and LA phonons in 2D MoS$_2$
obtained with the partitioning scheme for different values of the
cutoff~\cite{footnote2}. The dashed lines show the analytic forms for the
deformation potential and piezoelectric interactions in Eqs.~\eqref{eq:M_DP}
and~\eqref{eq:M_PE} with the parameters listed in Tab.~\ref{tab:parameters}.
While the variation in the deformation potential interaction is relatively
insignificant, the piezoelectric interaction is more sensitive to the chosen
cutoff. The best agreement between the analytic expression for the piezoelectric
interaction in Eq.~\eqref{eq:M_PE} and the first-principles results is obtained
for $r_\text{cut}=6.0$~{\AA}.

It should be emphasized that the finite value of the first-principles
piezoelectric interaction in the limit $q\rightarrow 0$ in Fig.~\ref{fig:M_fit}
is an artifact inherent to a supercell method. The finite real-space range of
the el-ph interaction in supercell methods naturally sets a lower limit for the
magnitude of the phonon wave vector $q_\text{min}$ at which long-range
interactions can be obtained reliably. It is given by $q_\text{min} = 2\pi /
L_\text{cell}$ where $L_\text{cell}$ is the size of the supercell (measured as
the diameter of a sphere that can be contained within the supercell). In the
calculations presented here $L_\text{cell}\sim 45$~{\AA} implying that
$q_\text{min} \sim 0.14$~\AA$^{-1}\sim 0.07\times2\pi/a$.

The real-space partitioning scheme outlined here can also be applied in other
first-principles calculations of the el-ph interaction based on
e.g. Wannier functions~\cite{Louie:e-ph}.

\subsection{Normal and umklapp processes}
\label{app:normalumklapp}

In order identify the normal and umklapp processes in the el-ph interaction, we
start by noticing that the gradients of the potential $\nabla_{\alpha l} V
\equiv \mathbf{f}_\alpha(\br - \mathbf{R}_l)$ in Eq.~\eqref{eq:deltaV} are
localized functions in real space (see Fig.~\ref{fig:supercell_matrix}), where
$\mathbf{f}_\alpha$ denotes an atom-specific function. Due to the periodicity of
the lattice, the gradient in unit cell $l$ is related to the gradient in the
reference cell through a translation by the lattice vector $\mathbf{R}_l$. We
now express the gradients of the potential in terms of their Fourier series,
\begin{equation}
  \label{eq:gradV_Fourier}
  \nabla_{\alpha l} V(\br) = \sum_{\bm{\kappa}} e^{i \bm{\kappa} \cdot (\br - \mathbf{R}_l)} 
                             \mathbf{f}_{\bm{\kappa}}^\alpha ,
\end{equation}
where $\bm{\kappa}=(\bm{\kappa}_\parallel,\bm{\kappa}_\perp)$ is a
three-dimensional Fourier variable. It is here important to distinguish between
the projections $\bm{\kappa}_\parallel/\bm{\kappa}_\perp$ of $\bm{\kappa}$ onto
the periodic/nonperiodic directions of the lattice. The sum over the unit cell
index $l$ in Eq.~\eqref{eq:deltaV} now only contains two exponential factors,
\begin{equation}
  \label{eq:delta_kappa_qplusG}
  \sum_l e^{i (\bq - \bm{\kappa}_\parallel) \cdot \mathbf{R}_l} 
      = N \delta_{\bm{\kappa}_\parallel, \bq + \mathbf{G}} ,
\end{equation}
where the reciprocal lattice vector $\mathbf{G}$ and the phonon wave vector
$\bq$ by definition have the dimensionality of the lattice. This restricts the
projection $\bm{\kappa}_\parallel$ to values $\bq + \mathbf{G}$. Inserting in
Eq.~\eqref{eq:deltaV}, we find for the phonon-induced potential
\begin{align}
  \label{eq:deltaV_normalumklapp}
  \delta V_{\bq\lambda}(\br) & = \sum_\mathbf{G} 
      e^{i (\bq + \mathbf{G}) \cdot \br_\parallel} \sum_{\bm{\kappa}_\perp \alpha} 
      e^{i \bm{\kappa}_\perp \cdot \br_\perp}
      \hat{\mathbf{e}}_{\bq\lambda}^\alpha \cdot
      \mathbf{f}_{(\bq+\mathbf{G},\bm{\kappa}_\perp)}^\alpha \nonumber \\
      & \equiv \sum_\mathbf{G} 
        e^{i (\bq + \mathbf{G}) \cdot \br_\parallel} 
        \delta V_{\bq + \mathbf{G}}^\lambda(\br_\perp) .
\end{align}
First of all, we note that this allows us write the phonon-induced potential on
the Bloch form in Eq.~\eqref{eq:deltaV_Fourier}. Secondly, the separation into
normal ($\mathbf{G}=\mathbf{0}$) and umklapp ($\mathbf{G}\neq\mathbf{0}$)
processes is now formally straight-forward.
\begin{figure}[!t]
  \includegraphics[width=0.32\linewidth]{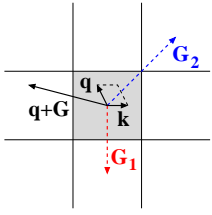}
  \hfill
  \caption{(Color online) Schematic illustration of an umklapp process involving
    a Fourier component of the scattering potential at $\bq + \mathbf{G}$. The
    square lattice denotes the reciprocal lattice with the shaded cell
    indicating the first Brillouin zone. The dashed arrows show a pair of
    reciprocal lattice vectors $\mathbf{G}_{1/2}$ from the electronic Bloch
    functions that conserve crystal momentum by bringing $\bq + \mathbf{G}$ back
    onto $\bq$ in the first Brillouin zone.}
\label{fig:umklapp}
\end{figure}

It is important to note that in first-principles calculations of the el-ph
interaction, the term ``umklapp process'' has a more general meaning than the one
encountered in conventional textbook discussions of the
subject~\cite{Madelung}. In order to illustrate this, it is instructive to
evaluate the matrix element $M_{\bk\bq}^\lambda$ in Eq.~\eqref{eq:M_first} with
the electronic Bloch functions expanded as $\psi_\bk(\br) = \sum_\mathbf{G}
e^{i(\bk+\mathbf{G})\cdot \br_\parallel} u_{\bk+\mathbf{G}}(\br_\perp)$, where
$u_{\bk+\mathbf{G}}$ are the Fourier components of the periodic part of the
Bloch functions. This yields
\begin{align}
  \label{eq:M_umklapp}
  M_{\bk\bq}^\lambda & = \bra{\mathbf{k+q}} \delta V_{\bq\lambda}(\br) \ket{\bk}
  \nonumber \\
  & = \sum_{\mathbf{G}} \sum_{\mathbf{G}_1 \mathbf{G}_2}
  u_{\bk+\bq+\mathbf{G}_1}^* \delta V_{\bq+\mathbf{G}}^\lambda
  u_{\bk+\mathbf{G}_2}  \nonumber \\
  & \quad \times \underbrace{\int\! d\br_\parallel \, e^{i (\mathbf{G} +
      \mathbf{G}_1 + \mathbf{G}_2 - \mathbf{G}_3) \cdot \br_\parallel}}_{=
    \delta_{\mathbf{G}_3, \mathbf{G} + \mathbf{G}_1 + \mathbf{G}_2 }} ,
\end{align}
where $\mathbf{G}_{1/2}$ are the reciprocal lattice vectors from the Bloch
functions, $\mathbf{G}_3$ is a reciprocal lattice vector that shifts $\bk+\bq$
to the first Brillouin zone in case it falls outside, and the $\br_\perp$
dependence has been integrated out. The $\delta$ function resulting from the
integral in the last line can be regarded as a generalized statement of
conservation of crystal momentum taking into account the reciprocal lattice
vectors from the Bloch functions. It differs from the standard textbook
definition where $\mathbf{G}_{1/2}=\mathbf{0}$ and $\mathbf{G} = \mathbf{G}_3$
implying that umklapp processes only contribute to the matrix element if
$\mathbf{G}_3 \neq \mathbf{0}$; i.e. $\bk+\bq\notin
\text{1. BZ}$~\cite{Madelung}. For the generalized conservation of crystal
momentum this is not the case. Here, umklapp processes involving all Fourier
components $\mathbf{G}$ of the scattering potential contribute regardless of the
value of $\mathbf{G}_3$. This is illustrated in Fig.~\ref{fig:umklapp} which
shows an umklapp process ($\mathbf{G}\neq\mathbf{0}$) for $\mathbf{G}_3 =
\mathbf{0}$ and a $(\mathbf{G}_1, \mathbf{G}_2)$ pair that conserves crystal
momentum. For intravalley scattering where $\bk + \bq$ is \emph{always} inside
the first Brillouin zone (if not, the first Brillouin zone can chosen such that
this is the case) and hence $\mathbf{G}_3 = \mathbf{0}$, coupling to phonons via
umklapp processes takes place through the type of process shown in
Fig.~\ref{fig:umklapp}.

In the following section we outline a Fourier filtering method which allows for
a numerical separation of normal and umklapp processes in the supercell method.

\subsubsection{Fourier filtering method}

In practice, the matrix elements of the el-ph interaction are evaluated using
Eq.~\eqref{eq:M_first}. The expression for the phonon-induced potential change
in Eq.~\eqref{eq:deltaV_normalumklapp} is therefore not directly applicable for
the separation of the normal and umklapp processes. Instead we note that the sum
over unit cell indices $l$ in Eq.~\eqref{eq:delta_kappa_qplusG} provides an
automatic selection of the Fourier components in $\nabla_0 V$ that contribute in
Eq.~\eqref{eq:deltaV_normalumklapp}. The Fourier expansion of $\nabla_0 V$ can
therefore be used directly in the calculation of the matrix element in
Eq.~\eqref{eq:M_first}. Writing the Fourier expansion as
\begin{align}
  \label{eq:gradV_normalumklapp}
  \nabla_0 V(\br) & = \sum_{\bm{\kappa_\parallel}\in 1. \text{BZ}} 
      e^{i \bm{\kappa_\parallel} \cdot \br_\parallel} 
      \mathbf{f}_{\bm{\kappa}_\parallel} (\br_\perp)  \nonumber \\
    & \quad + \sum_{\bm{\kappa_\parallel}\notin 1. \text{BZ}} 
      e^{i \bm{\kappa_\parallel} \cdot \br_\parallel} 
      \mathbf{f}_{\bm{\kappa}_\parallel} (\br_\perp) ,
\end{align}
the two terms with $\bm{\kappa}_\parallel$ lying inside and outside the
Brillouin zone (BZ) of the crystal lattice define the normal and umklapp
contribution to the el-ph interaction, respectively.

Numerically, the atomic gradients $\nabla_0 V$ are represented on a
three-dimensional real-space grid in the supercell having length $L_i$ and $N_i$
number of grid points in the direction of the lattice vector $\mathbf{a}_i$.
The resulting grid spacing is $\Delta x_i=L_i/N_i$. The values of the gradients
on the grid are denoted $\mathbf{f}_{ijk}$. The Fourier expansion is obtained
using the fast Fourier transform (FFT),
\begin{equation}
  \mathbf{f}_{\kappa_i\kappa_j\kappa_k} = \text{FFT}(\mathbf{f}_{ijk}) ,
\end{equation}
with the corresponding Fourier space grid in the direction of the primitive
reciprocal lattice vector $\mathbf{b}_i$ given by $\kappa_i = -
\kappa_{i,\text{max}} + n \Delta\kappa_i$ with $n=0,1,2,\ldots, N_i-1$,
$\Delta\kappa_i = 2\pi/ L_i$ and $\kappa_{i,\text{max}} = \pi / \Delta x_i$. As
the supercell has a real-space grid spacing significantly smaller and a size
significantly larger than the lattice constant $a$, i.e. $\Delta x_i \ll a$ and
$L_i \gg a$, respectively, we have that $\kappa_{i,\text{max}} \gg \pi/a$ and
$\Delta\kappa_i \ll \pi/a$.

The normal and umklapp processes can now be separated by Fourier filtering
$\mathbf{f}_{\kappa_i\kappa_j\kappa_k}$. This is done by zeroing the Fourier
components at $\bm{\kappa}=(\kappa_i,\kappa_j,\kappa_k)$ with
$\bm{\kappa}_\parallel$ lying outside or inside the Brillouin zone. Using a
tilde to denote the filtered quantities, we have
\begin{equation}
  \label{eq:normal_umklapp}
  \mathbf{\tilde{f}}_{\kappa_i\kappa_j\kappa_k} \rightarrow
  \left\{
    \begin{array}{l}
      \text{normal,}\;  \neq 0 \;\text{if}\; \bm{\kappa_\parallel}\in 1. \text{BZ} \\
      \text{umklapp,} \; \neq 0 \;\text{if}\; \bm{\kappa_\parallel}\notin 1. \text{BZ}
    \end{array}
  \right. .
\end{equation}
Applying the inverse FFT (IFFT)
\begin{equation}
    \mathbf{\tilde{f}}_{ijk} = \text{IFFT}(\mathbf{\tilde{f}}_{\kappa_i\kappa_j\kappa_k}) ,
\end{equation}
we obtain the filtered gradients in real space which can be used in the
numerical evaluation of the matrix element in Eq.~\eqref{eq:M_first}. Since the
gradients of the potential $\nabla_0 V$ are real valued, the imaginary part of
$\mathbf{\tilde{f}}_{ijk}$ can be discarded.


\begin{thebibliography}{71}
\expandafter\ifx\csname natexlab\endcsname\relax\def\natexlab#1{#1}\fi
\expandafter\ifx\csname bibnamefont\endcsname\relax
  \def\bibnamefont#1{#1}\fi
\expandafter\ifx\csname bibfnamefont\endcsname\relax
  \def\bibfnamefont#1{#1}\fi
\expandafter\ifx\csname citenamefont\endcsname\relax
  \def\citenamefont#1{#1}\fi
\expandafter\ifx\csname url\endcsname\relax
  \def\url#1{\texttt{#1}}\fi
\expandafter\ifx\csname urlprefix\endcsname\relax\def\urlprefix{URL }\fi
\providecommand{\bibinfo}[2]{#2}
\providecommand{\eprint}[2][]{\url{#2}}

\bibitem[{\citenamefont{Novoselov et~al.}(2005)\citenamefont{Novoselov, Jiang,
  Schedin, Booth, Khotkevich, Morozov, and Geim}}]{Geim:2D}
\bibinfo{author}{\bibfnamefont{K.~S.} \bibnamefont{Novoselov}},
  \bibinfo{author}{\bibfnamefont{D.}~\bibnamefont{Jiang}},
  \bibinfo{author}{\bibfnamefont{F.}~\bibnamefont{Schedin}},
  \bibinfo{author}{\bibfnamefont{T.~J.} \bibnamefont{Booth}},
  \bibinfo{author}{\bibfnamefont{V.~V.} \bibnamefont{Khotkevich}},
  \bibinfo{author}{\bibfnamefont{S.~V.} \bibnamefont{Morozov}},
  \bibnamefont{and} \bibinfo{author}{\bibfnamefont{A.~K.} \bibnamefont{Geim}},
  \bibinfo{journal}{PNAS} \textbf{\bibinfo{volume}{102}},
  \bibinfo{pages}{10451} (\bibinfo{year}{2005}).

\bibitem[{\citenamefont{Geim and Novoselov}(2007)}]{Geim:Graphene}
\bibinfo{author}{\bibfnamefont{A.~K.} \bibnamefont{Geim}} \bibnamefont{and}
  \bibinfo{author}{\bibfnamefont{K.~S.} \bibnamefont{Novoselov}},
  \bibinfo{journal}{Nature Mat.} \textbf{\bibinfo{volume}{6}},
  \bibinfo{pages}{183} (\bibinfo{year}{2007}).

\bibitem[{\citenamefont{Neto et~al.}(2009)\citenamefont{Neto, Guinea, Peres,
  Novoselov, and Geim}}]{RMP:Graphene}
\bibinfo{author}{\bibfnamefont{A.~H.~C.} \bibnamefont{Neto}},
  \bibinfo{author}{\bibfnamefont{F.}~\bibnamefont{Guinea}},
  \bibinfo{author}{\bibfnamefont{N.~M.~R.} \bibnamefont{Peres}},
  \bibinfo{author}{\bibfnamefont{K.~S.} \bibnamefont{Novoselov}},
  \bibnamefont{and} \bibinfo{author}{\bibfnamefont{A.~K.} \bibnamefont{Geim}},
  \bibinfo{journal}{Rev. Mod. Phys.} \textbf{\bibinfo{volume}{81}},
  \bibinfo{pages}{109} (\bibinfo{year}{2009}).

\bibitem[{\citenamefont{{Das Sarma} et~al.}(2011)\citenamefont{{Das Sarma},
  Adam, Hwang, and Rossi}}]{Sarma:RMP}
\bibinfo{author}{\bibfnamefont{S.}~\bibnamefont{{Das Sarma}}},
  \bibinfo{author}{\bibfnamefont{S.}~\bibnamefont{Adam}},
  \bibinfo{author}{\bibfnamefont{E.~H.} \bibnamefont{Hwang}}, \bibnamefont{and}
  \bibinfo{author}{\bibfnamefont{E.}~\bibnamefont{Rossi}},
  \bibinfo{journal}{Rev. Mod. Phys.} \textbf{\bibinfo{volume}{83}},
  \bibinfo{pages}{407} (\bibinfo{year}{2011}).

\bibitem[{\citenamefont{Wang et~al.}(2012)\citenamefont{Wang, {Kalantar-Zadeh},
  Kis, Coleman, and Strano}}]{Strano:NNanoReview}
\bibinfo{author}{\bibfnamefont{Q.~H.} \bibnamefont{Wang}},
  \bibinfo{author}{\bibfnamefont{K.}~\bibnamefont{{Kalantar-Zadeh}}},
  \bibinfo{author}{\bibfnamefont{A.}~\bibnamefont{Kis}},
  \bibinfo{author}{\bibfnamefont{J.~N.} \bibnamefont{Coleman}},
  \bibnamefont{and} \bibinfo{author}{\bibfnamefont{M.~S.}
  \bibnamefont{Strano}}, \bibinfo{journal}{Nature Nano.}
  \textbf{\bibinfo{volume}{7}}, \bibinfo{pages}{699} (\bibinfo{year}{2012}).

\bibitem[{\citenamefont{Chhowalla et~al.}(2013)\citenamefont{Chhowalla, Shin,
  Eda, Li, Loh, and Zhang}}]{Zhang:NChemReview}
\bibinfo{author}{\bibfnamefont{M.}~\bibnamefont{Chhowalla}},
  \bibinfo{author}{\bibfnamefont{H.~S.} \bibnamefont{Shin}},
  \bibinfo{author}{\bibfnamefont{G.}~\bibnamefont{Eda}},
  \bibinfo{author}{\bibfnamefont{L.-J.} \bibnamefont{Li}},
  \bibinfo{author}{\bibfnamefont{K.~P.} \bibnamefont{Loh}}, \bibnamefont{and}
  \bibinfo{author}{\bibfnamefont{H.}~\bibnamefont{Zhang}},
  \bibinfo{journal}{Nature Chem.} \textbf{\bibinfo{volume}{5}},
  \bibinfo{pages}{263} (\bibinfo{year}{2013}).

\bibitem[{\citenamefont{Podzorov et~al.}(2004)\citenamefont{Podzorov,
  Gershenson, Kloc, Zeis, and Bucher}}]{Bucher:High}
\bibinfo{author}{\bibfnamefont{V.}~\bibnamefont{Podzorov}},
  \bibinfo{author}{\bibfnamefont{M.~E.} \bibnamefont{Gershenson}},
  \bibinfo{author}{\bibfnamefont{C.}~\bibnamefont{Kloc}},
  \bibinfo{author}{\bibfnamefont{R.}~\bibnamefont{Zeis}}, \bibnamefont{and}
  \bibinfo{author}{\bibfnamefont{E.}~\bibnamefont{Bucher}},
  \bibinfo{journal}{Appl. Phys. Lett.} \textbf{\bibinfo{volume}{84}},
  \bibinfo{pages}{3301} (\bibinfo{year}{2004}).

\bibitem[{\citenamefont{Mak et~al.}(2010)\citenamefont{Mak, Lee, Hone, Shan,
  and Heinz}}]{Heinz:ThinMoS2}
\bibinfo{author}{\bibfnamefont{K.~F.} \bibnamefont{Mak}},
  \bibinfo{author}{\bibfnamefont{C.}~\bibnamefont{Lee}},
  \bibinfo{author}{\bibfnamefont{J.}~\bibnamefont{Hone}},
  \bibinfo{author}{\bibfnamefont{J.}~\bibnamefont{Shan}}, \bibnamefont{and}
  \bibinfo{author}{\bibfnamefont{T.~F.} \bibnamefont{Heinz}},
  \bibinfo{journal}{Phys. Rev. Lett.} \textbf{\bibinfo{volume}{105}},
  \bibinfo{pages}{136805} (\bibinfo{year}{2010}).

\bibitem[{\citenamefont{Splendiani et~al.}(2010)\citenamefont{Splendiani, Sun,
  Zhang, Li, Kim, Chim, Galli, and Wang}}]{Wang:Emerging}
\bibinfo{author}{\bibfnamefont{A.}~\bibnamefont{Splendiani}},
  \bibinfo{author}{\bibfnamefont{L.}~\bibnamefont{Sun}},
  \bibinfo{author}{\bibfnamefont{Y.}~\bibnamefont{Zhang}},
  \bibinfo{author}{\bibfnamefont{T.}~\bibnamefont{Li}},
  \bibinfo{author}{\bibfnamefont{J.}~\bibnamefont{Kim}},
  \bibinfo{author}{\bibfnamefont{C.-Y.} \bibnamefont{Chim}},
  \bibinfo{author}{\bibfnamefont{G.}~\bibnamefont{Galli}}, \bibnamefont{and}
  \bibinfo{author}{\bibfnamefont{F.}~\bibnamefont{Wang}},
  \bibinfo{journal}{Nano. Lett.} \textbf{\bibinfo{volume}{10}},
  \bibinfo{pages}{1271} (\bibinfo{year}{2010}).

\bibitem[{\citenamefont{Radisavljevic et~al.}(2011)\citenamefont{Radisavljevic,
  Radenovic, Brivio, Giacometti, and Kis}}]{Kis:MoS2Transistor}
\bibinfo{author}{\bibfnamefont{B.}~\bibnamefont{Radisavljevic}},
  \bibinfo{author}{\bibfnamefont{A.}~\bibnamefont{Radenovic}},
  \bibinfo{author}{\bibfnamefont{J.}~\bibnamefont{Brivio}},
  \bibinfo{author}{\bibfnamefont{V.}~\bibnamefont{Giacometti}},
  \bibnamefont{and} \bibinfo{author}{\bibfnamefont{A.}~\bibnamefont{Kis}},
  \bibinfo{journal}{Nature Nano.} \textbf{\bibinfo{volume}{6}},
  \bibinfo{pages}{147} (\bibinfo{year}{2011}).

\bibitem[{\citenamefont{Korn et~al.}(2011)\citenamefont{Korn, Heydrich, Hirmer,
  Schmutzler, and Sch{\"u}ller}}]{Schuller:Photocarrier}
\bibinfo{author}{\bibfnamefont{T.}~\bibnamefont{Korn}},
  \bibinfo{author}{\bibfnamefont{S.}~\bibnamefont{Heydrich}},
  \bibinfo{author}{\bibfnamefont{M.}~\bibnamefont{Hirmer}},
  \bibinfo{author}{\bibfnamefont{J.}~\bibnamefont{Schmutzler}},
  \bibnamefont{and}
  \bibinfo{author}{\bibfnamefont{C.}~\bibnamefont{Sch{\"u}ller}},
  \bibinfo{journal}{Appl. Phys. Lett.} \textbf{\bibinfo{volume}{99}},
  \bibinfo{pages}{102109} (\bibinfo{year}{2011}).

\bibitem[{\citenamefont{Zhang et~al.}(2012)\citenamefont{Zhang, Ye, Matsuhashi,
  and Iwasa}}]{Iwasa:Ambipolar}
\bibinfo{author}{\bibfnamefont{Y.}~\bibnamefont{Zhang}},
  \bibinfo{author}{\bibfnamefont{J.}~\bibnamefont{Ye}},
  \bibinfo{author}{\bibfnamefont{Y.}~\bibnamefont{Matsuhashi}},
  \bibnamefont{and} \bibinfo{author}{\bibfnamefont{Y.}~\bibnamefont{Iwasa}},
  \bibinfo{journal}{Nano. Lett.} \textbf{\bibinfo{volume}{12}},
  \bibinfo{pages}{1136} (\bibinfo{year}{2012}).

\bibitem[{\citenamefont{Xiao et~al.}(2012)\citenamefont{Xiao, Liu, Feng, Xu,
  and Yao}}]{Yao:SpinValley}
\bibinfo{author}{\bibfnamefont{D.}~\bibnamefont{Xiao}},
  \bibinfo{author}{\bibfnamefont{G.-B.} \bibnamefont{Liu}},
  \bibinfo{author}{\bibfnamefont{W.}~\bibnamefont{Feng}},
  \bibinfo{author}{\bibfnamefont{X.}~\bibnamefont{Xu}}, \bibnamefont{and}
  \bibinfo{author}{\bibfnamefont{W.}~\bibnamefont{Yao}},
  \bibinfo{journal}{Phys. Rev. Lett.} \textbf{\bibinfo{volume}{108}},
  \bibinfo{pages}{196802} (\bibinfo{year}{2012}).

\bibitem[{\citenamefont{Cao et~al.}(2012)\citenamefont{Cao, Wang, Han, Ye, Zhu,
  Shi, Niu, Tan, Wang, Liu et~al.}}]{Feng:ValleySelective}
\bibinfo{author}{\bibfnamefont{T.}~\bibnamefont{Cao}},
  \bibinfo{author}{\bibfnamefont{G.}~\bibnamefont{Wang}},
  \bibinfo{author}{\bibfnamefont{W.}~\bibnamefont{Han}},
  \bibinfo{author}{\bibfnamefont{H.}~\bibnamefont{Ye}},
  \bibinfo{author}{\bibfnamefont{C.}~\bibnamefont{Zhu}},
  \bibinfo{author}{\bibfnamefont{J.}~\bibnamefont{Shi}},
  \bibinfo{author}{\bibfnamefont{Q.}~\bibnamefont{Niu}},
  \bibinfo{author}{\bibfnamefont{P.}~\bibnamefont{Tan}},
  \bibinfo{author}{\bibfnamefont{E.}~\bibnamefont{Wang}},
  \bibinfo{author}{\bibfnamefont{B.}~\bibnamefont{Liu}}, \bibnamefont{et~al.},
  \bibinfo{journal}{Nature Commun.} \textbf{\bibinfo{volume}{3}},
  \bibinfo{pages}{887} (\bibinfo{year}{2012}).

\bibitem[{\citenamefont{Mak et~al.}(2012)\citenamefont{Mak, He, Shan, and
  Heinz}}]{Heinz:Control}
\bibinfo{author}{\bibfnamefont{K.~F.} \bibnamefont{Mak}},
  \bibinfo{author}{\bibfnamefont{K.}~\bibnamefont{He}},
  \bibinfo{author}{\bibfnamefont{J.}~\bibnamefont{Shan}}, \bibnamefont{and}
  \bibinfo{author}{\bibfnamefont{T.~F.} \bibnamefont{Heinz}},
  \bibinfo{journal}{Nature Nano.} \textbf{\bibinfo{volume}{7}},
  \bibinfo{pages}{494} (\bibinfo{year}{2012}).

\bibitem[{\citenamefont{Zeng et~al.}(2012)\citenamefont{Zeng, Dai, Yao, Xiao,
  and Cui}}]{Cui:ValleyPolarization}
\bibinfo{author}{\bibfnamefont{H.}~\bibnamefont{Zeng}},
  \bibinfo{author}{\bibfnamefont{J.}~\bibnamefont{Dai}},
  \bibinfo{author}{\bibfnamefont{W.}~\bibnamefont{Yao}},
  \bibinfo{author}{\bibfnamefont{D.}~\bibnamefont{Xiao}}, \bibnamefont{and}
  \bibinfo{author}{\bibfnamefont{X.}~\bibnamefont{Cui}},
  \bibinfo{journal}{Nature Nano.} \textbf{\bibinfo{volume}{7}},
  \bibinfo{pages}{490} (\bibinfo{year}{2012}).

\bibitem[{\citenamefont{Ayari et~al.}(2007)\citenamefont{Ayari, Cobas,
  Ogundadegbe, and Fuhrer}}]{Fuhrer:Ultrathin}
\bibinfo{author}{\bibfnamefont{A.}~\bibnamefont{Ayari}},
  \bibinfo{author}{\bibfnamefont{E.}~\bibnamefont{Cobas}},
  \bibinfo{author}{\bibfnamefont{O.}~\bibnamefont{Ogundadegbe}},
  \bibnamefont{and} \bibinfo{author}{\bibfnamefont{M.~S.}
  \bibnamefont{Fuhrer}}, \bibinfo{journal}{J. Appl. Phys.}
  \textbf{\bibinfo{volume}{101}}, \bibinfo{pages}{014507}
  (\bibinfo{year}{2007}).

\bibitem[{\citenamefont{Matte et~al.}(2010)\citenamefont{Matte, Gomathi, Manna,
  Late, Datta, Pati, and Rao}}]{Pati:Analogues}
\bibinfo{author}{\bibfnamefont{H.~S. S.~R.} \bibnamefont{Matte}},
  \bibinfo{author}{\bibfnamefont{A.}~\bibnamefont{Gomathi}},
  \bibinfo{author}{\bibfnamefont{A.~K.} \bibnamefont{Manna}},
  \bibinfo{author}{\bibfnamefont{D.~J.} \bibnamefont{Late}},
  \bibinfo{author}{\bibfnamefont{R.}~\bibnamefont{Datta}},
  \bibinfo{author}{\bibfnamefont{S.~K.} \bibnamefont{Pati}}, \bibnamefont{and}
  \bibinfo{author}{\bibfnamefont{C.~N.~R.} \bibnamefont{Rao}},
  \bibinfo{journal}{Angew. Chem.} \textbf{\bibinfo{volume}{122}},
  \bibinfo{pages}{4153} (\bibinfo{year}{2010}).

\bibitem[{\citenamefont{Liu et~al.}(2012)\citenamefont{Liu, Zhang, Lee, Lin,
  Chang, Su, Chang, Li, Shi, Zhang et~al.}}]{Li:LargeArea}
\bibinfo{author}{\bibfnamefont{K.-K.} \bibnamefont{Liu}},
  \bibinfo{author}{\bibfnamefont{W.}~\bibnamefont{Zhang}},
  \bibinfo{author}{\bibfnamefont{Y.-H.} \bibnamefont{Lee}},
  \bibinfo{author}{\bibfnamefont{Y.-C.} \bibnamefont{Lin}},
  \bibinfo{author}{\bibfnamefont{M.-T.} \bibnamefont{Chang}},
  \bibinfo{author}{\bibfnamefont{C.-Y.} \bibnamefont{Su}},
  \bibinfo{author}{\bibfnamefont{C.-S.} \bibnamefont{Chang}},
  \bibinfo{author}{\bibfnamefont{H.}~\bibnamefont{Li}},
  \bibinfo{author}{\bibfnamefont{Y.}~\bibnamefont{Shi}},
  \bibinfo{author}{\bibfnamefont{H.}~\bibnamefont{Zhang}},
  \bibnamefont{et~al.}, \bibinfo{journal}{Nano. Lett.}
  \textbf{\bibinfo{volume}{12}}, \bibinfo{pages}{1538} (\bibinfo{year}{2012}).

\bibitem[{\citenamefont{Liu and Ye}(2012)}]{Peide:Dual}
\bibinfo{author}{\bibfnamefont{H.}~\bibnamefont{Liu}} \bibnamefont{and}
  \bibinfo{author}{\bibfnamefont{P.~D.} \bibnamefont{Ye}},
  \bibinfo{journal}{IEEE Electron Devices Letters}
  \textbf{\bibinfo{volume}{33}}, \bibinfo{pages}{546} (\bibinfo{year}{2012}).

\bibitem[{\citenamefont{Kim et~al.}(2012)\citenamefont{Kim, Konar, Hwang, Lee,
  Lee, Yang, Jung, Kim, Yoo, Choi et~al.}}]{Kim:HighMobility}
\bibinfo{author}{\bibfnamefont{S.}~\bibnamefont{Kim}},
  \bibinfo{author}{\bibfnamefont{A.}~\bibnamefont{Konar}},
  \bibinfo{author}{\bibfnamefont{W.}~\bibnamefont{Hwang}},
  \bibinfo{author}{\bibfnamefont{J.~H.} \bibnamefont{Lee}},
  \bibinfo{author}{\bibfnamefont{J.}~\bibnamefont{Lee}},
  \bibinfo{author}{\bibfnamefont{J.}~\bibnamefont{Yang}},
  \bibinfo{author}{\bibfnamefont{C.}~\bibnamefont{Jung}},
  \bibinfo{author}{\bibfnamefont{H.}~\bibnamefont{Kim}},
  \bibinfo{author}{\bibfnamefont{J.}~\bibnamefont{Yoo}},
  \bibinfo{author}{\bibfnamefont{J.}~\bibnamefont{Choi}}, \bibnamefont{et~al.},
  \bibinfo{journal}{Nature Commun.} \textbf{\bibinfo{volume}{3}},
  \bibinfo{pages}{1011} (\bibinfo{year}{2012}).

\bibitem[{\citenamefont{Fallahazad and Tutuc}(2012)}]{Tutuc:FieldEffect}
\bibinfo{author}{\bibfnamefont{S.~L.~B.} \bibnamefont{Fallahazad}}
  \bibnamefont{and} \bibinfo{author}{\bibfnamefont{E.}~\bibnamefont{Tutuc}},
  \bibinfo{journal}{Appl. Phys. Lett.} \textbf{\bibinfo{volume}{101}},
  \bibinfo{pages}{223104} (\bibinfo{year}{2012}).

\bibitem[{\citenamefont{Fuhrer and Hone}(2013)}]{Hone:Measurement}
\bibinfo{author}{\bibfnamefont{M.~S.} \bibnamefont{Fuhrer}} \bibnamefont{and}
  \bibinfo{author}{\bibfnamefont{J.}~\bibnamefont{Hone}},
  \bibinfo{journal}{Nature Nano.} \textbf{\bibinfo{volume}{8}},
  \bibinfo{pages}{146} (\bibinfo{year}{2013}).

\bibitem[{\citenamefont{Radisavljevic and Kis}(2013)}]{Kis:Reply}
\bibinfo{author}{\bibfnamefont{B.}~\bibnamefont{Radisavljevic}}
  \bibnamefont{and} \bibinfo{author}{\bibfnamefont{A.}~\bibnamefont{Kis}},
  \bibinfo{journal}{Nature Nano.} \textbf{\bibinfo{volume}{8}},
  \bibinfo{pages}{147} (\bibinfo{year}{2013}).

\bibitem[{\citenamefont{Jena and Konar}(2007)}]{Konar:Engineering}
\bibinfo{author}{\bibfnamefont{D.}~\bibnamefont{Jena}} \bibnamefont{and}
  \bibinfo{author}{\bibfnamefont{A.}~\bibnamefont{Konar}},
  \bibinfo{journal}{Phys. Rev. Lett.} \textbf{\bibinfo{volume}{98}},
  \bibinfo{pages}{136805} (\bibinfo{year}{2007}).

\bibitem[{\citenamefont{Kaasbjerg
  et~al.}(2012{\natexlab{a}})\citenamefont{Kaasbjerg, Thygesen, and
  Jacobsen}}]{Kaasbjerg:MoS2}
\bibinfo{author}{\bibfnamefont{K.}~\bibnamefont{Kaasbjerg}},
  \bibinfo{author}{\bibfnamefont{K.~S.} \bibnamefont{Thygesen}},
  \bibnamefont{and} \bibinfo{author}{\bibfnamefont{K.~W.}
  \bibnamefont{Jacobsen}}, \bibinfo{journal}{Phys. Rev. B}
  \textbf{\bibinfo{volume}{85}}, \bibinfo{pages}{115317}
  (\bibinfo{year}{2012}{\natexlab{a}}).

\bibitem[{\citenamefont{Yoon et~al.}(2011)\citenamefont{Yoon, Ganapathi, and
  Salahuddin}}]{Salahuddin:HowGood}
\bibinfo{author}{\bibfnamefont{Y.}~\bibnamefont{Yoon}},
  \bibinfo{author}{\bibfnamefont{K.}~\bibnamefont{Ganapathi}},
  \bibnamefont{and}
  \bibinfo{author}{\bibfnamefont{S.}~\bibnamefont{Salahuddin}},
  \bibinfo{journal}{Nano. Lett.} \textbf{\bibinfo{volume}{11}},
  \bibinfo{pages}{3768} (\bibinfo{year}{2011}).

\bibitem[{\citenamefont{Popov et~al.}(2012)\citenamefont{Popov, Seifert, and
  Tom\'anek}}]{Tomanek:Designing}
\bibinfo{author}{\bibfnamefont{I.}~\bibnamefont{Popov}},
  \bibinfo{author}{\bibfnamefont{G.}~\bibnamefont{Seifert}}, \bibnamefont{and}
  \bibinfo{author}{\bibfnamefont{D.}~\bibnamefont{Tom\'anek}},
  \bibinfo{journal}{Phys. Rev. Lett.} \textbf{\bibinfo{volume}{108}},
  \bibinfo{pages}{156802} (\bibinfo{year}{2012}).

\bibitem[{\citenamefont{Hwang and {Das Sarma}}(2008{\natexlab{a}})}]{Sarma:100}
\bibinfo{author}{\bibfnamefont{E.~H.} \bibnamefont{Hwang}} \bibnamefont{and}
  \bibinfo{author}{\bibfnamefont{S.}~\bibnamefont{{Das Sarma}}},
  \bibinfo{journal}{Phys. Rev. B} \textbf{\bibinfo{volume}{77}},
  \bibinfo{pages}{235437} (\bibinfo{year}{2008}{\natexlab{a}}).

\bibitem[{\citenamefont{Price}(1984)}]{Price:BG}
\bibinfo{author}{\bibfnamefont{P.~J.} \bibnamefont{Price}},
  \bibinfo{journal}{Solid State Commun.} \textbf{\bibinfo{volume}{51}},
  \bibinfo{pages}{607} (\bibinfo{year}{1984}).

\bibitem[{\citenamefont{Stormer et~al.}(1990)\citenamefont{Stormer, Pfeiffer,
  Baldwin, and West}}]{West:ObservationOfBG}
\bibinfo{author}{\bibfnamefont{H.~L.} \bibnamefont{Stormer}},
  \bibinfo{author}{\bibfnamefont{L.~N.} \bibnamefont{Pfeiffer}},
  \bibinfo{author}{\bibfnamefont{K.~W.} \bibnamefont{Baldwin}},
  \bibnamefont{and} \bibinfo{author}{\bibfnamefont{K.~W.} \bibnamefont{West}},
  \bibinfo{journal}{Phys. Rev. B} \textbf{\bibinfo{volume}{41}},
  \bibinfo{pages}{1278} (\bibinfo{year}{1990}).

\bibitem[{\citenamefont{Efetov and Kim}(2010)}]{Kim:Controlling}
\bibinfo{author}{\bibfnamefont{D.~K.} \bibnamefont{Efetov}} \bibnamefont{and}
  \bibinfo{author}{\bibfnamefont{P.}~\bibnamefont{Kim}},
  \bibinfo{journal}{Phys. Rev. Lett.} \textbf{\bibinfo{volume}{105}},
  \bibinfo{pages}{256805} (\bibinfo{year}{2010}).

\bibitem[{\citenamefont{Hwang and {Das
  Sarma}}(2008{\natexlab{b}})}]{Sarma:Acoustic}
\bibinfo{author}{\bibfnamefont{E.~H.} \bibnamefont{Hwang}} \bibnamefont{and}
  \bibinfo{author}{\bibfnamefont{S.}~\bibnamefont{{Das Sarma}}},
  \bibinfo{journal}{Phys. Rev. B} \textbf{\bibinfo{volume}{77}},
  \bibinfo{pages}{115449} (\bibinfo{year}{2008}{\natexlab{b}}).

\bibitem[{\citenamefont{Castro et~al.}(2010)\citenamefont{Castro, Ochoa,
  Katsnelson, Gorbachev, Elias, Novoselov, Geim, and Guinea}}]{Guinea:Flexural}
\bibinfo{author}{\bibfnamefont{E.~V.} \bibnamefont{Castro}},
  \bibinfo{author}{\bibfnamefont{H.}~\bibnamefont{Ochoa}},
  \bibinfo{author}{\bibfnamefont{M.~I.} \bibnamefont{Katsnelson}},
  \bibinfo{author}{\bibfnamefont{R.~V.} \bibnamefont{Gorbachev}},
  \bibinfo{author}{\bibfnamefont{D.~C.} \bibnamefont{Elias}},
  \bibinfo{author}{\bibfnamefont{K.~S.} \bibnamefont{Novoselov}},
  \bibinfo{author}{\bibfnamefont{A.~K.} \bibnamefont{Geim}}, \bibnamefont{and}
  \bibinfo{author}{\bibfnamefont{F.}~\bibnamefont{Guinea}},
  \bibinfo{journal}{Phys. Rev. Lett.} \textbf{\bibinfo{volume}{105}},
  \bibinfo{pages}{266601} (\bibinfo{year}{2010}).

\bibitem[{\citenamefont{Kaasbjerg
  et~al.}(2012{\natexlab{b}})\citenamefont{Kaasbjerg, Thygesen, and
  Jacobsen}}]{Kaasbjerg:Unraveling}
\bibinfo{author}{\bibfnamefont{K.}~\bibnamefont{Kaasbjerg}},
  \bibinfo{author}{\bibfnamefont{K.~S.} \bibnamefont{Thygesen}},
  \bibnamefont{and} \bibinfo{author}{\bibfnamefont{K.~W.}
  \bibnamefont{Jacobsen}}, \bibinfo{journal}{Phys. Rev. B}
  \textbf{\bibinfo{volume}{85}}, \bibinfo{pages}{165440}
  (\bibinfo{year}{2012}{\natexlab{b}}).

\bibitem[{foo({\natexlab{a}})}]{footnote1}
\bibinfo{note}{The chemical potential of a 2DEG is related to the carrier
  density $n$ via $$ \mu(T) = k_\text{B}T \ln{\left( e^{\frac{n}{N_c}} - 1
  \right) } $$ where $N_c = g_s g_v m^* / 2 \pi\hbar^2 k_\text{B}T$ is the
  effective density of states at the band edge. For the valley-degenerate
  ($g_v=2$) conduction band of MoS$_2$, the effective density of states is $N_c
  \approx 4 \times 10^{10}\, T$~cm$^{-2}$ with $T$ measured in K. In the
  degenerate high-density limit $n \gg N_c$, the Fermi level becomes $E_F
  \approx 2\pi \hbar^2 n/g_sg_vm^* \sim 2.5\, n~\mathrm{meV}$ ($k_F=\sqrt{4\pi
  n/g_s g_v} \sim 0.02 \sqrt{n}\, \pi / a$) with $n$ measured in units of
  $10^{12}$~cm$^{-2}$.}

\bibitem[{\citenamefont{Das et~al.}(2008)\citenamefont{Das, Pisana,
  Chakraborty, Piscanec, Saha, Waghmare, Novoselov, Krishnamurthy, Geim,
  Ferrari et~al.}}]{Sood:Monitoring}
\bibinfo{author}{\bibfnamefont{A.}~\bibnamefont{Das}},
  \bibinfo{author}{\bibfnamefont{S.}~\bibnamefont{Pisana}},
  \bibinfo{author}{\bibfnamefont{B.}~\bibnamefont{Chakraborty}},
  \bibinfo{author}{\bibfnamefont{S.}~\bibnamefont{Piscanec}},
  \bibinfo{author}{\bibfnamefont{S.~K.} \bibnamefont{Saha}},
  \bibinfo{author}{\bibfnamefont{U.~V.} \bibnamefont{Waghmare}},
  \bibinfo{author}{\bibfnamefont{K.~S.} \bibnamefont{Novoselov}},
  \bibinfo{author}{\bibfnamefont{H.~R.} \bibnamefont{Krishnamurthy}},
  \bibinfo{author}{\bibfnamefont{A.~K.} \bibnamefont{Geim}},
  \bibinfo{author}{\bibfnamefont{A.~C.} \bibnamefont{Ferrari}},
  \bibnamefont{et~al.}, \bibinfo{journal}{Nature Nano.}
  \textbf{\bibinfo{volume}{3}}, \bibinfo{pages}{210} (\bibinfo{year}{2008}).

\bibitem[{\citenamefont{Chakraborty et~al.}(2012)\citenamefont{Chakraborty,
  Bera, Muthu, Bhowmick, Waghmare, and Sood}}]{Sood:Symmetry}
\bibinfo{author}{\bibfnamefont{B.}~\bibnamefont{Chakraborty}},
  \bibinfo{author}{\bibfnamefont{A.}~\bibnamefont{Bera}},
  \bibinfo{author}{\bibfnamefont{D.~V.~S.} \bibnamefont{Muthu}},
  \bibinfo{author}{\bibfnamefont{S.}~\bibnamefont{Bhowmick}},
  \bibinfo{author}{\bibfnamefont{U.~V.} \bibnamefont{Waghmare}},
  \bibnamefont{and} \bibinfo{author}{\bibfnamefont{A.~K.} \bibnamefont{Sood}},
  \bibinfo{journal}{Phys. Rev. B} \textbf{\bibinfo{volume}{85}},
  \bibinfo{pages}{161403} (\bibinfo{year}{2012}).

\bibitem[{\citenamefont{Kawamura and {Das Sarma}}(1990)}]{Sarma:Hetero1}
\bibinfo{author}{\bibfnamefont{T.}~\bibnamefont{Kawamura}} \bibnamefont{and}
  \bibinfo{author}{\bibfnamefont{S.}~\bibnamefont{{Das Sarma}}},
  \bibinfo{journal}{Phys. Rev. B} \textbf{\bibinfo{volume}{42}},
  \bibinfo{pages}{3725} (\bibinfo{year}{1990}).

\bibitem[{\citenamefont{Kn{\"a}bchen}(1997)}]{Knabchen:SurfaceAcousticII}
\bibinfo{author}{\bibfnamefont{A.}~\bibnamefont{Kn{\"a}bchen}},
  \bibinfo{journal}{Phys. Rev. B} \textbf{\bibinfo{volume}{55}},
  \bibinfo{pages}{6701} (\bibinfo{year}{1997}).

\bibitem[{\citenamefont{Mori and Ando}(1989)}]{Ando:ElphHetero}
\bibinfo{author}{\bibfnamefont{N.}~\bibnamefont{Mori}} \bibnamefont{and}
  \bibinfo{author}{\bibfnamefont{T.}~\bibnamefont{Ando}},
  \bibinfo{journal}{Phys. Rev. B} \textbf{\bibinfo{volume}{40}},
  \bibinfo{pages}{6175} (\bibinfo{year}{1989}).

\bibitem[{\citenamefont{Chen et~al.}(2008)\citenamefont{Chen, Jang, Xiao,
  Ishigami, and Fuhrer}}]{Fuhrer:GrapheneSiO2}
\bibinfo{author}{\bibfnamefont{J.-H.} \bibnamefont{Chen}},
  \bibinfo{author}{\bibfnamefont{C.}~\bibnamefont{Jang}},
  \bibinfo{author}{\bibfnamefont{S.}~\bibnamefont{Xiao}},
  \bibinfo{author}{\bibfnamefont{M.}~\bibnamefont{Ishigami}}, \bibnamefont{and}
  \bibinfo{author}{\bibfnamefont{M.~S.} \bibnamefont{Fuhrer}},
  \bibinfo{journal}{Nature Nano.} \textbf{\bibinfo{volume}{3}},
  \bibinfo{pages}{206} (\bibinfo{year}{2008}).

\bibitem[{\citenamefont{Fratini and Guinea}(2008)}]{Guinea:SubstrateLimited}
\bibinfo{author}{\bibfnamefont{S.}~\bibnamefont{Fratini}} \bibnamefont{and}
  \bibinfo{author}{\bibfnamefont{F.}~\bibnamefont{Guinea}},
  \bibinfo{journal}{Phys. Rev. B} \textbf{\bibinfo{volume}{77}},
  \bibinfo{pages}{195415} (\bibinfo{year}{2008}).

\bibitem[{\citenamefont{Konar et~al.}(2010)\citenamefont{Konar, Fang, and
  Jena}}]{Jena:HighKappa}
\bibinfo{author}{\bibfnamefont{A.}~\bibnamefont{Konar}},
  \bibinfo{author}{\bibfnamefont{T.}~\bibnamefont{Fang}}, \bibnamefont{and}
  \bibinfo{author}{\bibfnamefont{D.}~\bibnamefont{Jena}},
  \bibinfo{journal}{Phys. Rev. B} \textbf{\bibinfo{volume}{82}},
  \bibinfo{pages}{115452} (\bibinfo{year}{2010}).

\bibitem[{\citenamefont{Zhang et~al.}(2013)\citenamefont{Zhang, Xu, Badalyan,
  and Peeters}}]{Peeters:PiezoelectricSurface}
\bibinfo{author}{\bibfnamefont{S.~H.} \bibnamefont{Zhang}},
  \bibinfo{author}{\bibfnamefont{W.}~\bibnamefont{Xu}},
  \bibinfo{author}{\bibfnamefont{S.~M.} \bibnamefont{Badalyan}},
  \bibnamefont{and} \bibinfo{author}{\bibfnamefont{F.~M.}
  \bibnamefont{Peeters}}, \bibinfo{journal}{Phys. Rev. B}
  \textbf{\bibinfo{volume}{87}}, \bibinfo{pages}{075443}
  (\bibinfo{year}{2013}).

\bibitem[{\citenamefont{Min et~al.}(2012)\citenamefont{Min, Hwang, and {Das
  Sarma}}}]{Sarma:Interplay}
\bibinfo{author}{\bibfnamefont{H.}~\bibnamefont{Min}},
  \bibinfo{author}{\bibfnamefont{E.~H.} \bibnamefont{Hwang}}, \bibnamefont{and}
  \bibinfo{author}{\bibfnamefont{S.}~\bibnamefont{{Das Sarma}}},
  \bibinfo{journal}{Phys. Rev. B} \textbf{\bibinfo{volume}{86}},
  \bibinfo{pages}{085307} (\bibinfo{year}{2012}).

\bibitem[{\citenamefont{Leb{\`e}gue and Eriksson}(2009)}]{Eriksson:2D}
\bibinfo{author}{\bibfnamefont{S.}~\bibnamefont{Leb{\`e}gue}} \bibnamefont{and}
  \bibinfo{author}{\bibfnamefont{O.}~\bibnamefont{Eriksson}},
  \bibinfo{journal}{Phys. Rev. B} \textbf{\bibinfo{volume}{79}},
  \bibinfo{pages}{115409} (\bibinfo{year}{2009}).

\bibitem[{\citenamefont{Cheiwchanchamnangij and
  Lambrecht}(2012)}]{Lambrecht:Quasiparticle}
\bibinfo{author}{\bibfnamefont{T.}~\bibnamefont{Cheiwchanchamnangij}}
  \bibnamefont{and} \bibinfo{author}{\bibfnamefont{W.~R.~L.}
  \bibnamefont{Lambrecht}}, \bibinfo{journal}{Phys. Rev. B}
  \textbf{\bibinfo{volume}{85}}, \bibinfo{pages}{205302}
  (\bibinfo{year}{2012}).

\bibitem[{\citenamefont{Zhu et~al.}(2011)\citenamefont{Zhu, Cheng, and
  Schwingenschl{\"o}gl}}]{Schwing:GiantSO}
\bibinfo{author}{\bibfnamefont{Z.~Y.} \bibnamefont{Zhu}},
  \bibinfo{author}{\bibfnamefont{Y.~C.} \bibnamefont{Cheng}}, \bibnamefont{and}
  \bibinfo{author}{\bibfnamefont{U.}~\bibnamefont{Schwingenschl{\"o}gl}},
  \bibinfo{journal}{Phys. Rev. B} \textbf{\bibinfo{volume}{84}},
  \bibinfo{pages}{153402} (\bibinfo{year}{2011}).

\bibitem[{\citenamefont{Chen et~al.}(2007)\citenamefont{Chen, Ma, Cao, Zhang,
  and Zhang}}]{Zhang:2DSemiconductorSOC}
\bibinfo{author}{\bibfnamefont{L.}~\bibnamefont{Chen}},
  \bibinfo{author}{\bibfnamefont{Z.}~\bibnamefont{Ma}},
  \bibinfo{author}{\bibfnamefont{J.~C.} \bibnamefont{Cao}},
  \bibinfo{author}{\bibfnamefont{T.~Y.} \bibnamefont{Zhang}}, \bibnamefont{and}
  \bibinfo{author}{\bibfnamefont{C.}~\bibnamefont{Zhang}},
  \bibinfo{journal}{Appl. Phys. Lett.} \textbf{\bibinfo{volume}{91}},
  \bibinfo{pages}{102115} (\bibinfo{year}{2007}).

\bibitem[{\citenamefont{Biswas and Ghosh}(2013)}]{Ghosh:SO2DEG}
\bibinfo{author}{\bibfnamefont{T.}~\bibnamefont{Biswas}} \bibnamefont{and}
  \bibinfo{author}{\bibfnamefont{T.~K.} \bibnamefont{Ghosh}},
  \bibinfo{journal}{J. Phys.: Condens. Matter} \textbf{\bibinfo{volume}{25}},
  \bibinfo{pages}{035301} (\bibinfo{year}{2013}).

\bibitem[{\citenamefont{Kawamura and {Das Sarma}}(1992)}]{Sarma:Hetero2}
\bibinfo{author}{\bibfnamefont{T.}~\bibnamefont{Kawamura}} \bibnamefont{and}
  \bibinfo{author}{\bibfnamefont{S.}~\bibnamefont{{Das Sarma}}},
  \bibinfo{journal}{Phys. Rev. B} \textbf{\bibinfo{volume}{45}},
  \bibinfo{pages}{3612} (\bibinfo{year}{1992}).

\bibitem[{\citenamefont{Mortensen et~al.}(2005)\citenamefont{Mortensen, Hansen,
  and Jacobsen}}]{GPAW}
\bibinfo{author}{\bibfnamefont{J.~J.} \bibnamefont{Mortensen}},
  \bibinfo{author}{\bibfnamefont{L.~B.} \bibnamefont{Hansen}},
  \bibnamefont{and} \bibinfo{author}{\bibfnamefont{K.~W.}
  \bibnamefont{Jacobsen}}, \bibinfo{journal}{Phys. Rev. B}
  \textbf{\bibinfo{volume}{71}}, \bibinfo{pages}{035109}
  (\bibinfo{year}{2005}).

\bibitem[{\citenamefont{Larsen et~al.}(2009)\citenamefont{Larsen, Vanin,
  Mortensen, Thygesen, and Jacobsen}}]{GPAW1}
\bibinfo{author}{\bibfnamefont{A.~H.} \bibnamefont{Larsen}},
  \bibinfo{author}{\bibfnamefont{M.}~\bibnamefont{Vanin}},
  \bibinfo{author}{\bibfnamefont{J.~J.} \bibnamefont{Mortensen}},
  \bibinfo{author}{\bibfnamefont{K.~S.} \bibnamefont{Thygesen}},
  \bibnamefont{and} \bibinfo{author}{\bibfnamefont{K.~W.}
  \bibnamefont{Jacobsen}}, \bibinfo{journal}{Phys. Rev. B}
  \textbf{\bibinfo{volume}{80}}, \bibinfo{pages}{195112}
  (\bibinfo{year}{2009}).

\bibitem[{\citenamefont{Enkovaara et~al.}(2010)\citenamefont{Enkovaara,
  Rostgaard, Mortensen, Chen, Dulak, Ferrighi, Gavnholt, Glinsvad, Haikola,
  Hansen et~al.}}]{GPAW2}
\bibinfo{author}{\bibfnamefont{J.~.} \bibnamefont{Enkovaara}},
  \bibinfo{author}{\bibfnamefont{C.}~\bibnamefont{Rostgaard}},
  \bibinfo{author}{\bibfnamefont{J.~J.} \bibnamefont{Mortensen}},
  \bibinfo{author}{\bibfnamefont{J.}~\bibnamefont{Chen}},
  \bibinfo{author}{\bibfnamefont{M.}~\bibnamefont{Dulak}},
  \bibinfo{author}{\bibfnamefont{L.}~\bibnamefont{Ferrighi}},
  \bibinfo{author}{\bibfnamefont{J.}~\bibnamefont{Gavnholt}},
  \bibinfo{author}{\bibfnamefont{C.}~\bibnamefont{Glinsvad}},
  \bibinfo{author}{\bibfnamefont{V.}~\bibnamefont{Haikola}},
  \bibinfo{author}{\bibfnamefont{H.~A.} \bibnamefont{Hansen}},
  \bibnamefont{et~al.}, \bibinfo{journal}{J. Phys.: Condens. Matter}
  \textbf{\bibinfo{volume}{22}}, \bibinfo{pages}{253202}
  (\bibinfo{year}{2010}).

\bibitem[{\citenamefont{Mahan}(2010)}]{Mahan}
\bibinfo{author}{\bibfnamefont{G.~D.} \bibnamefont{Mahan}},
  \emph{\bibinfo{title}{Many-particle Physics}} (\bibinfo{publisher}{Springer},
  \bibinfo{year}{2010}), \bibinfo{edition}{3rd} ed.

\bibitem[{foo({\natexlab{b}})}]{footnote2}
\bibinfo{note}{The electron-phonon coupling has been calculated with DFT-LDA
  using a $17 \times 17$ supercell, a double-zeta polarized (DZP) basis for the
  electronic Bloch states, and 5~{\AA} of vacuum between the MoS$_2$ sheet and
  the cell boundaries in the direction perpendicular to the sheet. In this
  direction non-periodic boundary conditions must be applied in order to avoid
  spurious interlayer contributions to the long-range part of the
  electron-phonon interaction in the long-wavelength limit (which are present
  when periodic boundary conditions are applied). A real-space cutoff of
  $r_\text{cut} = 6.0$~{\AA} has been used to separate out the short-range
  deformation potential and long-range piezoelectric interactions.}

\bibitem[{\citenamefont{Madelung}(1996)}]{Madelung}
\bibinfo{author}{\bibfnamefont{O.}~\bibnamefont{Madelung}},
  \emph{\bibinfo{title}{Introduction to Solid State Physics}}
  (\bibinfo{publisher}{Springer}, \bibinfo{address}{Berlin},
  \bibinfo{year}{1996}).

\bibitem[{\citenamefont{Sai and Mele}(2003)}]{Mele:NTPiezo}
\bibinfo{author}{\bibfnamefont{N.}~\bibnamefont{Sai}} \bibnamefont{and}
  \bibinfo{author}{\bibfnamefont{E.~J.} \bibnamefont{Mele}},
  \bibinfo{journal}{Phys. Rev. B} \textbf{\bibinfo{volume}{68}},
  \bibinfo{pages}{241405} (\bibinfo{year}{2003}).

\bibitem[{foo({\natexlab{c}})}]{footnote3}
\bibinfo{note}{Due to a different choice of the primitive lattice vectors in
  App.~\ref{app:piezo} and our first-principles calculations (see
  Fig.~\ref{fig:mos2}), the angular dependencies of the matrix elements for the
  TA and LA phonons in Eq.~\eqref{eq:A_theta} are interchanged compared to the
  ones in Fig.~\ref{fig:M}.}

\bibitem[{\citenamefont{Duerloo et~al.}(2012)\citenamefont{Duerloo, Ong, and
  Reed}}]{Reed:Piezoelectricity}
\bibinfo{author}{\bibfnamefont{K.-A.~N.} \bibnamefont{Duerloo}},
  \bibinfo{author}{\bibfnamefont{M.~T.} \bibnamefont{Ong}}, \bibnamefont{and}
  \bibinfo{author}{\bibfnamefont{E.~J.} \bibnamefont{Reed}},
  \bibinfo{journal}{J. Phys. Chem. Lett.} \textbf{\bibinfo{volume}{3}},
  \bibinfo{pages}{2871} (\bibinfo{year}{2012}).

\bibitem[{foo({\natexlab{d}})}]{footnote4}
\bibinfo{note}{Ideally, this only holds for $\bk$ oriented along high-symmetry
  directions of the hexagonal lattice where $\abs{A_\lambda}^2$ is an even
  function of $\theta_{\bk,\bk'}$ such that the angular integration of the
  $\cos\theta_{\bk,\bk'}$ factor in Eq.~\eqref{eq:tau_acoustic} vanishes.}

\bibitem[{\citenamefont{Hybertsen and Louie}(1987)}]{Louie:Dielectric}
\bibinfo{author}{\bibfnamefont{M.~S.} \bibnamefont{Hybertsen}}
  \bibnamefont{and} \bibinfo{author}{\bibfnamefont{S.~G.} \bibnamefont{Louie}},
  \bibinfo{journal}{Phys. Rev. B} \textbf{\bibinfo{volume}{35}},
  \bibinfo{pages}{5585} (\bibinfo{year}{1987}).

\bibitem[{\citenamefont{Ando et~al.}(1982)\citenamefont{Ando, Fowler, and
  Stern}}]{Stern:2D}
\bibinfo{author}{\bibfnamefont{T.}~\bibnamefont{Ando}},
  \bibinfo{author}{\bibfnamefont{A.~B.} \bibnamefont{Fowler}},
  \bibnamefont{and} \bibinfo{author}{\bibfnamefont{F.}~\bibnamefont{Stern}},
  \bibinfo{journal}{Rev. Mod. Phys.} \textbf{\bibinfo{volume}{54}},
  \bibinfo{pages}{437} (\bibinfo{year}{1982}).

\bibitem[{\citenamefont{Maldague}(1978)}]{Maldague:ManyBody}
\bibinfo{author}{\bibfnamefont{P.~F.} \bibnamefont{Maldague}},
  \bibinfo{journal}{Surf. Sci.} \textbf{\bibinfo{volume}{73}},
  \bibinfo{pages}{296} (\bibinfo{year}{1978}).

\bibitem[{\citenamefont{Flensberg and Hu}(1995)}]{Flensberg:Plasmon}
\bibinfo{author}{\bibfnamefont{K.}~\bibnamefont{Flensberg}} \bibnamefont{and}
  \bibinfo{author}{\bibfnamefont{B.~Y.} \bibnamefont{Hu}},
  \bibinfo{journal}{Phys. Rev. B} \textbf{\bibinfo{volume}{52}},
  \bibinfo{pages}{14796} (\bibinfo{year}{1995}).

\bibitem[{\citenamefont{Yan et~al.}(2011)\citenamefont{Yan, Mortensen,
  Jacobsen, and Thygesen}}]{Jun:Response}
\bibinfo{author}{\bibfnamefont{J.}~\bibnamefont{Yan}},
  \bibinfo{author}{\bibfnamefont{J.~J.} \bibnamefont{Mortensen}},
  \bibinfo{author}{\bibfnamefont{K.~W.} \bibnamefont{Jacobsen}},
  \bibnamefont{and} \bibinfo{author}{\bibfnamefont{K.~S.}
  \bibnamefont{Thygesen}}, \bibinfo{journal}{Phys. Rev. B}
  \textbf{\bibinfo{volume}{83}}, \bibinfo{pages}{245122}
  (\bibinfo{year}{2011}).

\bibitem[{\citenamefont{Walukiewicz et~al.}(1984)\citenamefont{Walukiewicz,
  Ruda, Lagowski, and Gatos}}]{Gatos:Heterostructures}
\bibinfo{author}{\bibfnamefont{W.}~\bibnamefont{Walukiewicz}},
  \bibinfo{author}{\bibfnamefont{H.~E.} \bibnamefont{Ruda}},
  \bibinfo{author}{\bibfnamefont{J.}~\bibnamefont{Lagowski}}, \bibnamefont{and}
  \bibinfo{author}{\bibfnamefont{H.~C.} \bibnamefont{Gatos}},
  \bibinfo{journal}{Phys. Rev. B} \textbf{\bibinfo{volume}{30}},
  \bibinfo{pages}{4571} (\bibinfo{year}{1984}).

\bibitem[{foo({\natexlab{e}})}]{footnote5}
\bibinfo{note}{The field is here applied in the direction of the in-plane
  projection of the Mo-S bonds.}

\bibitem[{\citenamefont{Song and Dery}(2013)}]{Dery:SymmetryBased}
\bibinfo{author}{\bibfnamefont{Y.}~\bibnamefont{Song}} \bibnamefont{and}
  \bibinfo{author}{\bibfnamefont{H.}~\bibnamefont{Dery}}
  (\bibinfo{year}{2013}), \bibinfo{note}{arXiv:1302.3627}.

\bibitem[{\citenamefont{Giustino et~al.}(2007)\citenamefont{Giustino, Cohen,
  and Louie}}]{Louie:e-ph}
\bibinfo{author}{\bibfnamefont{F.}~\bibnamefont{Giustino}},
  \bibinfo{author}{\bibfnamefont{M.~L.} \bibnamefont{Cohen}}, \bibnamefont{and}
  \bibinfo{author}{\bibfnamefont{S.~G.} \bibnamefont{Louie}},
  \bibinfo{journal}{Phys. Rev. B} \textbf{\bibinfo{volume}{76}},
  \bibinfo{pages}{165108} (\bibinfo{year}{2007}).

\end{thebibliography}
\end{document}